\documentclass[twoside,12pt]{article}
\usepackage{epsfig}

\def\ga{\,\hbox{\hbox{$ > $}\kern -0.8em \lower
     1.0ex\hbox{$\sim$}}\,}
\def\la{\,\hbox{\hbox{$ < $}\kern -0.8em \lower
     1.0ex\hbox{$\sim$}}\,}
\def\eck#1{\left\lbrack #1 \right\rbrack}
\def\rund#1{\left( #1 \right)}
\def\ave#1{\langle #1 \rangle}
\newcommand{\be}{\begin{equation}}
\newcommand{\ee}{\end{equation}}
\newcommand{\bea}{\begin{eqnarray}}
\newcommand{\eea}{\end{eqnarray}}

\begin{document}

\title{ \vspace{1cm} Theory of core-collapse supernovae}
\author{H.-Th. Janka$^1$, K. Langanke$^{2,3}$, A. Marek$^1$, 
G. Mart\'{\i}nez-Pinedo$^2$, B. M\"uller$^1$ \\
$^1$ Max-Planck Institut f\"ur Astrophysik, Garching, Germany \\
$^2$Gesellschaft f\"ur Schwerionenforschung, Darmstadt, Germany \\
$^3$Institut f\"ur Kernphysik, Technische Universit\"at Darmstadt, Germany}
\maketitle

\begin{abstract}
Advances in our understanding and the modeling of stellar core-collapse 
and supernova explosions over the past 15 years are reviewed, concentrating
on the evolution of hydrodynamical simulations, the description of weak
interactions and nuclear equation of state effects, and new insights into
the nucleosynthesis occurring in the early phases of the explosion, in
particular the neutrino-p process. The latter is enabled by the 
proton-richness of the early ejecta, which was discovered because of
significant progress has been made in the treatment of neutrino transport
and weak interactions. This progress
has led to a new generation of sophisticated Newtonian 
and relativistic hydrodynamics simulations in spherical symmetry. Based
on these, it is now clear that the prompt bounce-shock
mechanism is not the driver of supernova explosions, and that the delayed
neutrino-heating mechanism can produce explosions without the aid of
multi-dimensional processes only if the progenitor star has an 
ONeMg core inside a very dilute He-core, i.e., has a mass in the 
8--10$\,M_\odot$ range. Hydrodynamic instabilities
of various kinds have indeed been recognized to occur in the supernova
core and to be of potential importance for the explosion. Neutrino-driven
explosions, however, have been seen in two-dimensional simulations with
sophisticated neutrino transport so far only when the star has a small
iron core and low density in the surrounding shells as being found
in stars near 10--11$\,M_\odot$. The explosion mechanism of more massive 
progenitors is still a puzzle. It might involve effects of three-dimensional
hydrodynamics or might point to the relevance of rapid rotation and
magnetohydrodynamics, or to still incompletely explored properties of 
neutrinos and the high-density equation of state.
\end{abstract}

Hardly any other astrophysical event is as complex and physically
diverse as the death of massive stars in a gravitational collapse and 
subsequent supernova explosion. All four known forces of nature 
are involved and play an important role in extreme regimes of
conditions. Relativistic plasma dynamics in a strong gravitational
field sets the stage, weak interactions govern the energy 
and lepton number loss of the system via the transport of neutrinos
from regions of very high opacities to the free-streaming regime,
electromagnetic and strong interactions determine the thermodynamic
properties, and nuclear and weak interactions change 
the composition of the stellar gas.
Supernova explosions thus offer a fascinating playground of physics
on most different scales of length and time and also provide a 
testbed for new or exotic phenomena. Naturally, these spectacular
astrophysical events have attracted --- and have deserved --- the interest 
and attention of researchers with very different backgrounds. To the
advantage of the field, also Hans Bethe has preserved for many years
his interest in the large diversity of physics problems posed by
supernovae. 

In this article we will discuss some of the progress that has been
made in the simulations of stellar core collapse and supernova 
explosions on the one hand, and in the description of the microphysics
input of such models on the other, since Hans Bethe published his grand
review article in 1990 \cite{Bethe90}. 
The field is very broad and has experienced
also a great expansion after long-soft gamma-ray bursts have been 
discovered to be linked to extraordinarily energetic explosions of
massive stars. A single article can therefore hardly cover all
interesting developments satisfactorily. For recent reviews of the 
theory and observations of stellar collapse events, focussing on 
different aspects, see 
Refs.~\cite{Mezzacappa05,Kotake06,WooJan05,WooBloo06}.

We will concentrate here on some 
selected topics, chosen according to our preference but also by
their links to contributions from Hans Bethe and his coworkers.
After giving a summary of the physical processes and 
evolutionary stages that lead from stellar core collapse to 
supernova explosion, we will briefly review the current status 
of modeling supernovae and of our present understanding of 
the explosion mechanism, in particular we will also highlight 
the open ends of the current research. In the
subsequent parts of our article we will
then concentrate on nuclear and weak interaction physics that
plays an important role for developing better numerical models,
and we will report new insights brought by such model
improvements about the nucleosynthesis during supernova explosions.

\begin{figure}
\center{\includegraphics[width=0.85\textwidth]{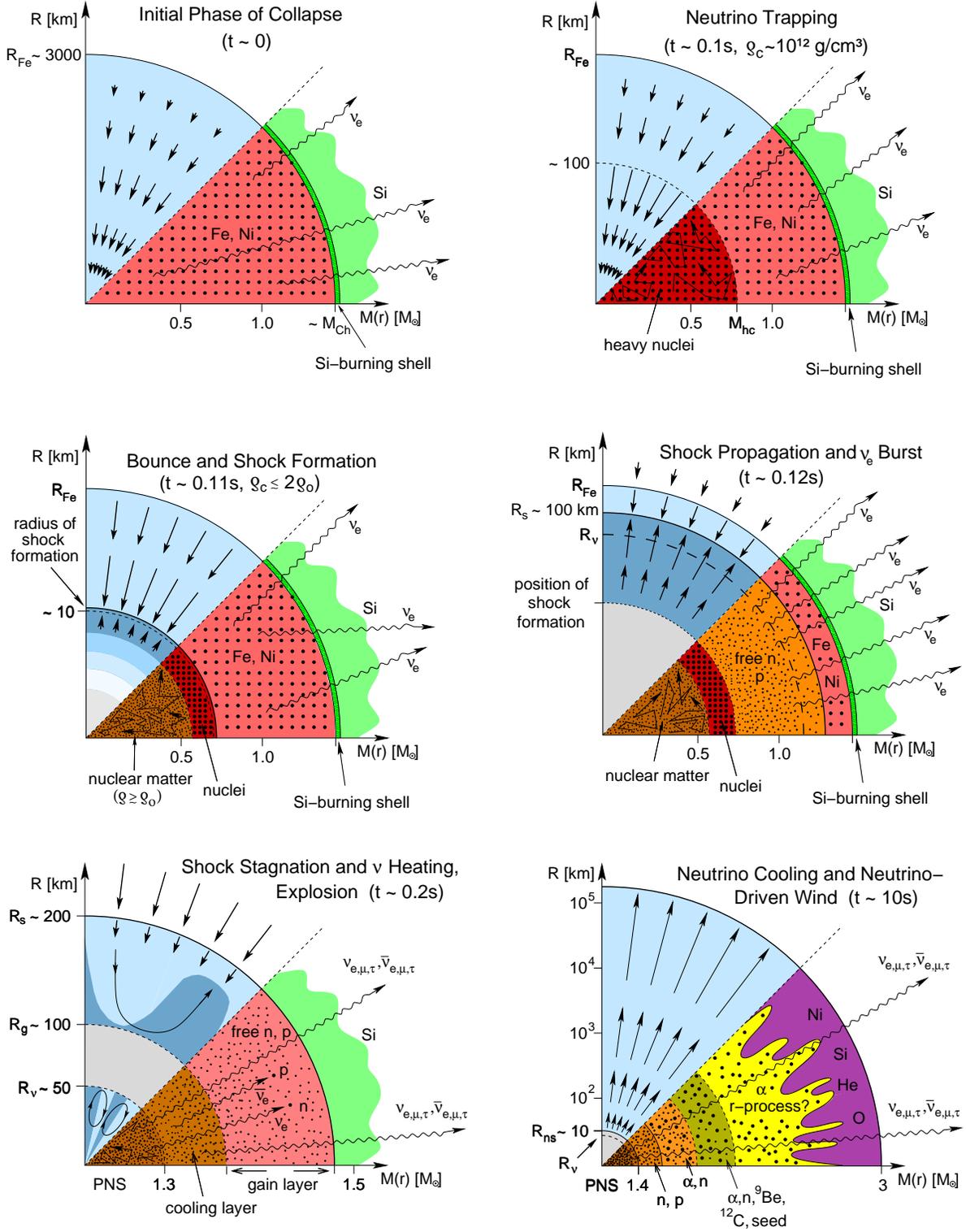}}
%
\caption{\small
Schematic representation of the evolutionary stages from stellar
core collapse through the onset of the supernova explosion to the
neutrino-driven wind during the neutrino-cooling phase of the
proto-neutron star (PNS). The panels display the dynamical conditions
in their upper half, with arrows representing velocity vectors.
The nuclear composition as well as the nuclear and weak processes
are indicated in the lower half of each panel.
The horizontal axis gives mass information.
$M_{\mathrm{Ch}}$ means the Chandrasekhar mass and $M_{\mathrm{hc}}$
the mass of the subsonically collapsing, homologous inner core.
The vertical axis shows corresponding radii, with $R_{\mathrm{Fe}}$,
$R_{\mathrm{s}}$, $R_{\mathrm{g}}$, $R_{\mathrm{ns}}$, and
$R_{\mathrm{\nu}}$ being the iron core radius, shock radius, gain
radius, neutron star radius, and neutrinosphere, respectively.
The PNS has maximum densities $\rho$ above
the saturation density of nuclear matter ($\rho_0$).}
\label{fig:snphases}
\end{figure}

\section{The current picture of stellar collapse and explosion}

At the end of hydrostatic burning, a massive star consists of concentric
shells that are the relics of its previous burning phases (hydrogen,
helium, carbon, neon, oxygen, silicon). Iron is the final stage of nuclear
fusion in hydrostatic burning, as the synthesis of any heavier element
from lighter elements does not release energy; rather, energy must be used 
up. When the iron core, formed in the center of the massive star, grows
by silicon shell burning to a mass around the Chandrasekhar mass limit of
about 1.44 solar masses, electron degeneracy pressure cannot longer
stabilize the core and it collapses. This starts
what is called a core-collapse supernova in course of which 
the star explodes and parts of the star's heavy-element core and of its
outer shells are ejected into the Interstellar Medium.


The onset of the collapse and the infall dynamics are very 
sensitive to the entropy and to the number of leptons
per baryon, $Y_e$ \cite{BBAL}.
In turn, these two quantities are mainly determined
by weak interaction processes, electron capture and $\beta$ decay.
First, in the early stage of the collapse, $Y_e$ decreases by 
electron capture on (Fe-peak) nuclei, reactions which are energetically
favorable when the electrons have Fermi energies of a few MeV at
the densities involved. This reduces the increase of the electron
pressure with density, thus accelerating the collapse, and shifts
the distribution of nuclei present in the core to more neutron-rich
material (Fig.~\ref{fig:snphases}, upper left panel). 
Second, many of the nuclei present can also $\beta$ decay.
While this process is quite unimportant compared to electron capture for
initial $Y_e$ values around 0.5, it becomes increasingly competative for
neutron-rich nuclei due to an increase in phase space related to larger
$Q_\beta$ values.

Electron capture, $\beta$ decay, and partial photodisintegration of 
iron-group nuclei to alpha particles cost the core
energy and reduce its electron density. As a consequence, the collapse
is accelerated. An important change in the physics of the collapse
occurs, as the density reaches $\rho_{\rm trap}\approx 10^{12}\,$g/cm$^3$
(Fig.~\ref{fig:snphases}, upper right panel). 
Then neutrinos are essentially trapped in the core, because 
their diffusion time (due to coherent conservative scattering on nuclei)
becomes larger than the collapse time \cite{Bethe90}.
After neutrino trapping, the collapse proceeds essentially homologously
\cite{Goldreich},
until nuclear densities ($\rho_0\approx10^{14}$ g/cm$^3$)
are reached. Since nuclear matter has a much lower compressibility, the
homologous core decelerates and bounces in response to the increased
nuclear matter pressure. This drives a shock wave
into the outer core, i.e. the region of the iron core which lies 
outside of the homologous core and 
in the meantime has continued to fall inwards at
supersonic speed. 

\subsection{Neutrino-driven explosions: spherically symmetric models}
\label{sec:1Dmodels}

The core bounce with the formation of a shock wave 
is the starting point of a sequence of events that ultimately 
triggers a supernova explosion (Fig.~\ref{fig:snphases}, 
middle left panel), but the exact mechanism of the explosion
and the crucial ingredients of this physically appealing scenario are
still uncertain and controversial.
If the shock wave is strong enough not only to stop
the collapse, but also to explode the outer burning shells of the star,
one speaks about the `prompt mechanism'. However, it appears as if the
energy available to the shock is not sufficient, and the shock
uses up its energy in the outer core mostly by the dissociation of
heavy nuclei into nucleons. This change in composition results in
even more energy loss, because the electron capture rate on free protons
is significantly larger than on neutron-rich nuclei due to the higher
$Q$-values of the latter. 
A large fraction of the neutrinos produced by these
electron captures behind the shock leave the star quickly in what is
called the neutrino burst at shock break-out, carrying away energy.
This leads to further neutronization of the matter. The shock is
weakened so much that it finally stalls and turns into an 
accretion shock at a radius between 100 and 200$\,$km, i.e., the 
matter downstream of the shock has negative velocities and continues
falling inward (Fig.~\ref{fig:snphases}, middle right panel). 
All state-of-the-art simulations of stellar core collapse
performed in Newtonian gravity \cite{Thomp03,Rampp00}, with an 
approximative treatment of general relativity 
\cite{Buras03,Buras06a,Buras06b}, in full general relativity
\cite{Mezzacappa01,Liebendoerfer01,Liebendoerfer03,Liebendoerfer04,Liebendoerfer05}, and
with stiff or soft nuclear equations of state currently available
for core-collapse simulations \cite{Janka04,Janka05a,Janka05b,Sumiyoshi05} 
(see also Sect.~\ref{sec:eos}) agree with the models of the
1980's and 1990's (e.g., \cite{Wilson86,Myra89,Bruenn93,Swesty94}) that 
the prompt shock is unable to trigger supernova 
explosions\footnote{Successful prompt explosions were found
in case of extremely small stellar iron cores \cite{Baron90} or
a very soft nuclear equation of state \cite{Baron87}, but 
it is possible that their true reason was the parameterized
equation of state or the
approximative treatment of neutrino transport used in the 
simulations.}.

After the core bounce, a compact remnant begins to form at the center
of the collapsing star, rapidly growing by the accretion of infalling 
stellar material until the explosion sets in. 
This nascent remnant --- the proto-neutron star --- will evolve
to a neutron star or may eventually collapse to a black hole,
depending on whether the progenitor star had a mass below or 
above roughly 25 solar masses. 
The newly born neutron star is initially still proton-rich and
contains a large number of degenerate electrons and neutrinos. The
latter are trapped because their mean free paths in the dense matter
are significantly shorter than the radius of the neutron star. It
takes a fraction of a second for the trapped neutrinos to diffuse
out \cite{Burrows90} (Fig.~\ref{fig:snphases}, lower panels). 
On their way to the neutrinosphere, the
neutrinos are down-scattered in energy space, thus converting their
initially high degeneracy energy to thermal energy of the stellar
medium \cite{BurLat86}. The further cooling of the hot
interior of the proto-neutron star then proceeds by neutrino-pair 
production and diffusive loss of neutrinos
of all three lepton flavors. After several tens of seconds the 
compact remnant becomes transparent to neutrinos and the neutrino
luminosity drops significantly \cite{Burrows88}.

In the explosion scenario by the `delayed neutrino-heating mechanism', 
the stalled shock wave can be revived by the neutrinos streaming
off the neutrinosphere. These neutrinos carry most
of the energy set free in the gravitational collapse of the stellar
core \cite{Burrows90} and deposit some of their energy in the layers
between the nascent neutron star and the stalled shock front mainly
by charged-current $\nu_e$ and $\bar\nu_e$ captures on free 
nucleons \cite{Wilson,BetWil85},
\begin{eqnarray}
\nu_e + n &\longrightarrow& e^- + p\ , \label{eq:nuabs}\\
\bar\nu_e + p &\longrightarrow& e^+ + n  \label{eq:nubarabs}
\end{eqnarray}
(Fig.~\ref{fig:snphases}, lower left panel).
This neutrino heating increases the pressure 
behind the shock and the heated layers begin to expand, creating
between shock front and neutron star surface a region of low density
but rather high temperature, the so-called
`hot bubble' \cite{Col89}. The persistent energy input by neutrinos
keeps the pressure high in this region and drives the shock outwards
again, eventually leading to a supernova explosion. This may take a
few 100 ms and requires that during this time interval a few percent 
of the radiated neutrino energy 
(or 10--20\% of the energy of electron neutrinos and antineutrinos)
are converted to thermal energy of nucleons, leptons, and photons.
The canonical explosion energy of a supernova 
is less than one percent of the total gravitational binding energy 
lost by the nascent neutron star in neutrinos.

The success of the delayed supernova mechanism turned out to be
sensitive to a delicate competition of neutrino cooling between
the neutrinosphere and the so-called `gain radius' on the one hand,
and neutrino heating between the gain radius und the shock on the
other (Fig.~\ref{fig:snphases}, lower left panel). The gain radius
is defined as the radial position where the neutrino heating rate
per nucleon,
\begin{equation}
Q_{\nu}^+ \approx  110\cdot \rund{
{L_{\nu_e,52}\langle\epsilon_{\nu_e,15}^2
\rangle\over r_7^2\,\,\ave{\mu}_{\nu_e}
}
\,Y_n\,+\,
{L_{\bar\nu_e,52}\langle\epsilon_{\bar\nu_e,15}^2\rangle\over r_7^2\,\,
\ave{\mu}_{\bar\nu_e}}\,Y_p }
\quad
\left\lbrack {{\rm MeV}\over {\rm s}\cdot N}\right\rbrack \ , 
\label{eq:nuheating}
\end{equation}
and the neutrino cooling rate per nucleon,
\begin{equation}
Q^-_{\nu} \approx 145\,
\rund{{k_{\mathrm{B}}T\over 2\,{\mathrm{MeV}}}}^{\! 6} 
\ \ \eck{{{\mathrm{MeV}}\over
{\mathrm{s}}\cdot N}} \ ,
\label{eq:nucooling}
\end{equation}
become equal. In the latter expression we have used the assumptions
that the sum of neutron and proton abundances is unity,
$Y_n+Y_p = 1$, and that the electron and positron degeneracy parameters,
$\eta_{e^\pm} = \mu_{e^\pm}/(k_{\mathrm{B}}T)$, are small:
$\eta_{e^-} = -\eta_{e^+}\equiv \eta_e\approx 0$.
The latter approximation is good in the shock-heated layers because the
electron number fraction $Y_e = n_e/n_{\mathrm{b}}$ ($n_e$ and 
$n_{\mathrm{b}}$ being the electron and baryon number density,
respectively) and thus the electron degeneracy is rather low
and $e^\pm$ pairs are abundant. In Eq.~(\ref{eq:nuheating}), $r_7$
is the radius in $10^7\,$cm. 
The neutrino luminosity $L_{\nu_i}$ (normalized to 
$10^{52}\,$erg$\,$s$^{-1}$ in Eq.~\ref{eq:nuheating}), 
the average squared neutrino energy $\ave{\epsilon_{\nu_i}^2}$
(in units of 15$\,$MeV in Eq.~\ref{eq:nuheating}),
and the mean value of the cosine of
the angle between the direction of the neutrino propagation and the
radial direction, $\ave{\mu_{\nu_i}}$, are calculated from the
neutrino phase space occupation function $f_{\nu_i}(\epsilon_{\nu},\mu)$
by
\begin{eqnarray}
L_{\nu_i} &=& 4\pi r^2c\,{2\pi \over (hc)^3}
\,\int\limits_0^\infty{\mathrm{d}}\epsilon_{\nu}\int\limits_{-1}^{+1}{\mathrm{d}
}\mu\,
\mu\,\epsilon_{\nu}^3\,f_{\nu_i}(\epsilon_{\nu},\mu)\ ,\label{eq:nulum}\\
\ave{\epsilon_{\nu_i}^2} &=&
\int\limits_0^\infty {\mathrm{d}}\epsilon_{\nu}\int\limits_{-1}^{+1}{\mathrm{d}}
\mu\,
\epsilon_{\nu}^5\,f_{\nu_i}
\rund{
\int\limits_0^\infty {\mathrm{d}}\epsilon_{\nu}\int\limits_{-1}^{+1}{\mathrm{d}}
\mu\,
\epsilon_{\nu}^3\,f_{\nu_i}}^{\!\! -1}\!\! ,\label{eq:avenue}\\
\ave{\mu_{\nu_i}} &=&
\int\limits_0^\infty {\mathrm{d}}\epsilon_{\nu}\int\limits_{-1}^{+1}{\mathrm{d}}
\mu\,
\mu\,\epsilon_{\nu}^3\,f_{\nu_i}
\rund{
\int\limits_0^\infty {\mathrm{d}}\epsilon_{\nu}\int\limits_{-1}^{+1}{\mathrm{d}}
\mu\,
\epsilon_{\nu}^3\,f_{\nu_i}}^{\!\! -1}\!\! .
\label{eq:fluxfactor}
\end{eqnarray}
The quantity $\ave{\mu_{\nu}}$ is also called flux factor and can be
understood as the ratio of the neutrino energy flux, $L_{\nu}/(4\pi r^2)$,
to the neutrino energy density times $c$. Typically,
it is close to 0.25 near the neutrinosphere of $\nu_e$ and $\bar\nu_e$ and
approaches unity when the neutrino distributions get more and more forward
peaked in the limit of free streaming with increasing distance from the
neutrinosphere. Angle-dependent transport, i.e., solving the Boltzmann
transport equation, is necessary to accurately determine the spectral
and angular distribution of the neutrinos.

Only if the heating is sufficiently strong, depending 
on the size of the neutrino luminosities and the hardness of the
neutrino spectra, an explosion can be triggered 
\cite{BetWil85,BurrGosh93,Janka01}. The effect is 
self-enhancing: strong heating of the matter accreted by the shock 
decelerates the infall and increases the time for matter to absorb
neutrino energy, thus raising the efficiency of energy deposition by
neutrinos. This positive feedback leads to further shock expansion
and can initiate the final runaway. Strong cooling has the opposite 
effect. It accelerates
the accretion flow through the gain layer and largely reduces the time
matter is exposed to heating, hence diminishing the chance for an
explosion \cite{Janka01}. The Livermore group found the mechanism
working successfully only when the neutrino luminosities from the 
neutron star were assumed to be boosted by neutron-finger convection
below the neutrinosphere (treated by a mixing length theory in the
spherically symmetric models) \cite{WilMay88,WilMay93}. The existence
of neutron-finger instabilities, however, was questioned by independent
analysis \cite{BruDin96,Bruenn04}. 

In addition to this conflicting situation in a crucial aspect of the
successful delayed explosion models, there were long-standing concerns
that other approximations or numerical deficiencies of the Livermore 
models, e.g.\ in the hydrodynamics scheme or resolution, the flux-limited
diffusion description of the neutrino transport, and the treatment of
neutrino-matter interactions, might have prevented explosions of the
models without the controversial neutron-finger convection. For this
reason, improved numerical algorithms for one-dimensional simulations
were developed. These are either based on 
implicit, adaptive-grid solvers for the hydrodynamics and Boltzmann
transport equations \cite{Liebendoerfer02,Liebendoerfer04} or 
use conservative, third-order schemes with a Riemann-solver for
the explicit integration of the hydrodynamics equations, coupled to an 
implicit solver for the moment equations of neutrino number, energy,
and momentum with a closure relation (variable Eddington-factor)
computed from a model Boltzmann equation \cite{Rampp02}. The latter
approach is basically Newtonian but employs an effective potential to
account for the effects of general relativistic gravity, and includes
corrections for gravitational redshift and time dilation in the 
neutrino transport. Core-collapse calculations with both codes were 
compared and show very good agreement \cite{Liebendoerfer05,Marek06}.

\begin{figure}
\center{\includegraphics[width=0.99\textwidth]{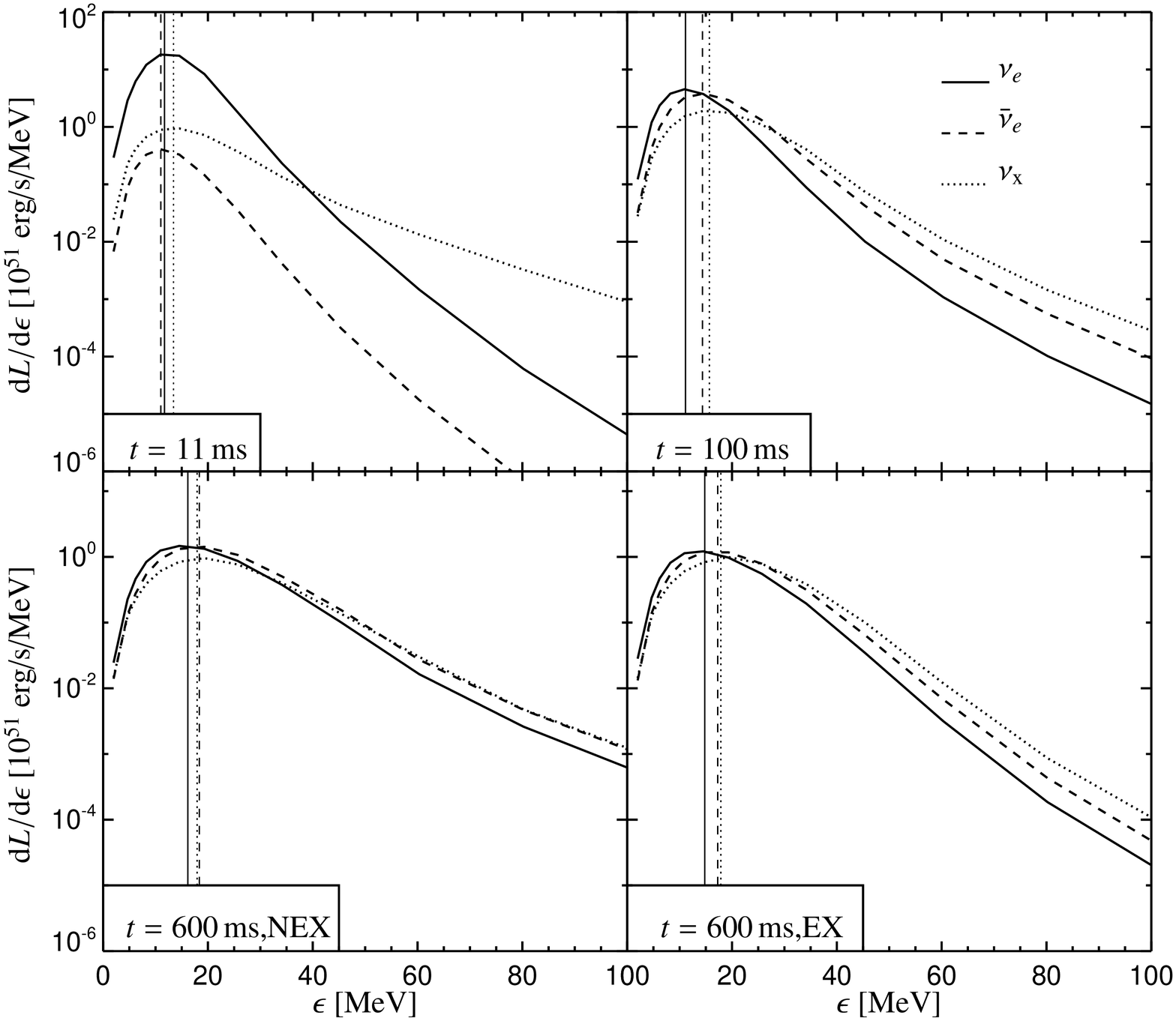}}
\caption{\small
Luminosity spectra for $\nu_e$, $\bar\nu_e$, and heavy-lepton
neutrinos $\nu_x$ for a spherically symmetric 20$\,M_\odot$ model.
The spectra are given for an observer at rest at 400$\,$km 
for three different postbounce times: Shortly after the peak
of the $\nu_e$ shock breakout burst 11$\,$ms after bounce, 
when the shock is near its maximum radius (around 100$\,$ms p.b.),
and at 600$\,$ms after bounce. The lower left
panel gives the spectra for a non-exploding model, the lower
right panel for a corresponding simulation in which an explosion
was artificially initiated at 230$\,$ms after core bounce, thus
stopping the accretion onto the nascent neutron star. The vertical
lines mark the locations of the mean neutrino energies, which are
defined as the ratio of energy flux to number flux, 
$\ave{\epsilon} = L_e/L_n$. Note the similarity of 
$\ave{\epsilon_{\bar\nu_e}}$ and $\ave{\epsilon_{\nu_x}}$ at
$t = 600\,$ms
($\nu_x$ stands for muon and tau neutrinos and antineutrinos).}
\label{fig:nuspectra}
\end{figure}

Considerable effort has also been spent on implementing weak 
interaction processes or improvements of the description of such
reactions that have been recognized to be of
importance in addition to the standard set of processes and their
canonical treatment as given by Bruenn \cite{Bruenn85}. Among these
extensions and improvements are shell-model based calculations for
$\beta$-processes of neutrinos and nuclei (see Sect.~\ref{sec:nuclbeta})
and for inelastic neutral-current neutrino-nuclei scatterings
(see Sect.~\ref{sec:nuclinel}), and the inclusion of ion-ion correlations
in coherent neutrino-nuclei scatterings
\cite{Horowitz97,Bruenn97,Itoh04,Marek05}.
Neutrino interactions with free nucleons have been corrected for
nucleon thermal motions and
recoil \cite{Schinder90} and the weak magnetism \cite{Horowitz02},
and have been generalized for including nucleon correlation effects
in high-density media \cite{Reddy98,Reddy99,BurrSaw98,BurrSaw99}.
Of particular importance for the transport of muon and tau neutrinos
are nucleon-nucleon bremsstrahlung, $NN \longrightarrow NN \nu\bar\nu$
\cite{Hannestad98,Thompson00}, and neutrino-antineutrino annihilation
to pairs of other flavors, $\nu_e\bar\nu_e \longrightarrow
\nu\bar\nu$ \cite{Buras03b}, which dominate $e^+e^-$-annihilation
in the production of $\nu_\mu\bar\nu_\mu$ and $\nu_\tau\bar\nu_\tau$
pairs \cite{Keil03}, and to a minor degree also scattering
reactions between neutrinos and antineutrinos of different flavors
such as $\nu_{\mu,\tau}\nu_e \longrightarrow \nu_{\mu,\tau}\nu_e$ and
$\nu_{\mu,\tau}\bar\nu_e \longrightarrow \nu_{\mu,\tau}\bar\nu_e$.
Including all these additional reactions and improvements makes the
radiated heavy-lepton neutrino spectra much more similar to the spectra
of $\nu_e$ and $\bar\nu_e$ than previously thought \cite{Keil03}.
The similarity of these spectra and the corresponding mean energies
at later postbounce times can be clearly seen in Fig.~\ref{fig:nuspectra}.

Spherically symmetric simulations with the new and more accurate
hydrodynamics and neutrino transport codes with or without the extended
set of neutrino reactions
\cite{Rampp00,Mezzacappa01,Liebendoerfer04,Buras06b,Thomp03,Sumiyoshi05},
disregarding neutron-finger instabilities and other multi-dimensional
effects, agree with the Livermore calculations that no delayed
explosions can be obtained 
for progenitors more massive than 10$\,M_\odot$.

\begin{figure}
\center{\includegraphics[width=0.49\textwidth]{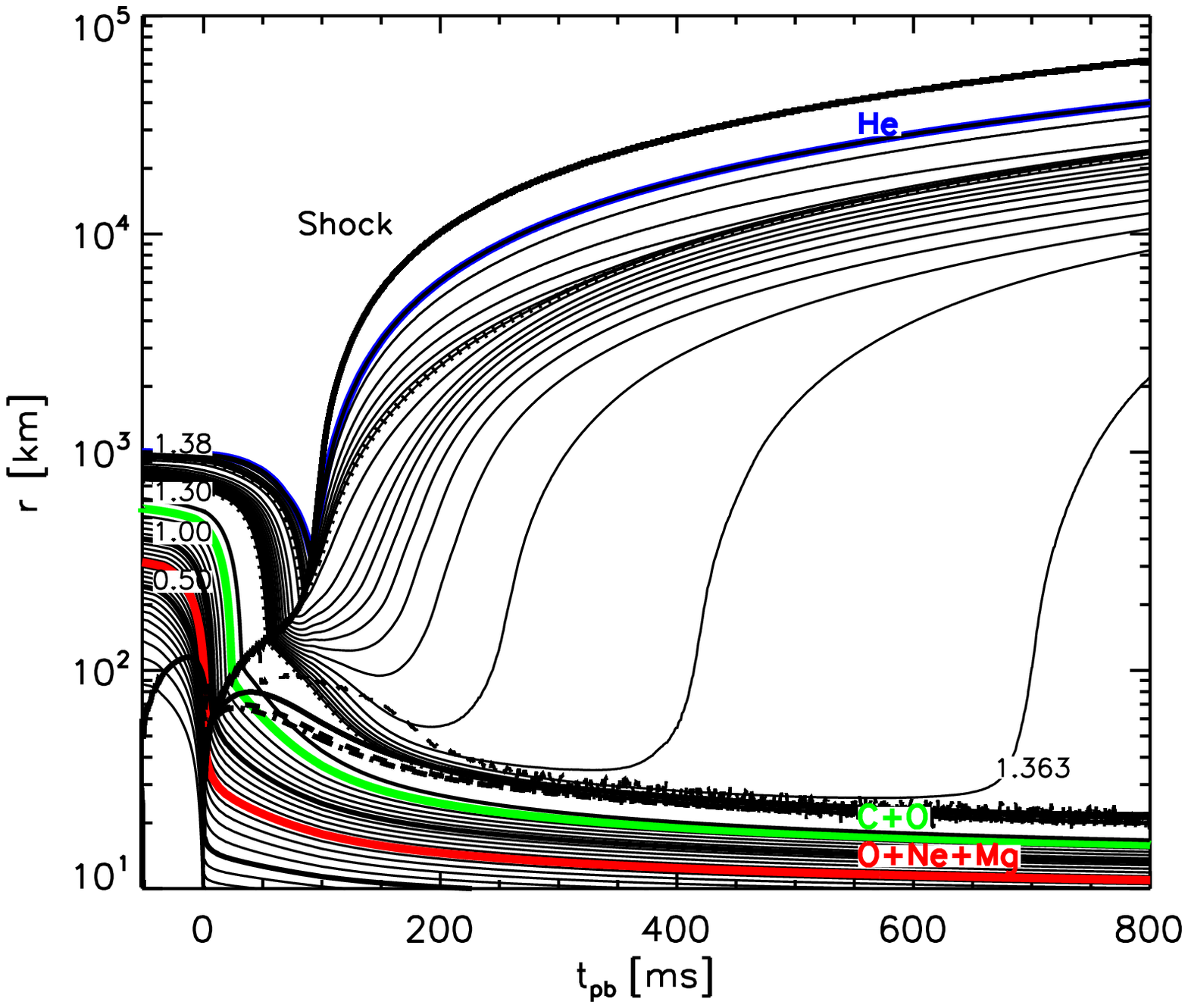}
        \includegraphics[width=0.49\textwidth]{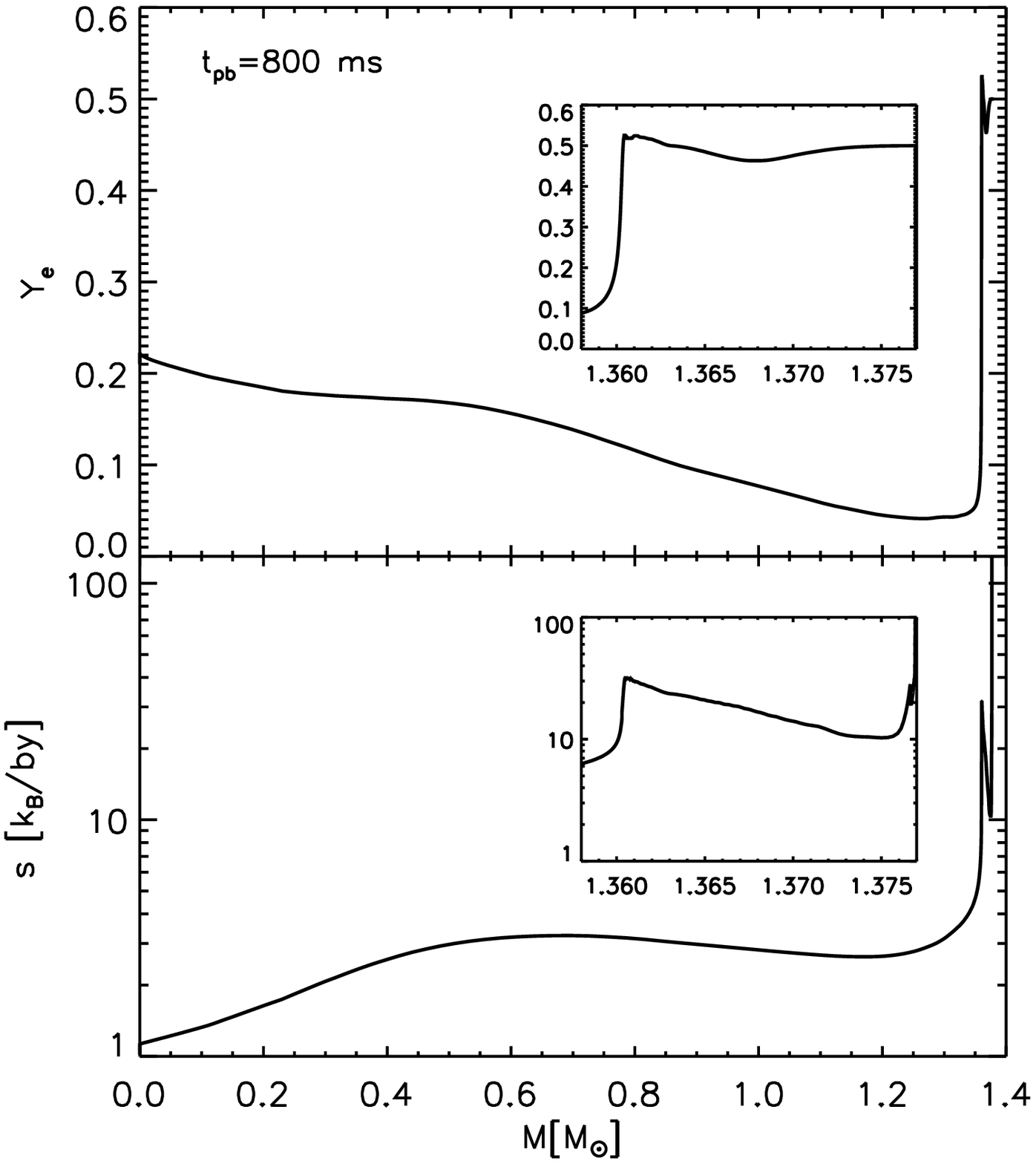}}
\caption{\small
Simulation of collapse and explosion of the ONeMg core
of an 8--10$\,M_\odot$ star \cite{Kitaura06,Janka05b}.
{\em Left:} 
The plot shows the time evolution of the mass shells with the inner
boundaries of the O+Ne+Mg shell, C+O shell, and He shell marked by thick
(colored) lines. The inner core of about 0.7$\,M_\odot$ contains
a dominant mass fraction of neon at the onset of collapse
\cite{Nomoto}. The explosion is driven by the baryonic wind
which is caused by neutrino heating around the nascent neutron star.
The thick solid, dashed, and dash-dotted lines mark the neutrinospheres
of $\nu_e$, $\bar\nu_e$, and heavy-lepton neutrinos, 
respectively, and the thin, dashed line is the gain radius that separates
the neutrino-cooling layer below from the neutrino heating layer above.
The black thick line
starting at $t = 0$ is the outward running supernova shock. Note that the
mass shells are chosen with finer spacing at larger radii in the collapsing
core.
{\em Right:} 
Electron fraction $Y_e$ and entropy $s$ versus enclosed mass
at the end of the explosion simulation of the ONeMg core. 
The neutrino-heated ejecta
have modest entropies of 10--30$\,k_{\mathrm B}$ per nucleon and
$Y_e$ slightly above or below 0.5. Much less neutron-rich material with 
less strong neutron excess than in previous models is expelled 
(figures taken from \cite{Kitaura06}).}
\label{fig:onemgexplosion}
\end{figure}

In contrast, in case of stars with birth masses of 8--10$\,M_\odot$, which 
develop instead of an iron core a core of ONeMg with a thin carbon shell,
surrounded by an extremely dilute and only loosely bound He-shell, 
neutrino heating was found
to power explosions in one-dimensional (1D) simulations even without
convective luminosity enhancement \cite{MayWil88}. Also this result 
of the Livermore group was confirmed qualitatively by more recent
models (\cite{Kitaura06}; Fig.~\ref{fig:onemgexplosion}), but the 
more sophisticated neutrino 
transport in the new simulations produces less energetic explosions
of only 1--$2\times 10^{50}\,$erg (=$\,$0.1--0.2$\,$bethe), 
in agreement with the estimated
kinetic energy of the filaments of the Crab nebula \cite{Davidson85}.
The latter has been suggested as the remnant of the explosion of a star
with ONeMg core because of its small carbon and oxygen abundances
and its helium overabundance \cite{Nomoto82}. Moreover, the early
supernova ejecta develop a proton-excess instead of the former 
neutron-richness (Fig.~\ref{fig:onemgexplosion}, right panel). 
This removes the problem of a vast overproduction of
some rare, very neutron-rich isotopes in this material (for more discussion
of the nucleosynthesis in proton-rich ejecta, see Sect.~\ref{sec:nucl}).

\subsection{Neutrino-driven explosions: multi-dimensional models}

Hans Bethe in his supernova review of 1990 \cite{Bethe90} already
pointed out that the hot-bubble region should be convectively
unstable (Fig.~\ref{fig:snphases}, lower left panel). 
A negative entropy gradient behind the accretion shock is 
generated because neutrino heating is much stronger near the gain
radius than at larger distances from the neutrinosphere. 
Multi-dimensional
hydrodynamic simulations indeed showed that violent convective
overturn develops in this layer and enhances the neutrino-energy
deposition there \cite{Her92,Her94,Bur95,JanMue94,JanMue96,Mez98}.
Neutrino-driven explosions were thus obtained even when 
spherically symmetric simulations failed 
\cite{Her94,Bur95,Fry99,FryWar02,FryWar04}. A second convectively
unstable region was found to exist inside the newly formed neutron
star \cite{Burrows87,Keil96,Dessart06,Buras06b}, also
indicated in the lower left panel of Fig.~\ref{fig:snphases}. It is,
however, probably less important for the explosion mechanism because
the enhancement of the neutrino emission is rather modest 
\cite{Buras06b}.

\begin{figure}
\center{\includegraphics[width=0.99\textwidth]{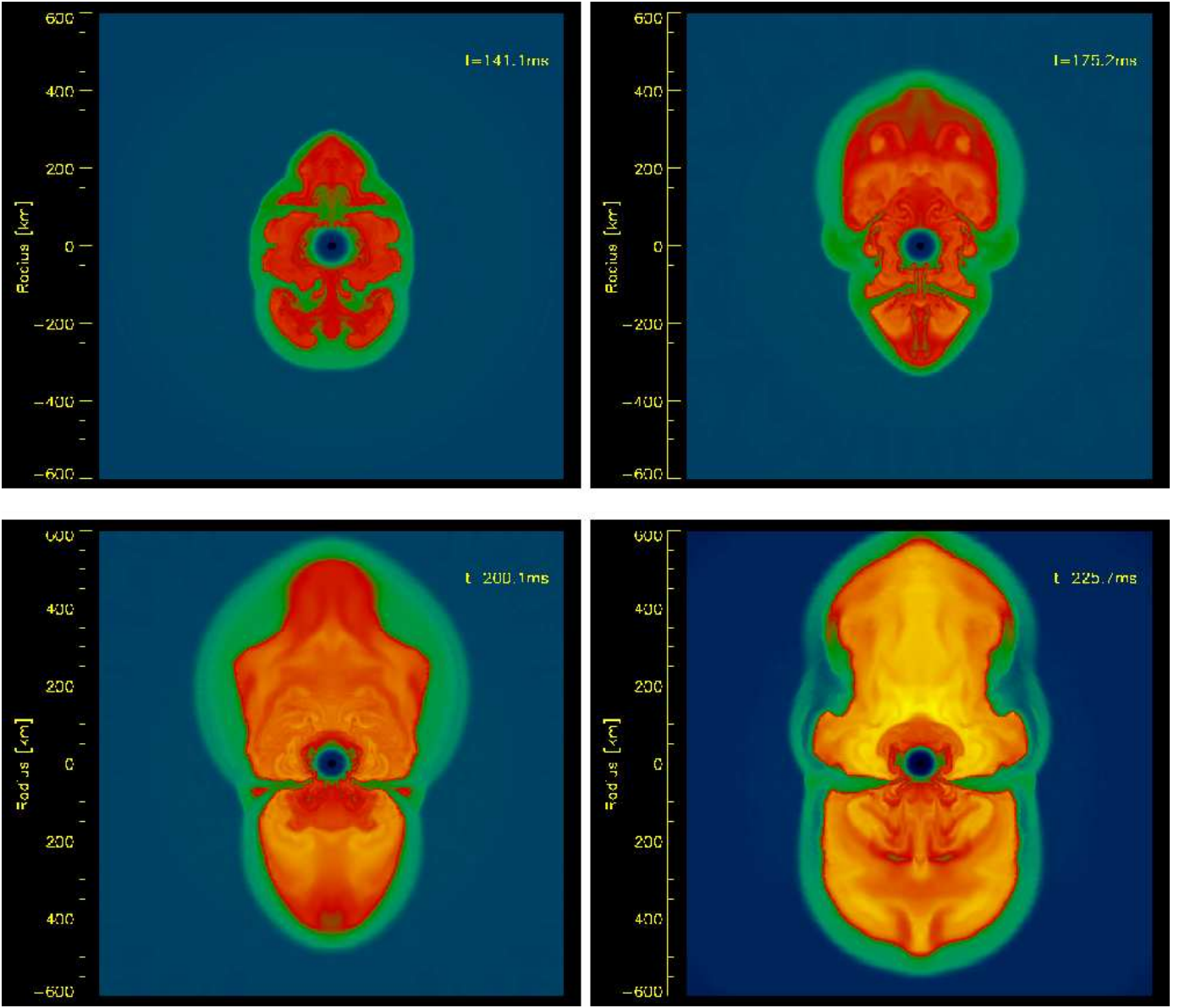}}
\caption{\small
Four stages (at postbounce times of 141.1$\,$ms,
175.2$\,$ms, 200.1$\,$ms, and 225.7$\,$ms) during the 
evolution of a (non-rotating), exploding two-dimensional
11.2$\,$M$_{\odot}$ model \cite{Buras06b}, visualized
in terms of the entropy. The scale is in km and the
entropies per nucleon vary
from about 5$\,k_{\mathrm B}$ (deep blue), 
to 10 (green), 15 (red and orange), up to more than 
25$\,k_{\mathrm B}$ (bright yellow). The dense neutron star
is visible as low-entropy ($\la 5\,k_{\mathrm B}$ per
nucleon) circle at the center.
The computation was performed in
spherical coordinates, assuming axial symmetry, and employing
the ``ray-by-ray plus'' variable Eddington factor
technique of Refs.~\cite{Rampp02,Buras06a}
for treating $\nu$ transport in multi-dimensional
supernova simulations. Equatorial
symmetry is broken on large scales soon after bounce, and
low-mode hydrodynamic instabilities (convective overturn
in combination with the SASI) begin to dominate the flow between
the neutron star and the strongly deformed supernova shock.
The model develops a --- probably rather weak --- explosion, the
energy of which was not determined before the simulation had
to be stopped because of CPU time limitations.}
\label{fig:112explosion}
\end{figure}

The success of the first 2D models with postshock convection
could not be reproduced by more recent calculations
\cite{Buras03,Buras06a,Buras06b,Burrows06a,Burrows06b} where
energy-dependent neutrino transport replaced the
simple grey diffusion schemes employed previously. 
With the currently most sophisticated treatment of neutrino
transport that is applied in multi-dimensional supernova 
simulations (at the same level of refinement in the treatment
of neutrino-matter interactions as in the state-of-the-art
1D models),
neutrino heating and postshock convection were found
to trigger a (probably rather weak) explosion only in case of an
11.2$\,M_\odot$ star (\cite{Buras06b}; see 
Fig.~\ref{fig:112explosion}).
The more massive progenitors investigated in
Ref.~\cite{Buras06b} had much higher densities and larger 
binding energies in the
shells surrounding the iron core. The ram pressure associated
with the high mass 
accretion rate of these infalling dense shells damped the 
shock expansion. Therefore no explosions were obtained for such 
stars during the simulated evolution periods of about 300$\,$ms
after core bounce, although a crucial criterion, the ratio of
advection timescale to heating 
timescale\footnote{This timescale ratio measures the
competing influence of two effects. On the one hand, gas accreted
by the stalled shock falls (`is advected') inward on a timescale
which measures its exposure to neutrino heating. On the other hand,
neutrino heating needs a certain time (the heating timescale) to
deposit sufficient energy for matter to become gravitationally unbound
\cite{Janka01b,Thompson05,Buras06b}.}, indicates that the 
critical condition for an explosion was not missed by much
(roughly a factor of two in the 2D, 90-degree simulations of 
Ref.~\cite{Buras06b}).

The onset of the explosion of the 11.2$\,M_\odot$ star
was aided by a recently discovered generic instability of the
accretion shock to non-radial deformation. This so-called
SASI (Standing Accretion Shock Instability; 
\cite{Blondin03}) shows highest growth rates
of the $l = 1,\,2$ modes (i.e., the dipole and quadrupole terms
of an expansion in spherical harmonics) and causes a bipolar
sloshing of the shock with pulsational strong
expansion and contraction. Since the shock is pushed farther
out and the time matter stays in the heating layer therefore   
increases, this strengthens the neutrino-energy deposition
and ultimately leads to a
globally asymmetric initiation of the explosion. The
rapid growth of this instability can be seeded by the 
perturbations resulting from the onset of convection in the
postshock layer, but occurs also in situations which are
convectively stable. It can be understood by an amplifying
`advective-acoustic cycle' in which vorticity
perturbations created at the shock are carried inward by the
accretion flow, create sound waves when the gas falls
onto the neutron star, and the sound waves propagating back
to the shock cause larger vorticity perturbations there,
closing the feedback loop 
\cite{Foglizzo,Foglizzo06}\footnote{Blondin and
Mezzacappa, however, interpret the SASI in their models
as a purely acoustic
phenomenon in which a standing pressure wave is amplified in
the acoustic cavity defined by the shock and the neutrinosphere
\cite{Mezzacappa05,Blondin06}.}.
The SASI was found to be present in all two-dimensional (2D)
core-collapse simulations
when the calculations were performed with a 180-degree grid 
and the onset of the explosion was delayed for a sufficiently 
long time (longer than in the 2D simulations with grey
diffusion), thus allowing the instability to develop
\cite{Scheck04,Scheck06,Buras06b,Burrows06a,Burrows06b}.
The presence of the SASI by itself, however, does not cause
the supernova explosion. Different from simulations without
neutrinos, where an accumulation of turbulent energy in the
postshock layer was found and led to shock expansion \cite{Blondin03}, 
this energy accumulation seems to be absent in the neutrino cooled and
heated environment of the supernova core. The role of the low-mode 
SASI in the successful 11.2$\,M_\odot$ model (and also in other simulations)
is more indirect. Being stronger than convection, it supports shock 
expansion and thus significantly improves the conditions for efficient
neutrino heating.

The potential importance of this instability for understanding
observable properties of supernovae and their remnants 
was demonstrated recently
by a large set of two- and three-dimensional simulations
in which neutrino-driven explosions were artificially 
initiated by chosing a sufficiently high value of the 
neutrino luminosity of the forming neutron star
\cite{Scheck04,Scheck06,JankaEtaCar}.
These simulations showed that the explosion asymmetry grows
stochastically from initial random seed perturbations, 
breaking the initial sphericity of the collapsing star on
a global scale. The highly anisotropic explosions due to 
a dominant $l = 1$-mode SASI-instability can lead to a 
large recoil of the newly born neutron star in the direction
opposite to the stronger mass ejection. While about half of
the computed models produced small neutron star kicks with
velocities of only around 100$\,$km$\,$s$^{-1}$ or less,
the mean value of the recoil velocity was found to be
several hundred kilometers per second and even cases
with more than 1000$\,$km$\,$s$^{-1}$ were obtained, in agreement 
with the measured proper motions of young pulsars. 
Three-dimensional simulations have revealed the possibility of
an unstable $l=1,\,m=1$ spiral SASI mode that can create a strong 
rotational flow in the vicinity of the accreting neutron star,
thus providing a new mechanism for the generation of neutron
star spin \cite{BlondinMezz06}.

Moreover, the globally
asymmetric onset of the explosion with sizable initial shock
deformation (axis ratios of more than 2:1 were obtained) triggers
strong hydrodynamic instabilities at the composition interfaces
of the progenitor when the shock propagates outward to
the stellar surface. This leads to large-scale element mixing
in the exploding star. Iron-group nuclei, silicon, and 
oxygen, as well as radioactive isotopes are carried from 
their production sites near the nascent neutron star into 
the hydrogen envelope, while hydrogen and helium are mixed deep
into the expanding ejecta. Supernova simulations in which the
SASI-deformed explosion was continuously followed from its 
initiation by the 
neutrino-heating mechanism until several hours later for 
the first time achieved to explain the anisotropies, element 
mixing, and high nickel velocities observed in Supernova 1987A 
without invoking artificial and ad hoc assumptions 
\cite{Kifonidis06}. A unified picture therefore seems to emerge
in which hydrodynamic instabilities in the supernova core 
during the first seconds of the explosion help starting the
explosion, create the seed for the ejecta asymmetries observed
later on, and are essential for explaining the measured 
kick velocities and rotation rates of pulsars.

\subsection{Alternative explosion scenarios and open ends}

But, of course, these conclusions are based on simulations in
which neutrino-driven explosions were obtained by a suitable
choice of the neutrino luminosities rather than fully 
self-consistent modeling. What is the missing ingredient 
why such self-consistent models for progenitor masses larger
than about 12$\,M_\odot$ have not produced explosions?
The situation is currently not clear.
Three-dimensional simulations with sophisticated
energy-dependent neutrino transport have not been performed yet
but are definitely indispensable to judge about the viability of the
neutrino-heating mechanism in the more massive progenitor stars,
in particular because the critical timescale ratio suggests that
the 2D models do not fail by much. 
In three dimensions the flow dynamics might reveal important
differences compared to axisymmetric models, in
particular with respect to the growth of convective and
Rayleigh-Taylor instabilities, the possibility of non-radial,
non-axisymmetric instability modes, or the development of
local fluid vortices. Three-dimensional simulations with 
energy-dependent neutrino transport, however,
have such high CPU-time demands that supercomputers ten to
hundred times more powerful than currently available will be 
needed, and numerical codes are necessary that are able to efficiently 
use thousands of processors. Moreover, although current 2D models
do not show a large luminosity boost by neutron star convection
\cite{Buras06b,Dessart06}, the question whether
convective or mixing processes below and around the neutrinosphere
could more strongly enhance the neutrinospheric neutrino emission 
and thus significantly strengthen
the neutrino heating behind the shock, must still be considered as
unsettled: Doubly diffusive instabilities \cite{Bruenn04},
neutrino-bubble instabilities \cite{Socrates05}, or magnetic
buoyancy instabilities \cite{Wilson05} deserve further investigation
by multi-dimensional modelling. 

Or do explosions require a different mechanism to be
at work, for example invoking very rapid rotation of the 
collapsing stellar core and the amplification of magnetic fields 
through compression and wrapping, $\alpha$-$\Omega$ dynamo action, 
or the magneto-rotational instability in the shear flow at the
periphery of a 
differentially rotating nascent neutron star (for a review, 
see \cite{Kotake06})? The fields could then convert
the free energy of the differential rotation 
of the forming compact remnant to kinetic energy of the supernova
ejecta either by magnetic forces (e.g., 
\cite{Wheeler02,Akiyama03,Yamada04,Ardeljan05,Obergaulinger06} 
and references
therein) or viscous heating behind the shock in addition to the 
energy input there by neutrinos \cite{Thompson05}. Such scenarios,
however, seem currently disfavored for ordinary supernovae because
of the slow rotation of their progenitors predicted by stellar
evolution calculations \cite{Heger05}. Instead, magnetohydrodynamic
effects are likely to be the mechanism by which accretion energy is
converted to the extreme energy output of hypernovae and the jetted
outflow in gamma-ray bursts.

Recently Burrows {\em et al.}, based on results from their 2D 
core-collapse simulations, proposed a new acoustic mechanism
for launching and powering the explosion 
\cite{Burrows06a,Burrows06b}. Due to anisotropic accretion,
the neutron star in their models is exited to strong g-mode
oscillations. The large-amplitude core motions then
create powerful sonic activity in the neutron star surroundings
by which energy is transported to the shock, driving the explosion.
The source of this energy is the gravitational binding energy
of the accreted gas, converted to sound by the rapidly ringing
neutron star, which thus acts like a transducer.
Burrows {\em et al.}\ attribute their discovery to the use of a 
new --- according to their arguments superior because linear 
momentum conserving --- treatment of the effects
of gravity in their code and the use of a
computational grid without the coordinate singularity that is
present at the stellar center in the commonly used polar grids.
They speculate that the combination of both aspects allows them
to see the excitation of the $l=1$ g-mode oscillations
in their 2D simulations, while other groups for numerical reasons
can not. This reasoning, however, is
questioned by the fact that the 2D models of the Garching
group show the presence of this mode and of the higher 
($l =2,\,3,\,...$) modes in the neutron star as well, all 
of them with similar, but a factor 10--100 smaller, 
amplitudes than in the calculations of Burrows {\em et al.}
Moreover, experiments with the code setup used for supernova
simulations by the Garching group reveal that a large-amplitude 
core g-mode, when artificially instigated, can very well be
followed over many periods. So the question arises, whether
a large-amplitude oscillation as seen by
Burrows {\em et al.}\ is really excited in the supernova core.
And, in particular, will it also be excited when simulations
are performed in three dimensions, i.e., without the enforced 
axial symmetry and associated
preferred axis-direction of 2D simulations? Moreover, 
the core g-mode in the calculations by Burrows {\em et al.}\ 
gains sufficient strength for powering
the explosion only at very late 
postbounce times ($t > 1\,$s; \cite{Burrows06b}). 
Neutron stars with large masses
are the consequence, and the acoustic mechanism will have
its chance only if no other mechanism achieves to explode the
star faster (as neutrino heating does, e.g., in case of the exploding
8--11$\,M_\odot$ models of Figs.~\ref{fig:onemgexplosion} and 
\ref{fig:112explosion}). And even then, the
core oscillation energy that is potentially available for the 
explosion might be rather low, possibly 
too low to account for explosion energies in agreement with supernova 
observations. 

Significant improvements of all numerical codes currently in use 
and again 3D simulations will be needed to finally
confirm or reject the exitation of strong 
g-mode $l=1$ oscillations in the neutron star.
Also further improvements of the neutrino transport methods will have
to be implemented in the long run 
\cite{Cardall,Livne04,Swesty06,Hubeny06}, because 
the best energy-dependent schemes that have been applied in full-scale
supernova simulations so far either do not couple the energy bins
\cite{Burrows06a,Burrows06b} or use approximations to the neutrino
transport in non-radial directions \cite{Buras03,Buras06a,Buras06b}.

Only recently, collective $\nu\bar\nu$-flavor conversion was discovered
to be of potential relevance between neutrinosphere and supernova
shock \cite{Duan,Hannestad06}. Different from matter-enhanced
Mikheyev-Smirnov-Wolfenstein (MSW) neutrino oscillations, which 
are suppressed in the dense medium of the
supernova core unless the neutrino mass difference is in the eV to 
keV range --- which requires the uncertain existence of a sterile 
neutrino ---, the new flavor transformation
process depends only on the presence of a dense neutrino radiation 
field. Its implications for the supernova explosion mechanism and
nucleosynthesis, however, are still unclear and subject of ongoing 
research.

Finally, but certainly of crucial importance, all further steps of
improving supernova simulations must be accompanied and supported
by the improvement of the microphysics that plays a decisive role in 
the supernova core. Electron captures on nuclei, for example,
govern the reduction 
of $Y_e$ during the core collapse phase and thus decide about the
shock formation radius and subsequent shock expansion.
Similarly, the nuclear equation of state, which is still
incompletely known, also influences strongly the shock formation
and initial shock strength by its incompressibility at core bounce.
The nuclear EoS, moreover, determines the compactness 
and thus the size and contraction of the nascent neutron star,
which in turn influence the shock evolution after bounce.
In the following, we will therefore discuss in more detail the
nuclear physics input needed for supernova models.

\section{Nuclear physics input}

\subsection{Composition of matter and equation of state}
\label{sec:eos}

On general grounds
the structure of nuclear matter is determined by a competition between
the short-ranged surface force and the long-range Coulomb force. The surface
energy is reduced with aggregation as this decreases the ratio of surface area
to volume. On the other hand, the Coulomb energy is reduced by dispersion
as this decreases the inter-proton repulsion. As a result, 
for densities smaller than about $\rho_0/10$, nucleonic matter consists
of individual nuclei, where $\rho_0=2.5 \times 10^{14}$ g/cm$^3$ is the
uniform nuclear matter saturation density. However, at larger densities
the equilibrium configuration of nuclear matter is achieved by
diverse geometrical forms in a competition of uniform nuclear matter
and the vacuum \cite{Baym71}.

During most of stellar evolution
the nuclear composition is determined by a network of nuclear reactions
between the nuclei present in the stellar environment. 
Electromagnetic reactions
of the type $(p,\gamma)$, ($\alpha,\gamma)$, and once free neutrons are
produced, also $(n,\gamma)$, play a particularly important role to
fuse nuclides to successively larger nuclei. As the stellar environment
has a finite temperature $T$, these reactions are in competition with
the inverse dissociation reactions ($(\gamma,p)$, $(\gamma,\alpha$),
$(\gamma,n$)) and for temperatures exceeding $T \approx Q/30$, where
$Q$ is the threshold energy (Q-value) for the dissociation process
to occur, a capture reaction and its inverse get into equilibrium.
Similarly also nuclear reactions mediated by the strong and electromagnetic
interaction get
into equilibrium with their inverse once the temperature is high enough
for the charged particles to effectively penetrate the Coulomb barrier.
For supernovae this has the important consequence that for
temperatures exceeding a few 100 keVs, all reactions mediated by
the electromagnetic and strong interaction are in equilibrium with their 
inverse and the nuclear composition becomes independent of the
rates for these reactions. It is given by Nuclear Statistical
Equilibrium (NSE) with the constraints that the  total mass and charge
of the composition is conserved. The latter condition reflects the fact
that charge can only be changed by reactions via the weak interaction
which, however, are not in equilibrium as neutrinos can leave the star
(until neutrino trapping is effectively reached).
Due to the two conserved quantities (mass, charge neutrality) there exist
two independent chemical potentials which are conventionally chosen as
$\mu_n$ and $\mu_p$ for neutrons and protons, respectively. In NSE
the chemical potential for a nucleus with charge number $Z$ and
mass number $A$ is then given by
\begin{equation}
 \mu(Z,A) = Z \mu_p + (A-Z) \mu_N.
\label{eq:chempot}
\end{equation}
The nuclei obey Boltzmann statistics. Thus the number density $n(Z,A)$
of a nucleus (mass $m(Z,A)$) is related to its chemical potential via

\begin{equation}
    \mu(Z,A) = m(Z,A) c^2 + k T \ln\left[\frac{n(Z,A)}{G(Z,A)}
    \left(\frac{2\pi\hbar^2}{m(Z,A)kT}\right)^{3/2}\right]
\label{eq:boltzmann}
\end{equation}
with the nuclear partition function
\begin{equation}
    G(Z,A) = \sum_i (2 J_i +1) e^{-E_i/kT},
\end{equation}
where the sum runs over all nuclear states with excitation energy $E_i$
and total angular momentum $J_i$. 
Upon inserting (\ref{eq:boltzmann}) into (\ref{eq:chempot}),
one obtains the Saha equation which relates the abundance
($Y(Z,A) \rho/ m_u =n(Z,A)$) 
of a nucleus to the proton and neutron abundances

\begin{equation}
    Y(Z,A) = \frac{G(Z,A) A^{3/2}}{2^A}(\rho N_A)^{A-1} Y_p^Z Y_n^N
    \left(\frac{2\pi\hbar^2}{m_u k T}\right)^{3/2(A-1)} e^{E_b(Z,A)/kT}
\end{equation}

with $E_b(Z,A)=(N m_n + Z m_p - M(Z,A)) c^2$
  and the constraints
$\sum_i Y_i A_i =1$ (conservation number nucleons) and
$\sum_i Y_i Z_i = Y_e$ (charge neutrality),
where the sum runs over all nuclei present in the composition and
$Y_e$ is the electron-to-nucleon ratio. Due to charge neutrality,
this is identical to the proton-to-nucleon ratio.

\begin{figure}
\begin{center}
    \includegraphics[width=0.45\columnwidth,angle=270]{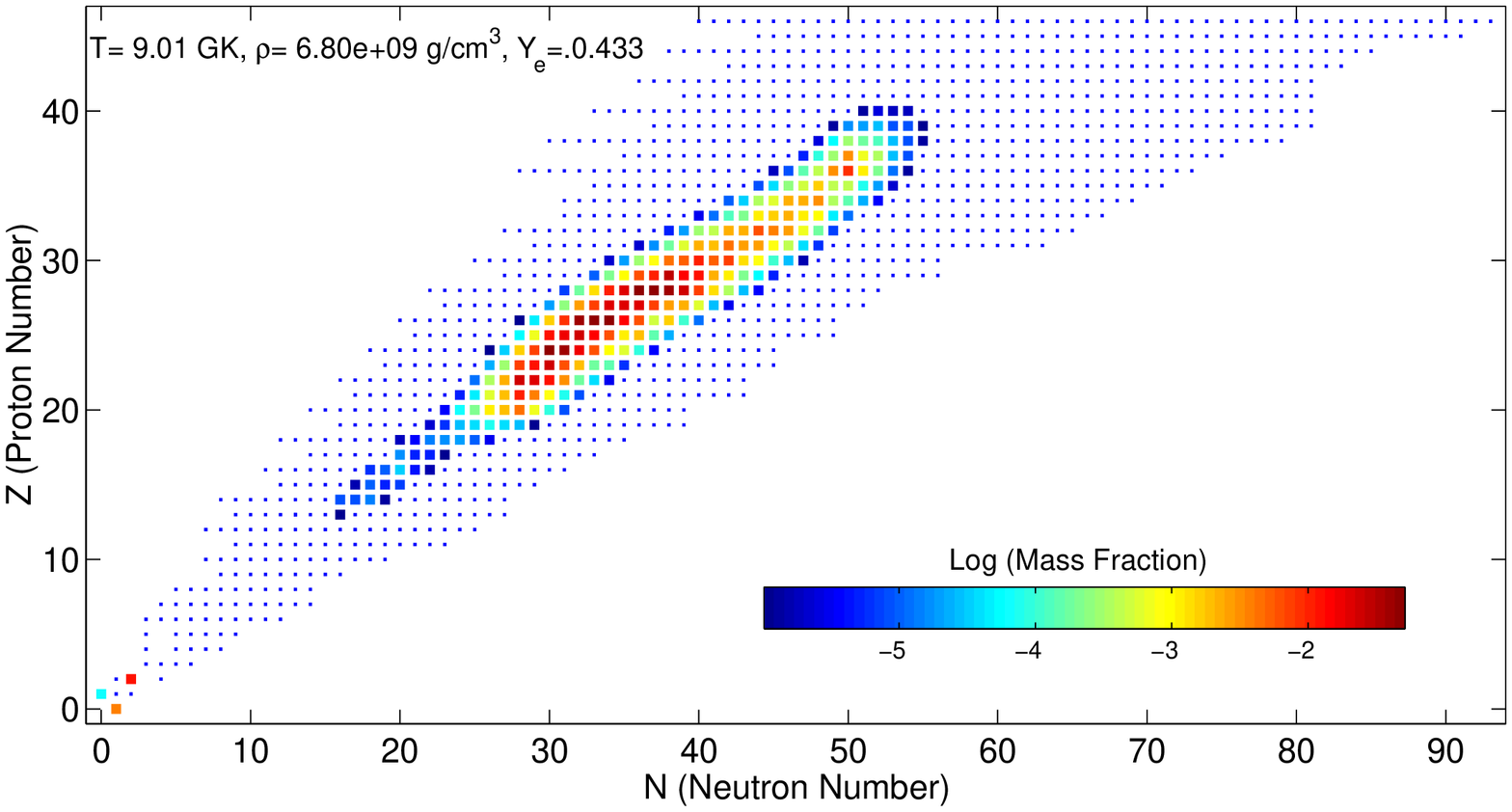}
    \includegraphics[width=0.45\columnwidth,angle=270]{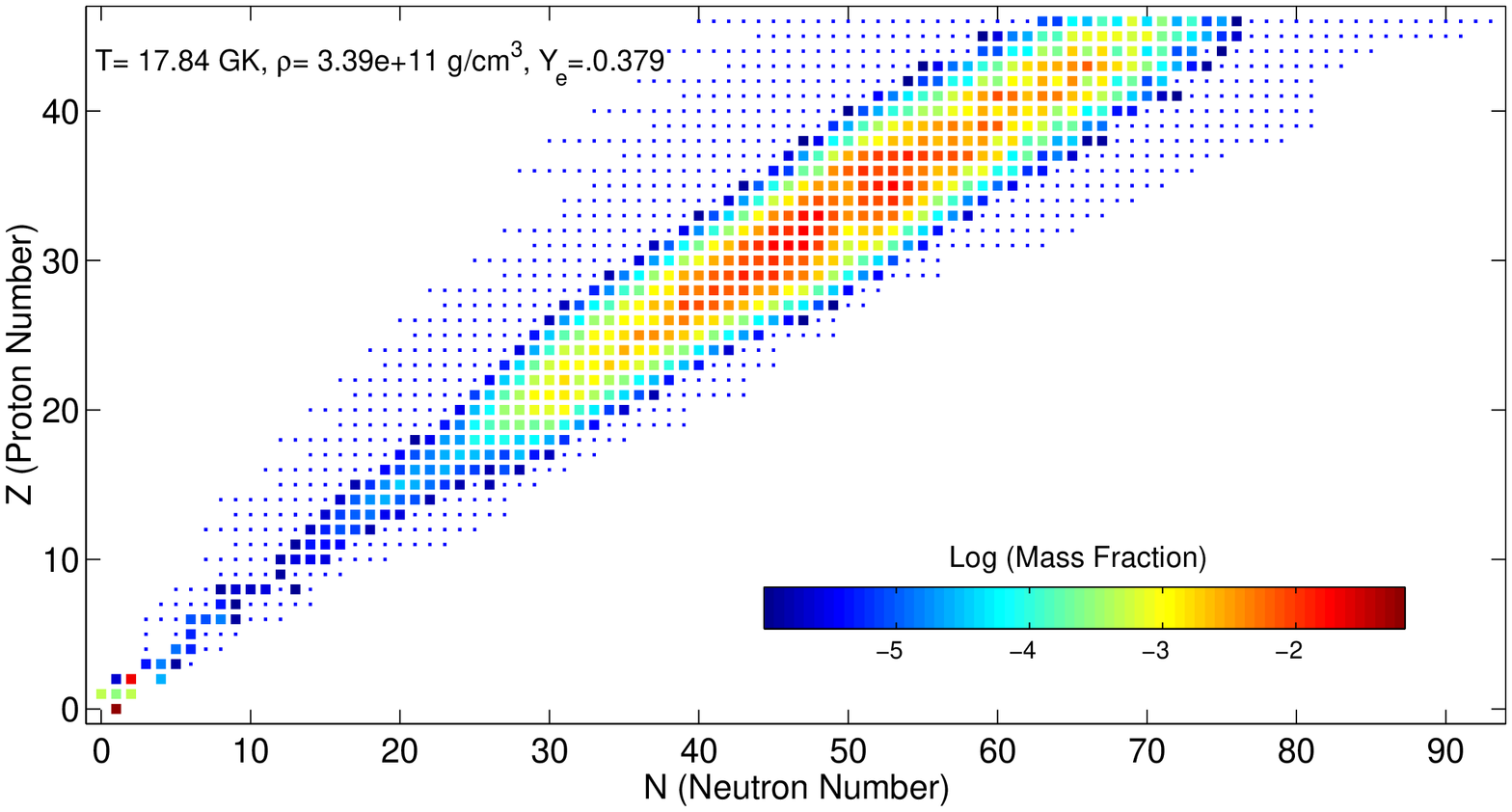}
\caption{\small
Abundance distributions of nuclei in Nuclear Statistical Equilibrium
at conditions which resemble the presupernova stage (top)
and the neutrino trapping phase (bottom) of core-collapse
simulations (courtesy of W.R. Hix).
  \label{fig:NSEabund}}
\end{center}
\end{figure}

The fact that temperatures are high enough in the late stage
of stellar evolution to drive matter into NSE facilitates
simulations significantly as the matter composition becomes
independent of the rates for reactions mediated by the electromagnetic 
and strong interaction. In practice, stellar evolution
is studied in two steps: The star is followed through its
various hydrostatic core and shell burning stages 
considering an appropriate nuclear network
and taking advantage of the fact that neutrinos can leave the star
unhindered and thus serve only as a source for energy losses.
The simulations stop once the star reaches the {\it 
presupernova} stage; i.e. the inner stellar 
core collapses with velocities in excess of 1000 km/s,
which is achieved when the central density and temperatures are of order
$10^{10}\,$g/cm$^3$ and $10^{10}\,$K (860~keV) \cite{WW95,Heger01a}. 
The presupernova models
(which assume spherical symmetry) become the input of the
actual supernova simulations, which explicitly assume that NSE is 
established. Nevertheless the nuclear composition changes quite drastically
during the collapse as temperature and density increase as the collapse
progresses 
and, importantly, the weak interaction is 
initially not in equilibrium. As we will discuss in the next section,
electron captures on nuclei and on free protons occur which
reduce $Y_e$ and the nuclear composition is shifted to nuclei
with larger neutron excess (smaller proton-to-nucleon ratio)
which favors heavier nuclei. The increasing temperature 
has the consequence that the number of nuclei present in the composition
with sizable abundances grows. This effect and the
shift to heavier and more neutron-rich
nuclei is 
confirmed in Fig.~\ref{fig:NSEabund} which shows the NSE abundance distribution
for two typical conditions during the collapse.
Note that with
increasing density, correlations among the nuclei and effects
of the surrounding plasma become increasingly relevant
\cite{Hix96,HixPhD,Bravo}.

At densities larger than about $\rho_0/10$, the matter composition
experiences a drastic change resulting in a variety of complex shapes.
This complexity arises as the nucleons try to be simultaneously 
correlated due to the
nuclear attraction and anti-correlated due to Coulomb repulsion.
Furthermore, surface tension favors spherical shapes, while the
Coulomb interaction often favors other geometrical shapes.
As a consequence the transition from
individual nuclei to uniform nuclear matter occurs in various steps:
from large spherical nuclei immersed into the vacuum, to rod-like geometry,
to slabs of uniform matter, to cylindrical bubbles (tubes), to 
spherical bubbles of vacuum immersed in uniform matter and finally to
uniform nuclear matter \cite{Ravenhall,Williams}. 
Reminded by the shapes of the various
transitions at subnuclear densities one often speaks of 
{\it nuclear pasta}. More recent studies of the ground state properties
of nuclear pasta can be found in \cite{Watanabe}; attempts of investigating
the dynamical properties of nuclear pasta are reported for example in 
\cite{Magierski,Khan,Horowitz}.

In the shock-heated region temperature gets so high that nuclear matter
is dissolved in free protons and neutrons.
 
The transition region of shape complexity requires also special care
in the derivation of the equation of state (EoS) which describes
the relevant thermodynamical quantities as function of 3 independent
input parameters, e.g. density, temperature and electron-to-nucleon ratio.
In the range of temperature and densities relevant for supernovae modelling
matter can be modelled as a mixture of electrons, positrons, photons and
nuclear constituents. The description is simplified by the justified
assumption that matter is in equilibrium with respect to electromagnetic and 
strong interactions (this assumption is not being made in presupernova
simulations). Although nuclear matter composition is then described by a Saha
equation, an explicite consideration of all possible nuclei (and nuclear
shapes) is computationally yet unfeasible. Thus nuclear matter is represented
by four species: free protons, free neutrons, alpha-particles (which
represent all light particles present) and an individual heavy nucleus
standing in as a representative for the ensemble of heavy nuclei present.
With changing inputs ($\rho$, $T$, and $Y_e$) the mass and charge number
of this average nucleus is readjusted. Burrows and Lattimer have shown
that such a single-nucleus average for the true ensemble of heavy nuclei
has little effect on the thermodynamical properties
\cite{Burrows86}. However, it is not
justified in the calculation of the weak interaction rates, as we will discuss
in the next section.

\begin{figure}
\center{\includegraphics[width=0.45\textwidth]{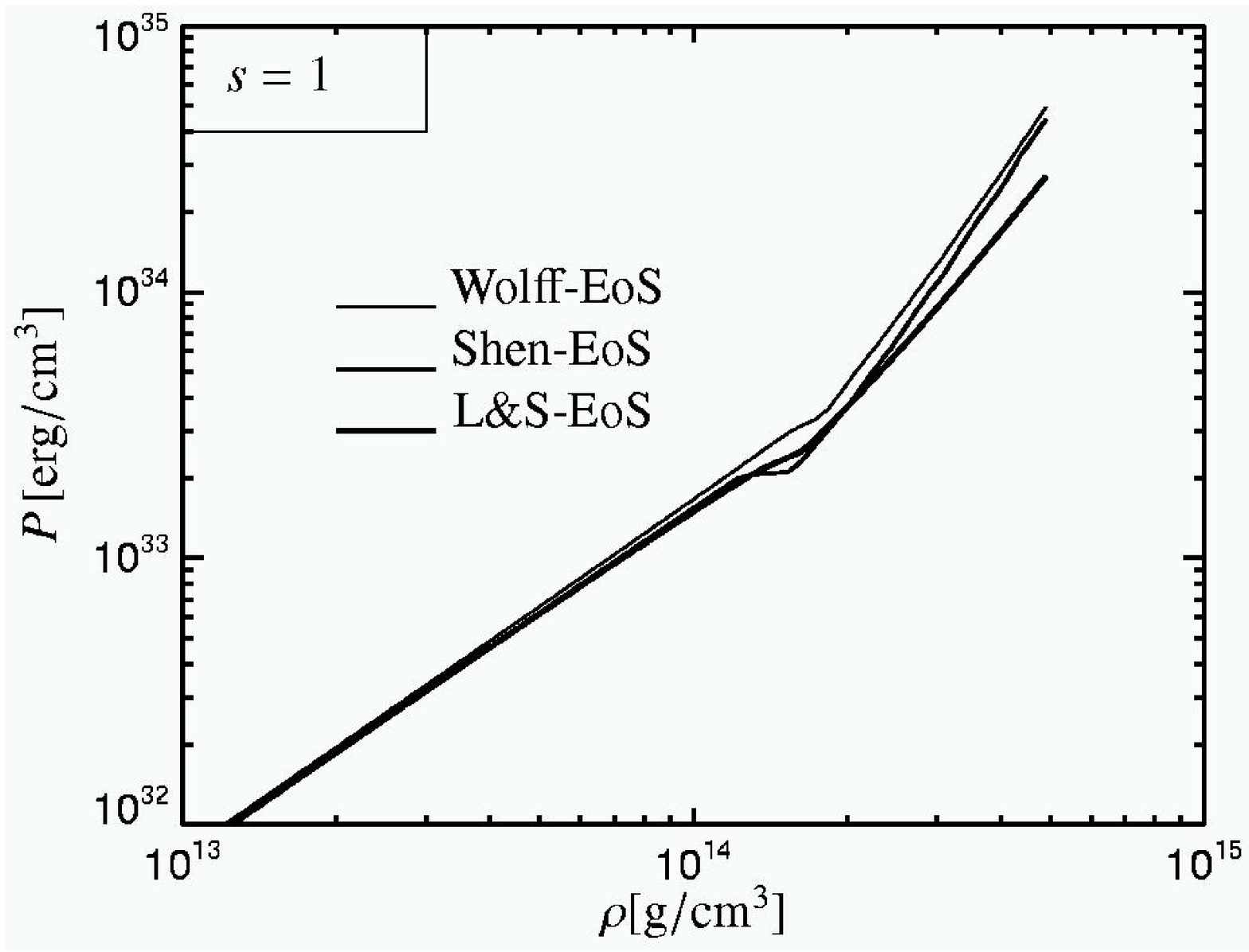}
        \includegraphics[width=0.45\textwidth]{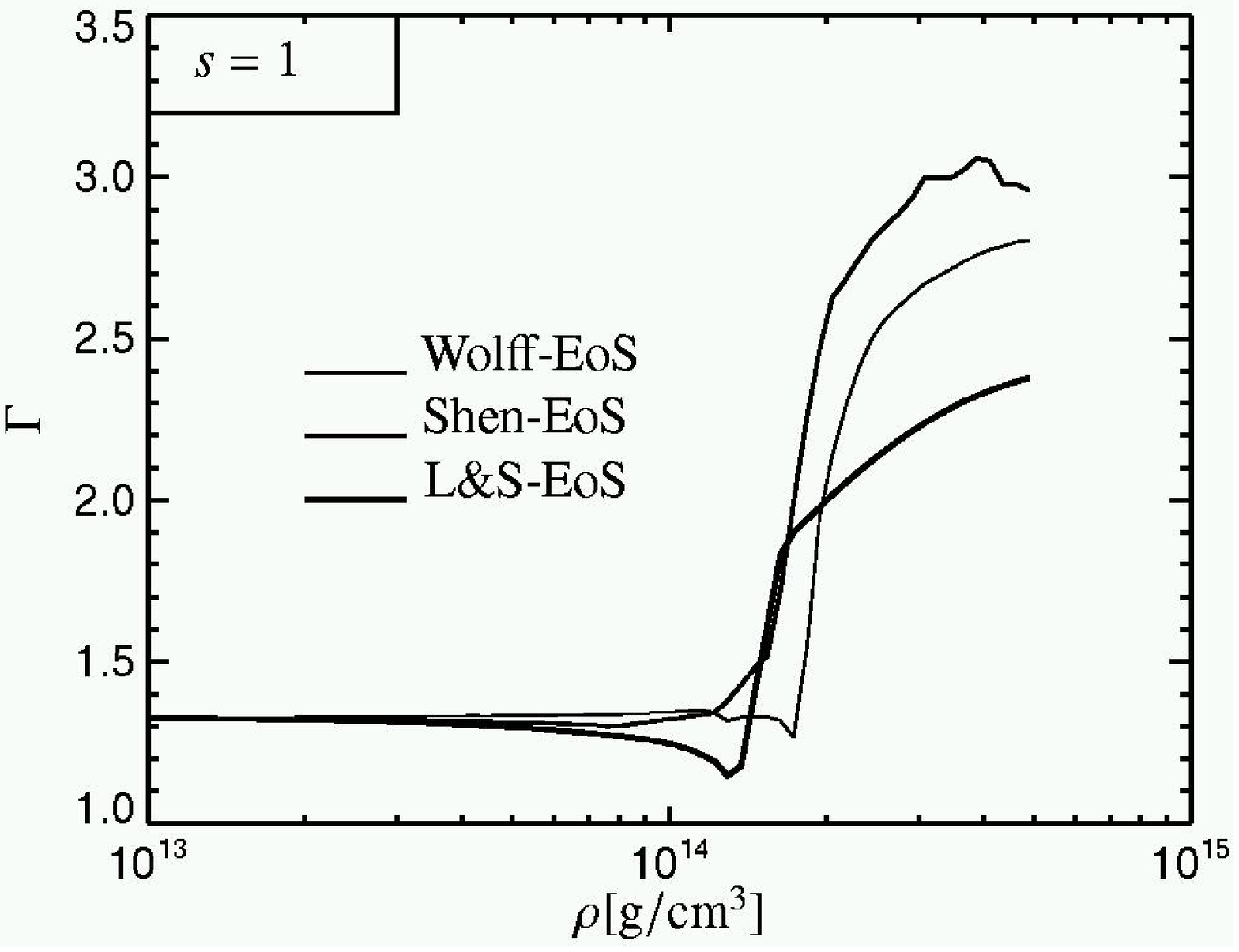}}\\
\center{\includegraphics[width=0.45\textwidth]{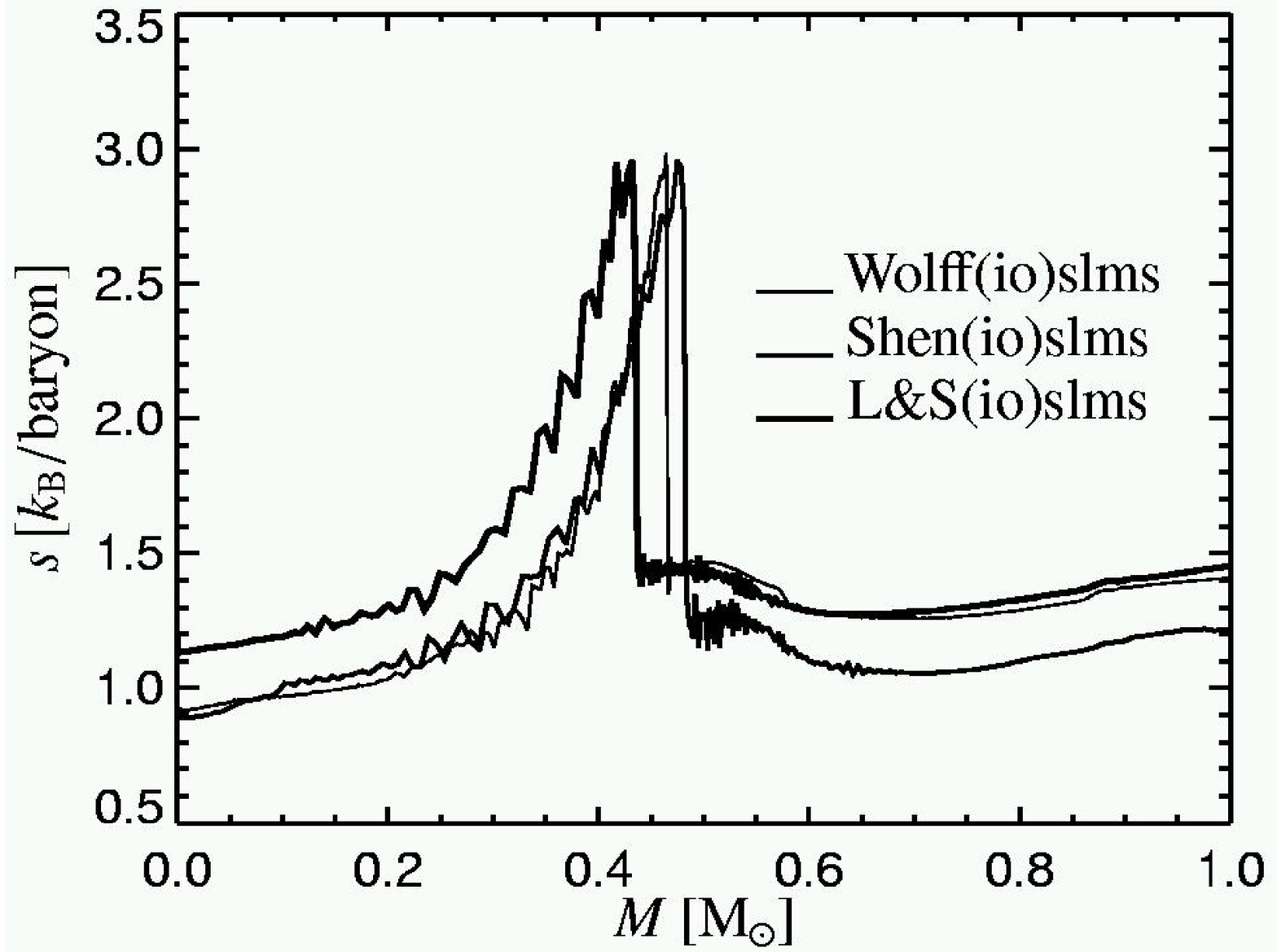}}
\caption{\small
Pressure and adiabatic index
$\Gamma \equiv \left (\partial\ln P/\partial\ln\rho\right )_s$
vs.\ mass density
for an entropy $s = 1\,k_{\mathrm B}$ per nucleon and
$Y_e = 0.4$, and entropy profile vs.\ enclosed mass at the 
time of shock formation for simulations with the EoSs of
Wolff \& Hillebrandt (``Wolff'', thin lines; \cite{Hillebrandt}),
Shen et al.\ (``Shen'', medium; \cite{Shen98}),
and Lattimer \& Swesty (``L\&S'', thick; \cite{Lattimer91}).
}
\label{fig:EoSmodels1}
\end{figure}

\begin{figure}
\center{\includegraphics[width=0.49\textwidth]{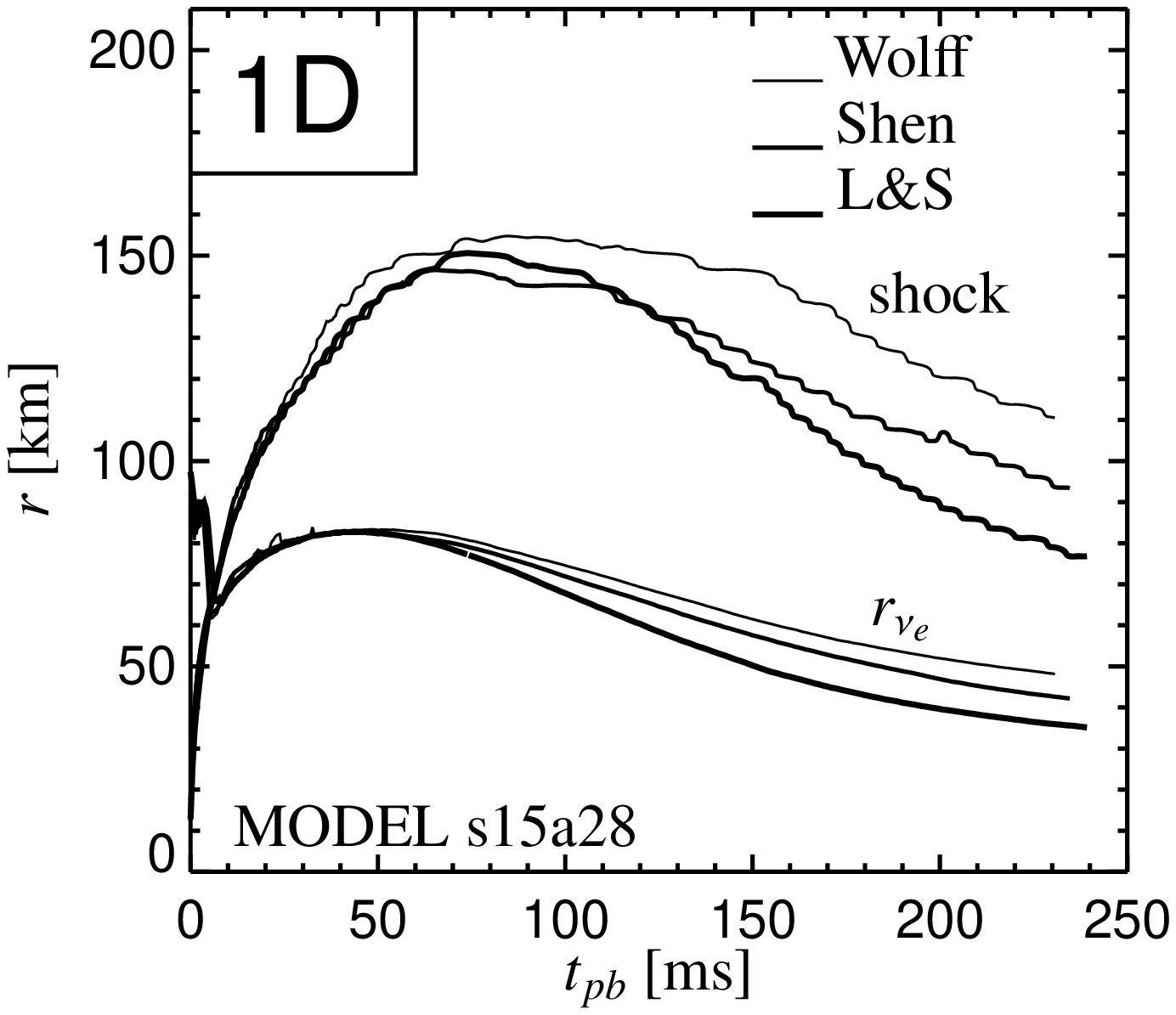}
        \includegraphics[width=0.49\textwidth]{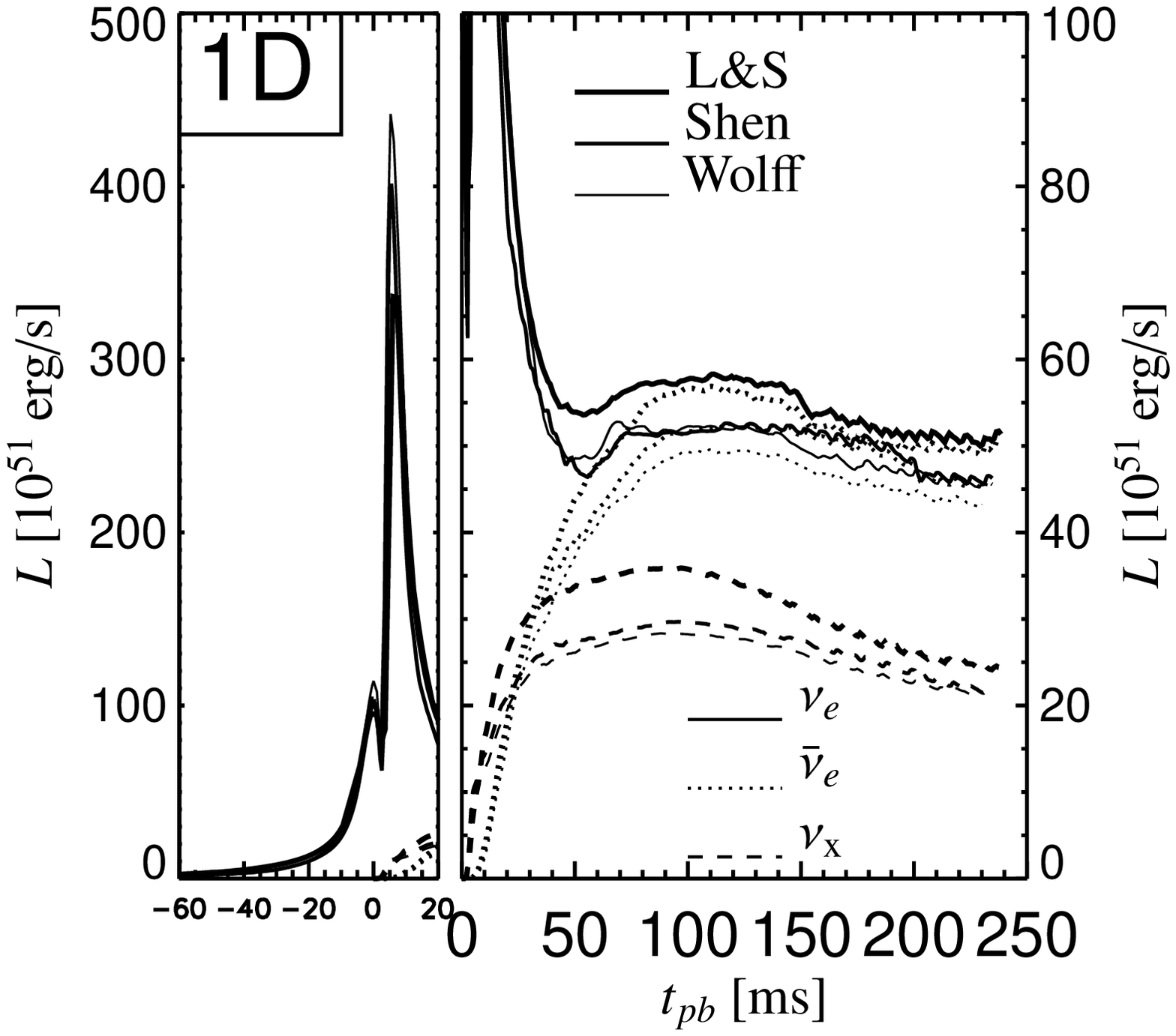}}\\
\center{\includegraphics[width=0.49\textwidth]{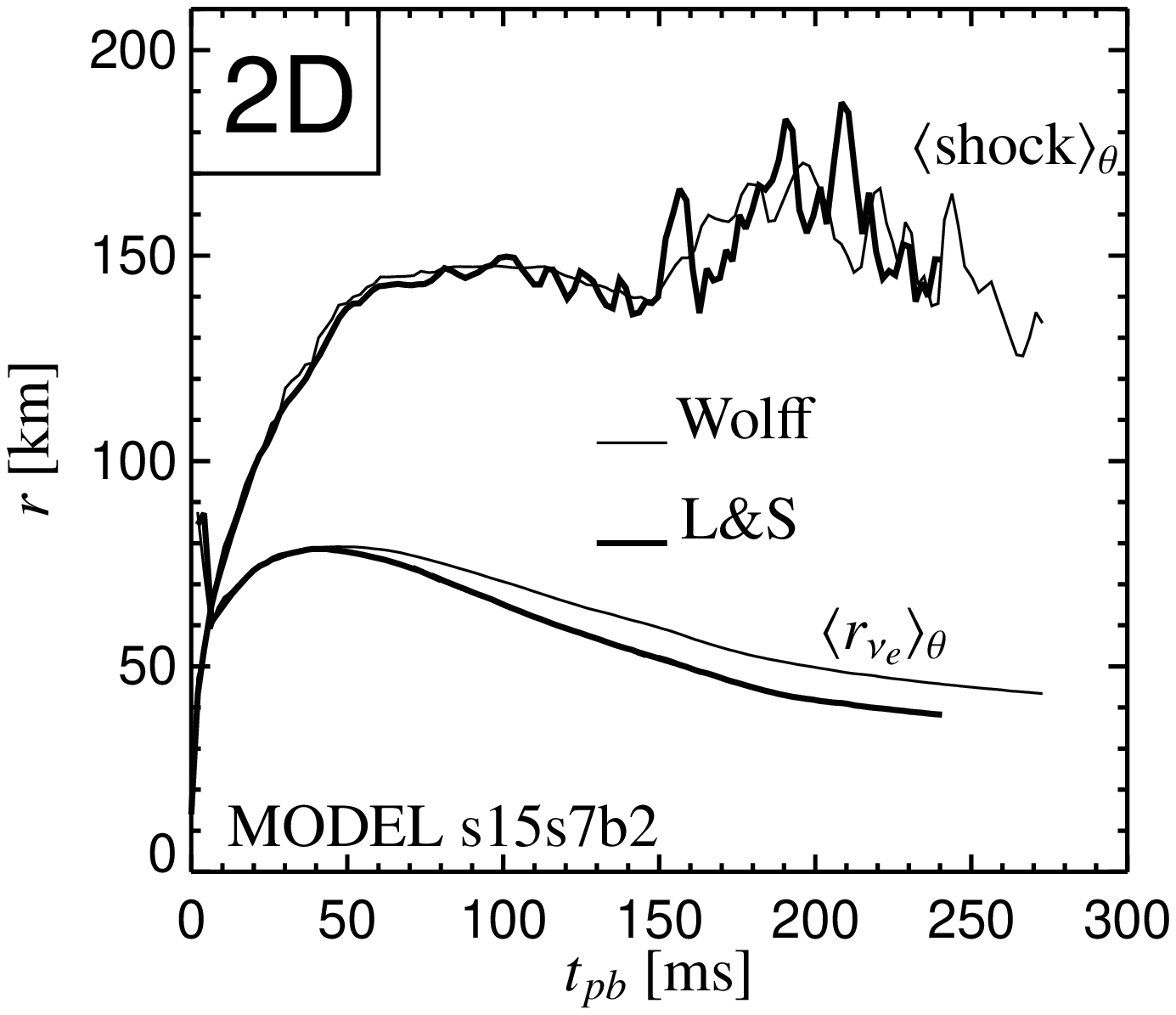}
        \includegraphics[width=0.49\textwidth]{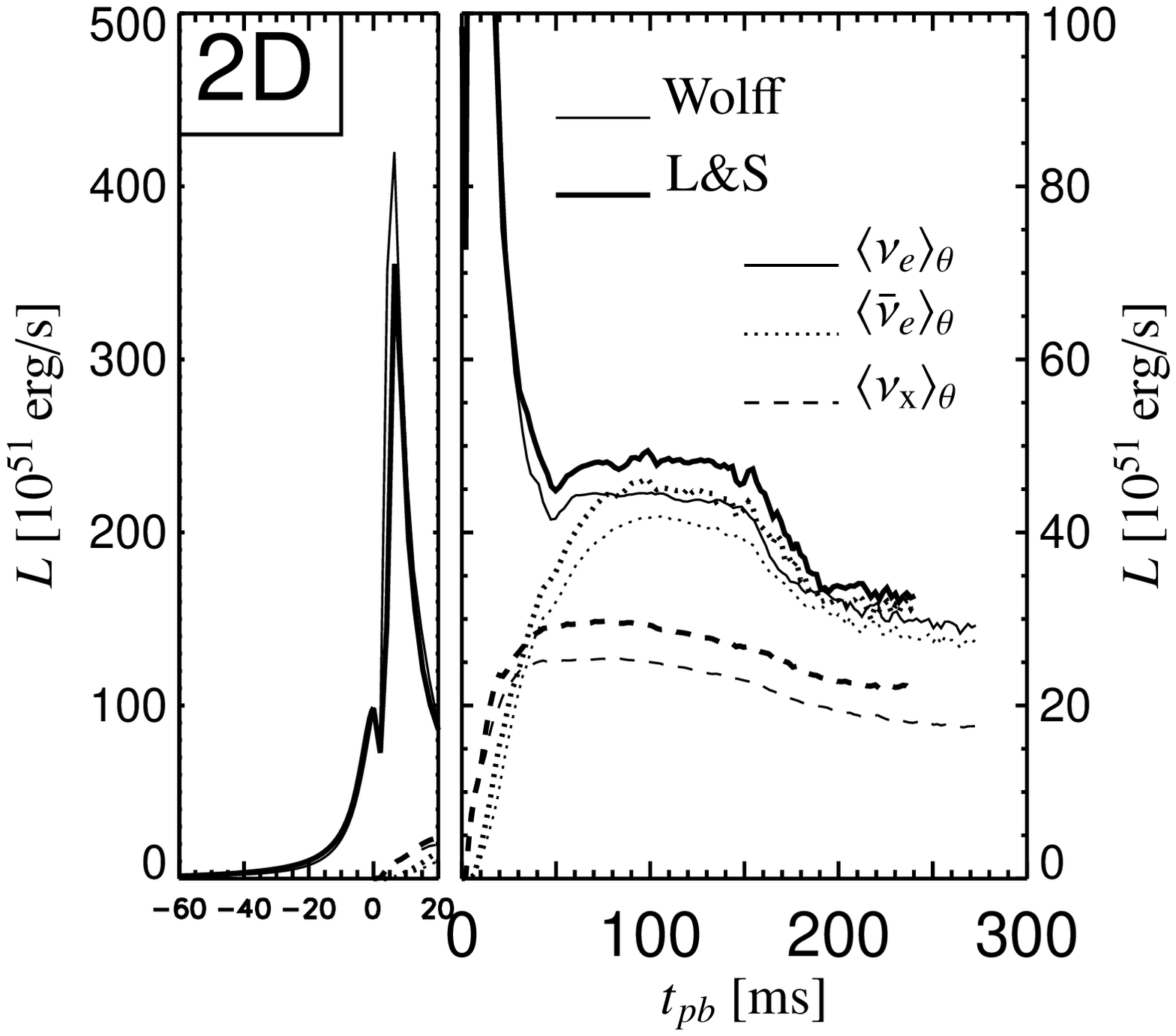}}
\caption{\small
{\em Top:} Results from 1D core-collapse simulations with the
equations of states (EoSs) ``Wolff'', ``Shen'', and ``L\&S'' 
(cf.\ Fig.~\ref{fig:EoSmodels1}) for the 15$\,M_\odot$ progenitor
model s15a28 from Heger et al.\ \cite{Heger01a}. 
{\em Bottom:} Results from 2D core-collapse simulations with the          
equations of states ``Wolff'' and ``L\&S'' for the 15$\,M_\odot$
progenitor model s15s7b2 from Woosley \& Weaver \cite{WW95},
using the Garching ``ray-by-ray plus'' variable Eddington factor
technique of Refs.~\cite{Rampp02,Buras06a}
for treating $\nu$ transport in multi-dimensional models.
The left panels show the shock positions and neutrinosphere
radii of $\nu_e$ as functions of postbounce time, the right
panels display the neutrino luminosities with the prompt 
$\nu_e$ burst in the left window and the postbounce
luminosities of $\nu_e$ (solid lines), $\bar\nu_e$ (dotted) and
heavy-lepton $\nu$'s and $\bar\nu$'s (dashed) in the right window.
The results of the 2D runs were averaged over latitudinal angles
$\theta$. The left plots demonstrate that the different neutron
star contraction for the three equations of state forces the 
stalled shock to retreat significantly differently during the 
postbounce evolution in 1D models, whereas
the postshock convection in the 2D simulations covers the EoS
influence completely despite the remaining (but somewhat
reduced) differences of the
neutron star radii as functions of time. Because of the latter fact
the neutrino luminosities reveal significant differences both
in the 1D and 2D cases, because the softer nuclear phase in 
particular of the L\&S EoS leads to higher neutrinospheric 
temperatures. Note that a detailed comparison of the upper and
lower plots is hampered by the use of different 15$\,M_\odot$
progenitors.
}
\label{fig:EoSmodels2}
\end{figure}

For more than a decade,
the standard EoS in supernova simulations has been the one of Lattimer
and Swesty \cite{Lattimer91}. 
These authors derive the EoS by minimization of the free
energy of the matter. The nuclear part is derived in the spirit
of the finite-temperature compressible liquid-drop model
\cite{Lattimer91} treating nuclei within the Wigner-Seitz
approximation, e.g. a heavy nucleus in a unit cell is surrounded by
a charge-neutral spherical cell made of a gas of neutrons, protons
and alpha-particles. Possible effects of neutron skins are ignored.
To describe the shape phase transition regime
between $\rho_0/10$ and $\rho_0$, a shape function is introduced
which correctly reproduces the limiting cases of nuclei (low density)
and bubbles (high density) and in-between approximates more
elaborate calculations performed by Ravenhall {\it et al.} at 
zero temperatures \cite{Ravenhall}. 
The parameters required in the functional form of the
free energy were adjusted either to experiment or to some standard
Skyrme interaction.
The electrons are modelled as non-interacting ultrarelativistic particles,
photons are treated as ideal Bose gas. For most of the supernova
density-temperature regime, the EoS is dominated by electrons, photons
or bulk nuclear matter. As stated above, the largest impact of the nuclear
constituents are found in the shape transition region, which likely is also
the region of largest uncertainty in the EoS. For supernova simulations
density regions larger than twice saturation density are of less interest. Thus,
the Lattimer-Swesty EoS does not attempt to describe potential phase
transitions to pion or kaon condensates or  quark-hadron phases of matter
and also does not include the effects of hyperons (e.g. \cite{Weber04}).

The Lattimer-Swesty EoS has recently been improved and is now applicable for 
temperatures $T=1-30$ MeV and densities between $10^7$ g/cm$^3$ and 6$\rho_0$.
The improved EoS now explicitly accounts for neutron skin effects and
is built on both non-relativistic potential models and relativistic field
models.

An alternative EoS has been presented by Shen {\it al.}
\cite{Shen98}, based on the 
Thomas-Fermi spherical cell model and deriving the matter energy from
relativistic field theoretical  (Brueckner-Hartree-Fock)
models. However, this EoS
does not consider non-spherical geometries and consequently the shape
transition region at sub-saturation densities is approximated by
a sudden switch from a nuclear composition to uniform nuclear matter.

The Garching group has used both the Lattimer-Swesty EoS (chosing
a soft version with an incompressibility modulus of bulk nuclear
matter of 180$\,$MeV and a symmetry energy of 29.3$\,$MeV) and
the harder Shen-EoS (with an incompressibility of 281$\,$MeV and a
symmetry energy of 36.9$\,$MeV)
in 1D as well as 2D supernova simulations, 
and compared the results with models in which the EoS of Hillebrandt
and Wolff \cite{Hillebrandt} was employed 
\cite{Janka04,Janka05b,Marek03,Marek06a}. The latter was 
constructed by a temperature-dependent Hartree-Fock calculation
with a Skyrme force for the nucleon-nucleon interaction. It has
an incompressibility parameter of 263$\,$MeV and a symmetry energy
of 32.9$\,$MeV. The three equations of state show clear differences
for example in the nuclear composition during the core-collapse
phase and in the adiabatic index
$\Gamma\equiv \left (\partial\ln P/\partial\ln\rho\right )_s$
around and above the phase transition to homogeneous nuclear
matter (Fig.~\ref{fig:EoSmodels1}). Due to a much larger abundance
of free protons and a longer collapse time, the L\&S EoS leads
to a considerably stronger deleptonization in the infall phase
and therefore a shock-formation position that is significantly inside
the one obtained with the Shen EoS (Fig.~\ref{fig:EoSmodels1}, right
panel). In none of the 1D simulations for different progenitors with
$M > M_\odot$ an explosion was obtained. 
Since the radius of the stalled shock follows closely the 
initial growth of the radius of the nascent neutron star due to
mass accretion and the following shrinking of the settling
neutron star, the radii of maximum shock expansion and subsequent
shock contraction also differ in 1D simulations with the three 
EoSs (Fig.~\ref{fig:EoSmodels2}, upper left panel). A more compact
and thus hotter neutron star also leads to higher neutrino
luminosities and larger mean energies of the neutrinos radiated
during the postbounce evolution (Fig.~\ref{fig:EoSmodels2}, 
right panels). Similar findings, although quantitatively 
somewhat different --- probably because of numerical reasons --- were
obtained by Sumiyoshi et al.\ \cite{Sumiyoshi05}, who also continued
their simulations for the L\&S and Shen-EoSs to a postbounce
time as late as 1$\,$s.

\subsection{Neutrino-nuclei beta-interactions}
\label{sec:nuclbeta}

Nuclear processes, mediated by the weak-interaction, play an essential
role during the collapse. While positron captures on nuclei and nuclear
$\beta^+$ decays are also considered in supernova simulations, the two
important weak processes are nuclear beta-decay
\begin{equation}
(Z,A) \rightarrow (Z+1,A) + e^- + \bar{\nu_e}
\end{equation}
and electron capture
\begin{equation}
(Z,A) + e^- \rightarrow (Z-1,A) + {\nu_e}.
\end{equation}
Both processes have in common that they create neutrinos and that they change
the number of electrons. Both of these properties are quite important:
the neutrinos leave the star nearly unhindered for core densities
smaller than a few $10^{11}$ g/cm$^3$ and carry away energy in this way
cooling the stellar core and keeping its entropy low
\cite{BBAL}. As electrons
dominate the matter pressure during most of the collapse (for 
densities less than $10^{13}$ g/cm$^3$), a change of the electron 
number density ( or better in the electron-to-baryon ratio $Y_e$)
directly influences the collapse dynamics.

Except at rather high densities, nuclear beta-decays and electron captures
are dominated by allowed transitions. For beta-decays (and nuclei with 
neutron excess),
these are Fermi transitions (which for each parent state can only proceed 
to a single state
in the daughter nucleus, the isobaric analog state) and Gamow-Teller 
transitions GT$_-$, in which a neutron is changed into a proton.
Under core-collapse conditions electron captures are dominated by
Gamow-Teller (GT) transitions GT$_+$, 
in which a proton is changed into a neutron.
Gamow-Teller transitions are mediated by the operator $\bf{\sigma} \tau_\pm$
where $\bf{\sigma}$ is the spin operator and $\tau_\pm$ are the
isospin raising or lowering operators, respectively.

As the stellar interior has finite temperature $T$, beta decays and electron
captures can occur from excited nuclear levels, where the thermal nuclear
ensemble is described by a Boltzmann distribution. Beta-decay 
$\lambda_\beta$ and electron
capture rates $\lambda_{ec}$  
can be derived in perturbation theory and the respective
formulas and derivations are presented in \cite{FFN2,Langanke00}.
One obtains 
\begin{equation}
  \label{eq:rate}
  \lambda_\beta = \frac{\ln 2}{K} \sum_i \frac{(2J_i+1)
    e^{-E_i/(kT)}}{G(Z,A,T)} \sum_j (B_{ij} (F)+B_{ij}({GT})) \Phi_{ij}^\alpha ,
\end{equation}
\begin{equation}
  \label{eq:beta}
  \lambda_{ec} = \frac{\ln 2}{K} \sum_i \frac{(2J_i+1)
    e^{-E_i/(kT)}}{G(Z,A,T)} \sum_j B_{ij} ({GT}) \Phi_{ij}^\alpha ,
\end{equation}
where 
$K$ can be determined
from superallowed Fermi transitions, $K=6146\pm
6$~s~\cite{Towner}.  $G(Z,A,T)=\sum_i (2J_i+1) \exp(-E_i/(kT))$ is the
partition function of the parent nucleus.  $B_{ij}$ is the reduced
transition probability of the nuclear transition, where we have 
restricted ourselves to allowed transitions only.

The Fermi transition strength is given by:

\begin{equation}
  \label{eq:fermidef}
  B_{ij}(F) = \frac{\langle j || \sum_k \bf{t}^k_\pm || i
  \rangle^2}{2 J_i +1}.
\end{equation}
If isospin is a good quantum number,  the Fermi
transition strength is concentrated in the isobaric analog state (IAS)
of the parent state. Equation~(\ref{eq:fermidef}) reduces to,
\begin{equation}
  \label{eq:fermiT}
  B_{ij}(F) = T(T+1)-T_{z_i} T_{z_j},
\end{equation}
where $j$ denotes the IAS of the state $i$. Here the reduction
in the overlap between nuclear wave functions due to isospin mixing
has been neglected as it
is estimated to be small ($\approx 0.5$\%~\cite{Towner}).

The GT strength is given by:

\begin{equation}
  \label{eq:bgt}
  B_{ij}(GT) = \left(\frac{g_A}{g_V}\right)^2_{\rm{eff}}
  \frac{\langle j||\sum_k \bf{\sigma}^k \bf{t}^k_\pm || i
  \rangle^2}{2 J_i +1},
\end{equation}
where the matrix element is reduced with respect to the spin operator
$\bf{\sigma}$ only (Racah convention~\cite{edmons}) and the sum runs
over all nucleons.  For the isospin rising and lowering operators,
$\bf{t}_\pm = (\bf{\tau}_x \pm i \bf{\tau}_y)/2$, we use the
convention $\bf{t}_+ p = n$; thus, `$+$' refers to electron capture
and `$-$' to  $\beta^-$
transitions.  Finally, $(g_A/g_V)_{\rm{eff}}$ is the effective ratio
of axial ($g_A$) and vector ($g_V$) 
coupling constants that takes into account the
observed quenching of the GT strength~\cite{osterfeld}. We
use~\cite{Martinez96}

\begin{equation}
  \label{eq:gaeff}
  \left(\frac{g_A}{g_V}\right)_{\rm{eff}} = 0.74
  \left(\frac{g_A}{g_V}\right)_{\rm{bare}},
\end{equation}
with $(g_A/g_V)_{\rm{bare}} = -1.2599(25)$~\cite{Towner}. If the
parent nucleus (with isospin $T$) has a neutron excess, then the
GT$_-$ operator can connect to states with isospin $T-1$, $T$, $T+1$
in the daughter, while GT$_+$ can only reach states with $T+1$. This
isospin selection is one reason why the GT$_+$ strength is more
concentrated in the daughter nucleus (usually within a few MeV around
the centroid of the GT resonance), while the GT$_-$ is spread over
10-15~MeV in the daughter nucleus and is significantly more
structured.
The difference of the total GT 
strengths in the $\beta^-$ and electron capture directions is fixed
to the threefold of the neutron excess $(N-Z)$ which is known as the
Ikeda sumrule \cite{Ikeda}.

The last factor in equation~(\ref{eq:rate}), $\Phi_{ij}^\alpha$, is
the phase space integral given by

  \label{eq:phase}
  \begin{eqnarray}
    \label{eq:phase-ec}
    \Phi_{ij}^{ec} = \int_{w_l}^\infty w & p & (Q_{ij}+w)^2
    F(Z,w) \nonumber \\
    & \times & S_e(w) (1-S_\nu(Q_{ij}+w)) dw,
  \end{eqnarray}
  \begin{eqnarray}
    \label{eq:phase-b-}
    \Phi_{ij}^{\beta^-} = \int_1^{Q_{ij}} w & p & (Q_{ij}-w)^2
    F(Z+1,w) \nonumber\\
    & \times & (1-S_e(w)) (1-S_\nu(Q_{ij}-w)) dw,
  \end{eqnarray}
where $w$ is the total, rest mass and kinetic, energy of the electron
or positron in units of $m_e c^2$, and $p=\sqrt{w^2-1}$ is the momentum
in units of $m_e c$. Then the total energy available in
$\beta$-decay, $Q_{ij}$, in units of $m_e c^2$ is given by

\begin{equation}
  \label{eq:qn}
    Q_{ij} = \frac{1}{m_e c^2} (M_p - M_d + E_i -E_j),
\end{equation}
where $M_p, M_d$ are the nuclear masses of the parent and daughter
nucleus, respectively, while $E_i, E_j$ are the excitation energies of
the initial and final states. 
$w_l$ is the capture threshold total energy, rest plus kinetic, in
units of $m_e c^2$ for positron (or electron) capture. Depending on
the value of $Q_{ij}$ in the corresponding electron (or positron)
emission one has $w_l=1$ if $Q_{ij} > -1$, or $w_l=|Q_{ij}|$ if
$Q_{ij} < -1$.  $S_e, S_p,$ and $S_\nu$ are the positron, electron,
and neutrino (or antineutrino) distribution functions, respectively.
For core-collapse conditions, electrons and
positrons are well described by Fermi-Dirac distributions, with
temperature $T$ and chemical potential $\mu$. For electrons,

\begin{equation}
  \label{eq:fermie}
  S_e = \frac{1}{\exp\left(\frac{E_e - \mu_e}{kT}\right)+1},
\end{equation}
with $E_e = w m_e c^2$. The positron distribution is defined
similarly with $\mu_p = - \mu_e$. The chemical potential, $\mu_e$, is
determined from the density inverting the relation

\begin{equation}
  \label{eq:inverme}
  \rho Y_e = \frac{1}{\pi^2 N_A}\left(\frac{m_e c}{\hbar}\right)^3
\int_0^\infty (S_e-S_p) p^2 dp,
\end{equation}
where $N_A$ is Avogadro's number. Note that the density of
electron-positron pairs has been removed in~(\ref{eq:inverme}) by
forming the difference $S_e-S_p$.

The remaining factor appearing in the phase space integrals is the
Fermi function, $F(Z,w)$, that corrects the phase space integral for
the Coulomb distortion of the electron or positron wave function near
the nucleus.  It can be approximated by
\begin{equation}
  \label{eq:Fzw}
  F(Z,w)=2(1+\gamma)(2pR)^{-2(1-\gamma)} \frac{|\Gamma (\gamma +
    i y)|^2}{|\Gamma (2\gamma+1)|^2} e^{\pi y},
\end{equation}
where $\gamma=\sqrt{1-(\alpha Z)^2}$, $y=\alpha Zw/p$, $\alpha$ is the
fine structure constant, and $R$ is the nuclear radius.

Laboratory studies of $\beta$-decays and electron captures can only observe
the GT strength within the $Q_\beta$ value, i.e. the part of the GT strength
which is energetically reachable. Unfortunately this is usually only a small
part of the GT strength distribution. A breakthrough in experimental
investigations of GT distributions has been achieved with the study
of charge-exchange reactions at intermediate projectile energies where
the reaction cross section at extreme forward angles is proportional
to the GT strength. Hence, using the pioneering probes like
(p,n) and (n,p) reactions it has been possible to measure the GT$_-$
and GT$_+$ strength
distributions, respectively, over a large excitation 
energy interval \cite{Gade,Triumf}.
These experiments revealed two quite decisive results: At first,
the GT strength is distributed over many states in the daughter nucleus
caused by the residual nuclear interaction which does not commute with
the 
$\bf{\sigma} \bf{t}_\pm$ operator. Secondly, the GT strength is 'quenched'.
As the GT operator does not act on the spatial nuclear wave functions it
cannot connect nuclear orbitals in 
different harmonic oscillator shells. However, it has been
found that so-called $0 \hbar \omega$ shell model calculations, which account
for all two-nucleon correlations of the valence nucleons within one major
oscillator shell, overestimate the measured GT strengths by a constant
factor (which lead to the introduction of the effective axial-vector
coupling constant as discussed above) \cite{brownq,Langanke95}. 
The origin of the quenching
is mainly due to higher-shell admixtures in the ground state wave function
which causes 
a portion of the GT strength to be shifted to moderately high excitation 
energies. In fact, by detailed (p,n) reaction studies on $^{90}$Zr GT$_-$
strength has been identified up to excitation energies 
of order 90 MeV \cite{Sakai}.

While the GT data from the (p,n) and (n,p) experiments have been
inexpensible for the understanding of GT distributions in nuclei, these
probes have relatively poor energy resolution, in particular the (n,p)
experiments. Nowadays high-resolution measurements of the GT strengths
are possible due to the development of ($^3$He,t) 
\cite{Fujita} and (d,$^2$He) \cite{Frekers} reaction 
probes which allow to determine the GT$_-$ and GT$_+$ strength 
with a resolutions of
30 keV and 150 keV, respectively. 
Although experimentally one can only determine
the GT strength distributions for nuclear ground states, these data serve
as precious constraints and tests for the nuclear models which are then
required to evaluate the GT strengths for the excited nuclear states which
are needed to calculate the $\beta$-decay and electron capture rates
at finite temperatures.

\begin{figure}[t]
\begin{center}
    \includegraphics[width=0.8\columnwidth]{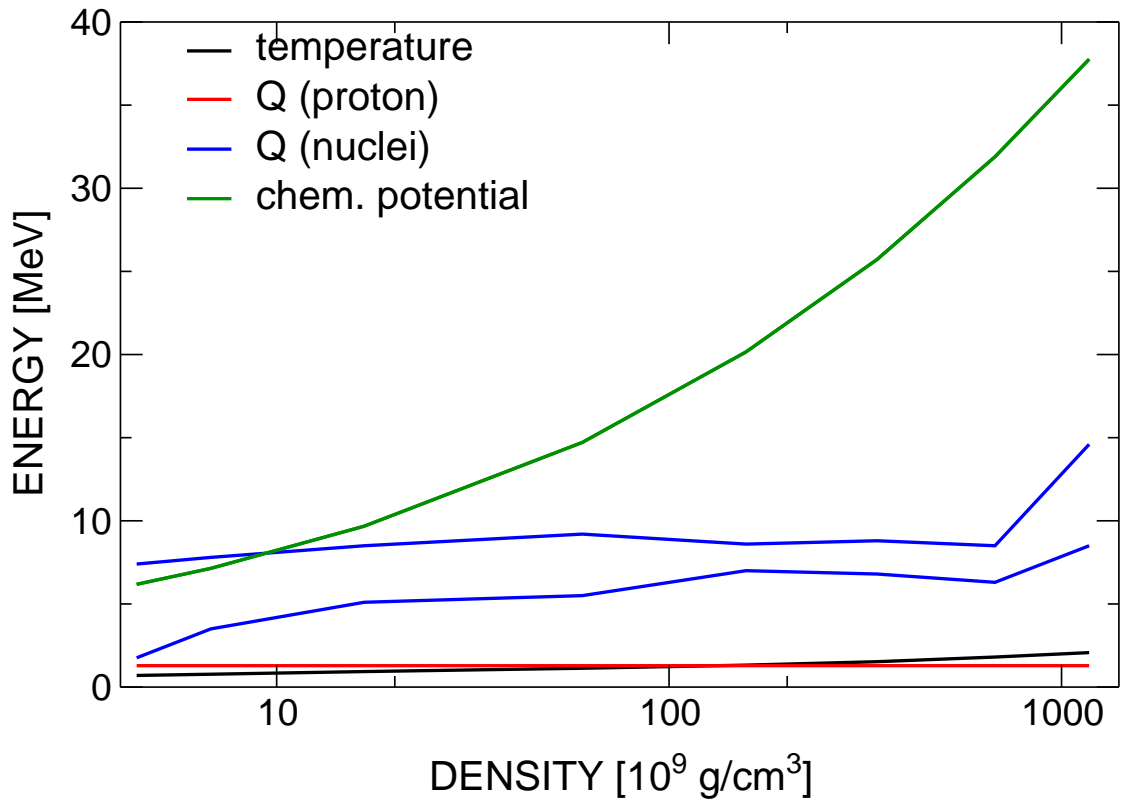}
 \leavevmode
\caption{\small
Sketch of the various energy scales related to electron capture on
protons and nuclei as a function of densitiy during a
supernova core collapse simulation. Shown are the chemical potential,
respectively Fermi energy, of electrons, the Q-values for electron
capture on free protons (constant), and the average Q-value for electron
capture on nuclei for the given composition at each density.
\label{fig:energyscales}}
\end{center}
\end{figure}

Before we discuss which reactions are important and how one can calculate the
relevant reaction rates, it is useful to compare the various energy scales
of the problem. This is done in Fig.~\ref{fig:energyscales} 
which shows the electron
chemical potential $\mu_e$ and typical core temperatures and nuclear Q-values
(defined as $\mu_n-\mu_p$, the difference of neutron and proton chemical 
potential) as functions of the central stellar density. Obviously
$\mu_e$, which is proportional to $\rho^{1/3}$, grows much faster
than the other energies. This has several important quantities:
\begin{enumerate}
\item
With growing density
nuclear beta-decays are increasingly hindered by the presence of
electrons which block the available decay phase space. For densities 
larger than $10^{10}$ g/cm$^3$ beta-decays are effectively irrelevant.
As we will show below,
this is quite fortuitious as reliable beta-decay rates are difficult
to calculate, but for supernova simulations are only required for nuclei
for which large-scale shell model calculations can be performed.
\item
For densities $\rho$ less than a few $10^{10}$ g/cm$^3$,
the electron
chemical potential is of the same order of magnitude as the nuclear
$Q$-value. Hence electron capture rates are quite sensitive to the
detailed GT$_+$ strength distribution. Fortunately under such conditions
electron capture mainly occurs on nuclei in the mass range $A \sim 60$
for which large-scale nuclear shell model calculations, which ---
as we will show below ---
describe the GT$_+$ distribution quite well,  can be performed.
\item
For even higher densities nuclei with mass numbers $A > 65$, for which
detailed shell model calculations are yet not possible due to computational
limitations, become quite abundant. However, when this happens during the
collapse the electron chemical potential is noticeably larger than the
$Q$-value. As a consequence it might be sufficient for a reliable
estimate of the capture rates to employ a nuclear model which 
describes the centroid and the total GT$_+$ strength reasonably well.
\end{enumerate}

\begin{figure}
\begin{center}
    \includegraphics[width=0.85\columnwidth]{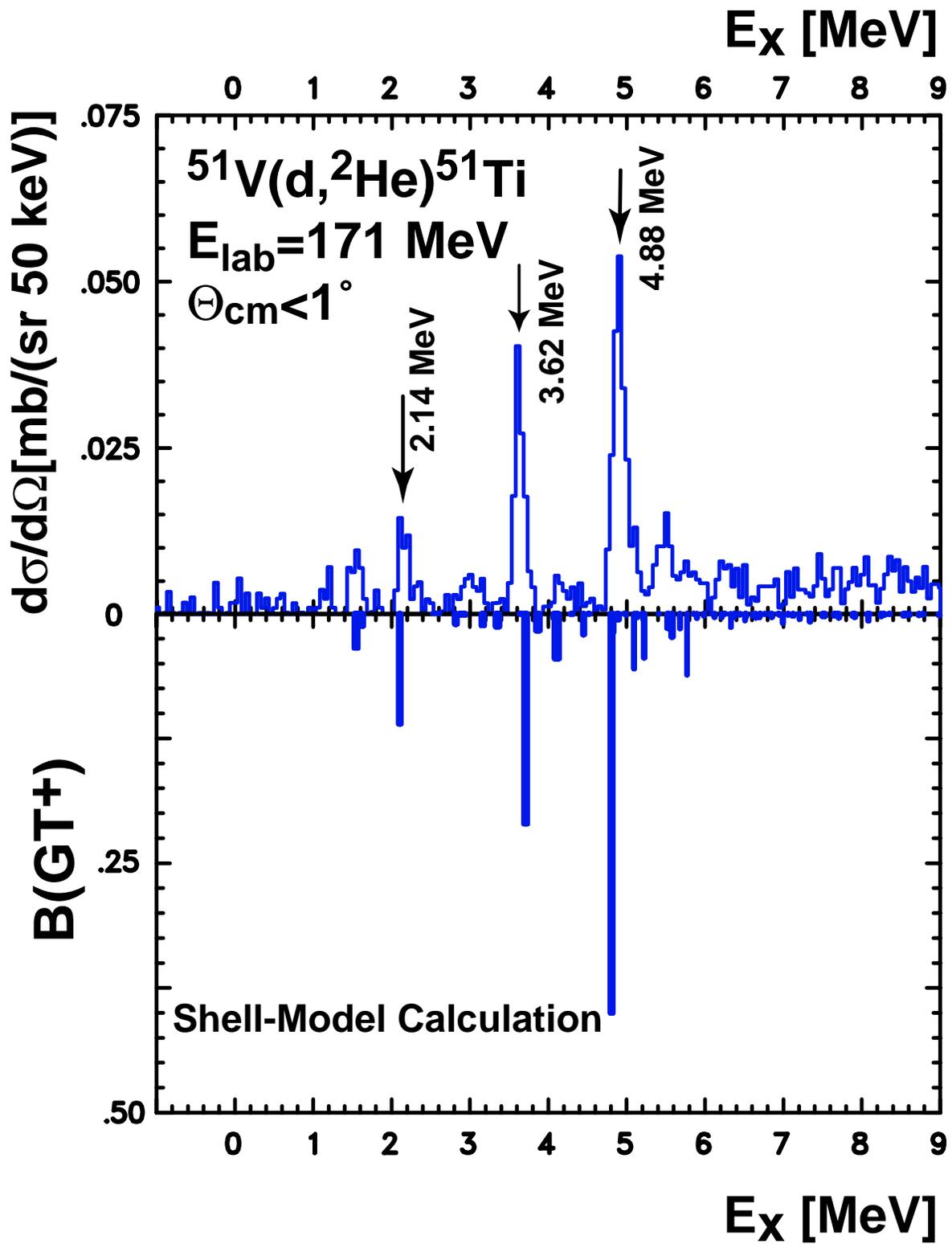}
 \leavevmode
\caption{\small
Comparison of the measured energy spectrum (left)
and the $^{51}$V(d,$^2$He)$^{51}$Ti cross section (right) at
forward angles (which is proportional to the GT$_+$ strength) with the
shell model predictions (the GT data are from \protect\cite{Baeumer}).
  \label{fig:v51}}
\end{center}
\end{figure}

When Fuller, Fowler and Newman (FFN) \cite{FFN2,FFN1,FFN3,FFN4} 
did their pioneering calculations of stellar
weak-interaction rates based on the independent particle model,
only little experimental guidance existed; 
in particular, no experimental information about the GT$_+$ strength 
distribution was available. This situation then changed drastically
when GT distributions, mainly due to the (p,n) work at the Indiana cyclotron
\cite{Gade} and to the (n,p) work performed at 
TRIUMF \cite{Williams,Alford,Roennquist,El-Kateb}, 
for several nuclei in the 
iron mass range became available. It had already become clear due to
studies of sd-shell nuclei \cite{brownq}, 
that a description of the GT distributions
requires a careful consideration of nucleon-nucleon correlations
and that the diagonalization shell model
would be the method of choice to do the job.
Although studies of mid-shell pf nuclei were initially prohibited by
computational limitations, these could be overcome by the development
of new codes like ANTOINE by Etienne Caurier and by growing computer power.
Finally, Caurier {\it et al.} then performed large-scale shell model
calculations for nuclei in the iron mass range which indeed reproduced
the available experimental GT$_-$ and GT$_+$ data quite 
well \cite{Caurier99}. Furthermore,
these studies also described the low-energy nuclear spectra well.
In summary, the study of Caurier {\it et al.} implied that the shell model
was indeed the tool to calculate the weak-interaction rates as needed
for late-stage presupernova core evolution for which the stellar composition is
dominated by nuclei with mass 
numbers $A=45-65$ \cite{Caurier99}. This task has been
executed in \cite{Langanke00} where the rates for 
electron and positron captures and for $\beta^+$ and $\beta^-$ decays
have been calculated for stellar conditions and for more than 100 nuclei
in the mass range $A=45-65$. The rates are based on the explicite
evaluation of the GT distributions for the lowest energy states of each 
nucleus, supplemented by the contributions of the `back-resonances'.
Calculated energies or GT transition strengths have been replaced by 
experimental data whenever possible. In recent years the shell model
GT distributions have been stringently tested when the (d,$^2$He) GT$_+$
with an order of magnitude better energy resolution than the (n,p) data became
available \cite{Frekers}. Although not perfect, the shell model results proved
to be quite accurate. Comparisons of electron capture rates calculated
with the experimental (d,$^2$He) and with the shell model GT$_+$ 
distributions (considering ground states only) generally agree within a factor
of 2 at the relevant temperature and density regimes. 
Fig.~\ref{fig:v51} compares the shell model spectrum and GT$_+$ distribution
for $^{51}$V with the experimental data.

\begin{figure}
\begin{center}
    \includegraphics[width=0.85\columnwidth]{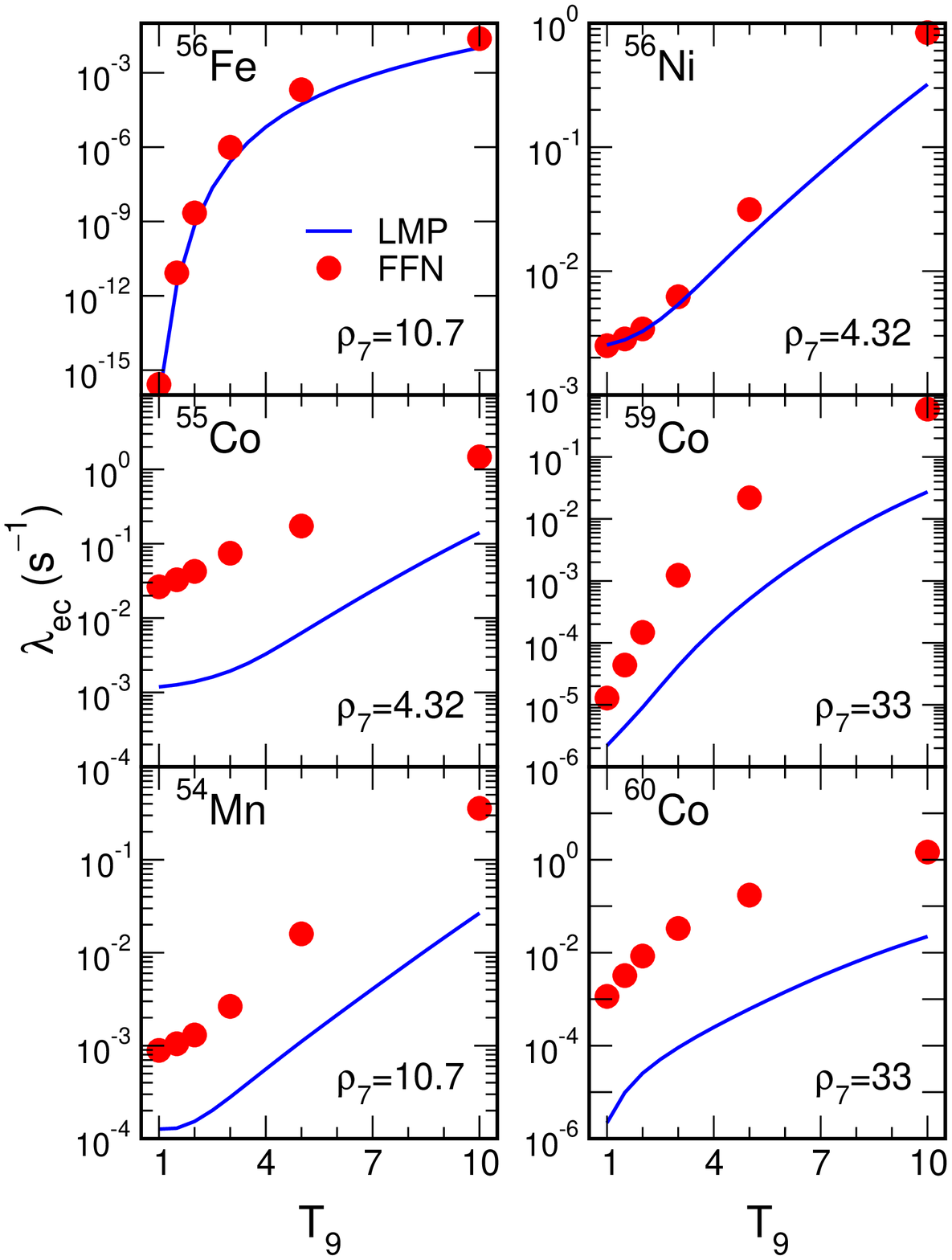}
 \leavevmode
\caption{\small
Shell model electron-capture rates as a function of
temperature ($T_9$ measures the temperature in $10^9$ K) and for
selected densities ($\rho_7$ defines the density in
$10^7$~g~cm$^{-3}$) and nuclei.  For comparison, the FFN rates are
given by the full points. (from \protect\cite{Langanke00})
\label{fig:ecrates}}
\end{center}
\end{figure}

Fig. \ref{fig:ecrates}  compares the shell model and FFN electron capture 
rates for several nuclei at temperatures and densities which are typical
for the core in late-stage presupernovae. Importantly the shell model rates
are systematically smaller than the FFN rates with quite significant
consequences for the presupernova and core-collapse evolutions, as 
is discussed below. The reasons for the differences between the FFN
and shell model rates are discussed in \cite{Langanke00}. They include
differences in the systematics of nuclear pairing effects (which have been
added empirically by FFN), but also improved experimental data which
had not been available when FFN derived their seminal stellar
weak-interaction rate tables. The differences between the FFN and shell model
$\beta_-$ rates are smaller than for electron capture and do not show
a systematic tendency. Comparisons are presented in 
\cite{Martinez99,Langanke00}.

\begin{figure}
\begin{center}
    \includegraphics[width=0.99\columnwidth]{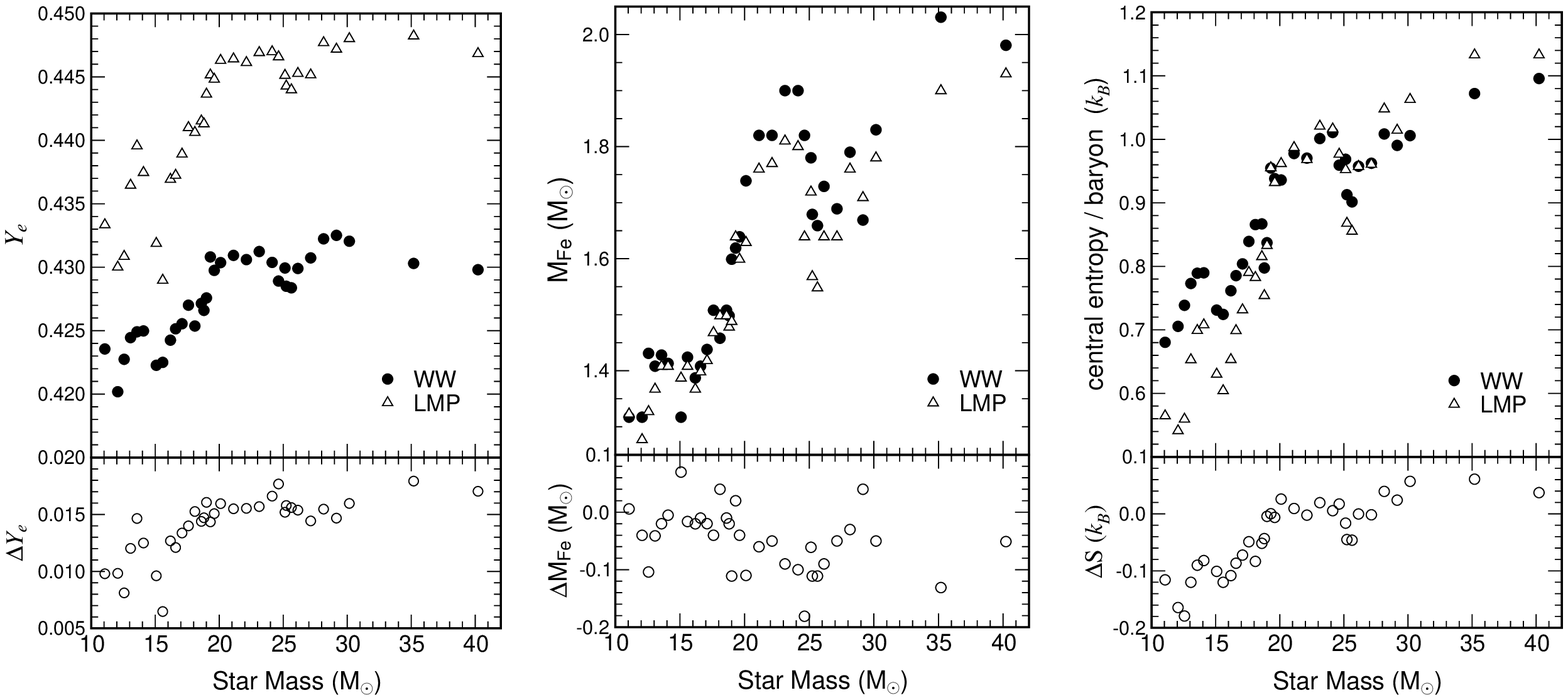}
 \leavevmode
\caption{\small
Comparison of the center values of $Y_e$ (left), the iron
core sizes (middle), and the central entropy (right)
for $11-40 M_\odot$ stars between the WW models,
which used the FFN rates,
and the  ones using the shell model weak interaction 
rates (LMP) (from \protect\cite{Heger01a}).
  \label{fig:supcore}}
\end{center}
\end{figure}

To study the influence of the shell model rates on presupernova models
Heger {\it et al.} \cite{Heger01,Heger01a}
have repeated the calculations of Weaver and Woosley
\cite{WW95} keeping the stellar physics, except for the weak rates,
as close to the original studies as possible.
Fig. \ref{fig:supcore} examplifies the consequences of the shell model
weak interaction rates for presupernova models in terms of the three
decisive quantities: the central $Y_e$ value and entropy and the iron
core mass.
The central values of $Y_e$  at the onset of core collapse
increased by 0.01--0.015 for the new rates. This is a significant effect.
We note that the new models also result in lower core entropies for
stars with $M \leq 20 M_\odot$, while for
$M \geq 20 M_\odot$, the new models actually have a
slightly larger entropy.
The iron core masses are generally smaller
in the new models where the effect is larger for more massive stars ($M
\ge 20 M_\odot$), while for the most common supernovae ($M \le 20 M_\odot$) the
reduction is by about 0.05 $M_\odot$.

Electron capture dominates the weak-interaction processes during
presupernova evolution. However, as already anticipated by
Aufderheide et al. \cite{Aufderheide}, during silicon burning, $\beta$ decay
(which increases $Y_e$) can compete and adds to the further cooling of
the star \cite{Heger01,Heger01a}. 
With increasing densities, $\beta$-decays are hindered as the
increasing Fermi energy of the electrons blocks the available phase
space for the decay. Thus, during collapse $\beta$-decays can be
neglected

We note that the shell model weak interaction rates predict the
presupernova evolution to proceed along a temperature-density-$Y_e$
trajectory where the weak processes are dominated by nuclei rather
close to stability. Thus it will be possible, after radioactive
ion-beam facilities become operational, to further constrain the shell
model calculations by measuring relevant beta decays and GT
distributions for unstable nuclei. Ref.~\cite{Heger01a}
identifies those nuclei
which dominate (defined by the product of abundance times rate) the
electron capture and beta decay during various stages of the final evolution
of a 15$M_\odot$, 25$M_\odot$ and 40$M_\odot$ star.

The continuous electron captures in the core do not only reduce
the electron-to-nucleon ratio, and hence the pressure which the 
electron degeneracy  can stem against the gravitational collapse,
it also shifts the nuclear composition in the core to more 
neutron-rich and heavier nuclei (see Figs. \ref{fig:NSEabund}.
As noted by Fuller, this tendency might
significantly reduce the electron captures once nuclei with neutron
numbers $N>40$ are abundant \cite{Fuller82}. 
The argument is based on the independent
particle model (IPM): GT transitions can only proceed to the same
nuclear orbital or to the spin-orbit partner in the same harmonic oscillator
shell. Consequently, for nuclei with charge-numbers $Z<40$, but
neutron numbers $N>40$ GT transitions are completely blocked by
the Pauli principle within the independent particle model \cite{Bruenn85}.
Hence it has been assumed for many years that electron captures during
the collapse phase occur on free protons rather than on nuclei, although
the abundance of free protons is noticeably smaller than the one of 
nuclei (e.g. \cite{Bethe90}). 
However, the IPM picture is not applicable. At first,
the core has a sizable finite temperature which can lead to thermal
unblocking of GT transitions \cite{Wambach84}. 
More importantly, the energy gap between the pf-shell 
($f_{5/2},p_{1/2}$) and the lowest orbital of the next (sdg) shell
($g_{9/2}$) is only 2--3 MeV. The residual interaction 
thus mixes nucleons in these orbitals leading to neutron holes
in the pf-shell or to proton excitations to sdg (mainly $g_{9/2}$)
orbitals. Both effects unblock GT$_+$ transitions. In fact, recent
experiments have observed a non-vanishing GT$_+$ strength
for $^{72}$Ge and $^{76}$Se \cite{Frekerspriv}
which should be exactly zero in the
IPM. 

An additional complication arizes for the calculation of the stellar
electron capture rates due to the finite temperature of the core.
At a typical core temperature of $T=1$ MeV, 
the internal excitation energy of a nucleus with
mass number $A \sim 80$ is about 8-10 MeV, i.e. much larger than
the energy gap between the pf- and sdg-shells. Hence, one expects
that at the stellar temperatures the unquenching of GT$_+$ transitions
is more effective than for the ground state. Unfortunately, 
diagonalization shell model calculations, which have been successfully
applied to the evaluation of electron capture rates for lighter
(i.e. pure pf-shell) nuclei cannot be employed for the required
multi-shell calculations due to the extreme dimensions of the
model spaces involved (for $^{72}$Ge a calculation within
the pf-sdg model space contains about $10^{24}$ configurations).
To overcome this problem, a hybrid model has been 
suggested \cite{Langankehyb}.
In the first step, the nucleus is described by a Slater determinant
with temperature-dependent occupation numbers for the valence orbitals.
These occupation numbers are determined within the Shell Model Monte Carlo
(SMMC) approach which allows the calculation of 
thermally-averaged properties of 
nuclei in extremely large model spaces (here the full pf-sdg shells)
and considering the relevant correlations among nucleons \cite{Koonin97}.
Although the SMMC allows in principle the calculation of GT 
strength distributions, such calculations are numerically quite challenging
and usually enable only the determination of the first 2 moments
of the distribution which is insufficient for the calculation of stellar
electron capture rates \cite{Radha97}. Therefore, the capture 
rates are derived in a second step
from the GT$_+$ strength distributions calculated 
within the Random Phase Approximation built on top of the temperature-dependent
Slater determinant. Using the hybrid model, capture rates have been determined
for about 200 nuclei and, combined with the rates for pf-shell 
nuclei \cite{Langanke00},
a rate table for the appropriate collapse temperatures, densities and
$Y_e$ values has been compiled. 

\begin{figure}
\begin{center}
    \includegraphics[width=0.5\columnwidth,angle=270]{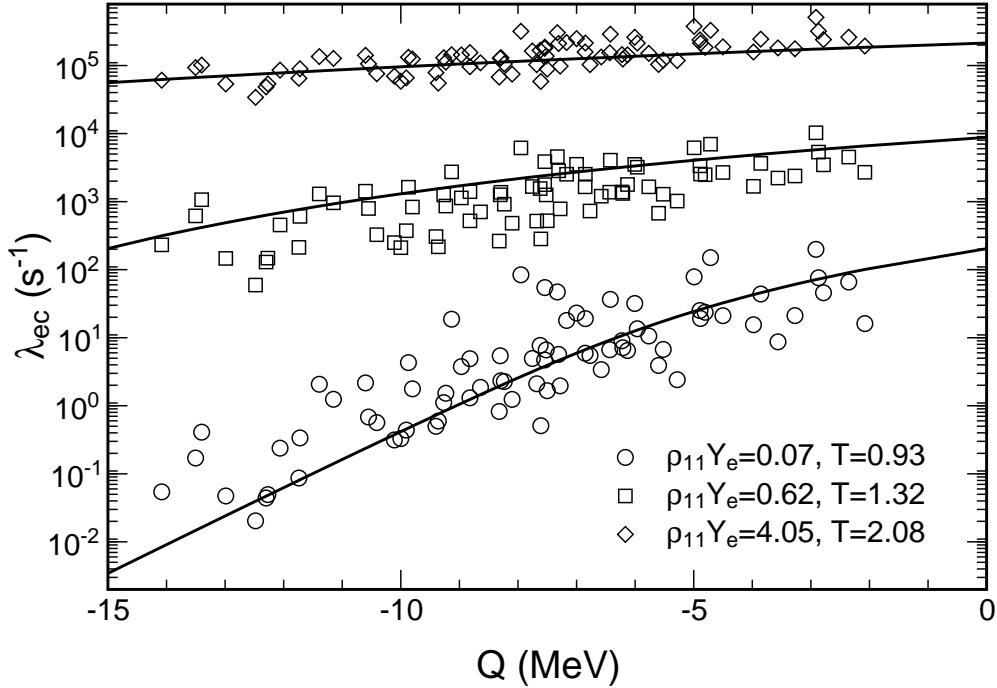}
\caption{\small
Electron capture rates on nuclei as function of $Q$-value
for three different stellar conditions. Temperature is measured in
MeV.  The solid lines represent the approximate $Q$-dependence
of the rates as defined in Eq.~(\ref{eq:qvalue}). (from \protect\cite{Langanke03})} 
\label{fig:qvalues}
\end{center}
\end{figure}

The Random Phase Approximation (RPA)
treats nucleon-nucleon correlations only on the
1-particle-1-hole level and hence misses higher-order correlations,
which are considered within the shell model and which are important
to reproduce details of the GT strength like fragmentation and the low-energy
tail of the strength
(which is particularly important for the calculation of $\beta$-decay rates). 
However, the RPA is known to give a reasonable
account of the centroid of strength distributions as well as of the total
GT$_+$ strength. Thus, one should be cautious to use the hybrid model
for stellar conditions for which the electron chemical potential
is similar to the nuclear Q-value as then the capture rate will be 
quite sensitive to the detailed GT$_+$ strength distribution
(shell model and hybrid model rates can then easily deviate by factors 2--3
\cite{Sampaiothesis}). Luckily, 
$\mu_e \sim Q$ occurs at core densities at which the core composition
is still dominated by nuclei $A \approx 55-65$ for which large-scale 
shell model rates exist \cite{Langanke00}. 
At higher densities, when nuclei with $A>65$, 
for which diagonalization shell model calculations are
not possible and the capture rates have to be estimated by the hybrid
model, are abundant, the electron chemical potentials is noticeably larger than
the typical nuclear $Q$-values. Under these conditions capture rates are
mainly
sensitive to the centroid and the total strength of the GT$_+$ distributions
which are reasonably well described within the RPA. At even higher
densities,
say $\rho>10^{11}$ g/cm$^3$, 
the capture rates on nuclei become quite
similar at larger densities, 
depending now
basically only on the total GT strength, but not its detailed
distribution.  This is demonstrated in Fig.~\ref{fig:qvalues} which
shows the hybrid model capture rates as function of $Q$-value at 3
different stellar conditions.  The $Q$-value dependence of the capture
rate for a transition from a parent state at excitation energy $E_i$
to a daughter state at $E_f$ ($\Delta E = E_f-E_i)$ is well
approximated by \cite{FFN4,Langanke03}

\begin{eqnarray}
\lambda (Q) & = & \ln(2) 10^{-3.588} (kT/m_e)^5 |M_{GT}|^2 \nonumber \\
& \times  & (F_4(\eta) - 2
\chi F_3(\eta) +
\chi^2 F_2 (\eta))
\label{eq:qvalue}
\end{eqnarray}

where $\chi = (Q-\Delta E)/kT$, $\eta=U_F +m_e + Q - \Delta E$ and $U_F$
is the kinetic electron chemical potential (the total includes the
electron rest mass $m_e$). The quantities
$F_k$ are the relativistic Fermi integrals of order $k$.

\begin{figure}
\begin{center}
    \includegraphics[width=0.80\columnwidth]{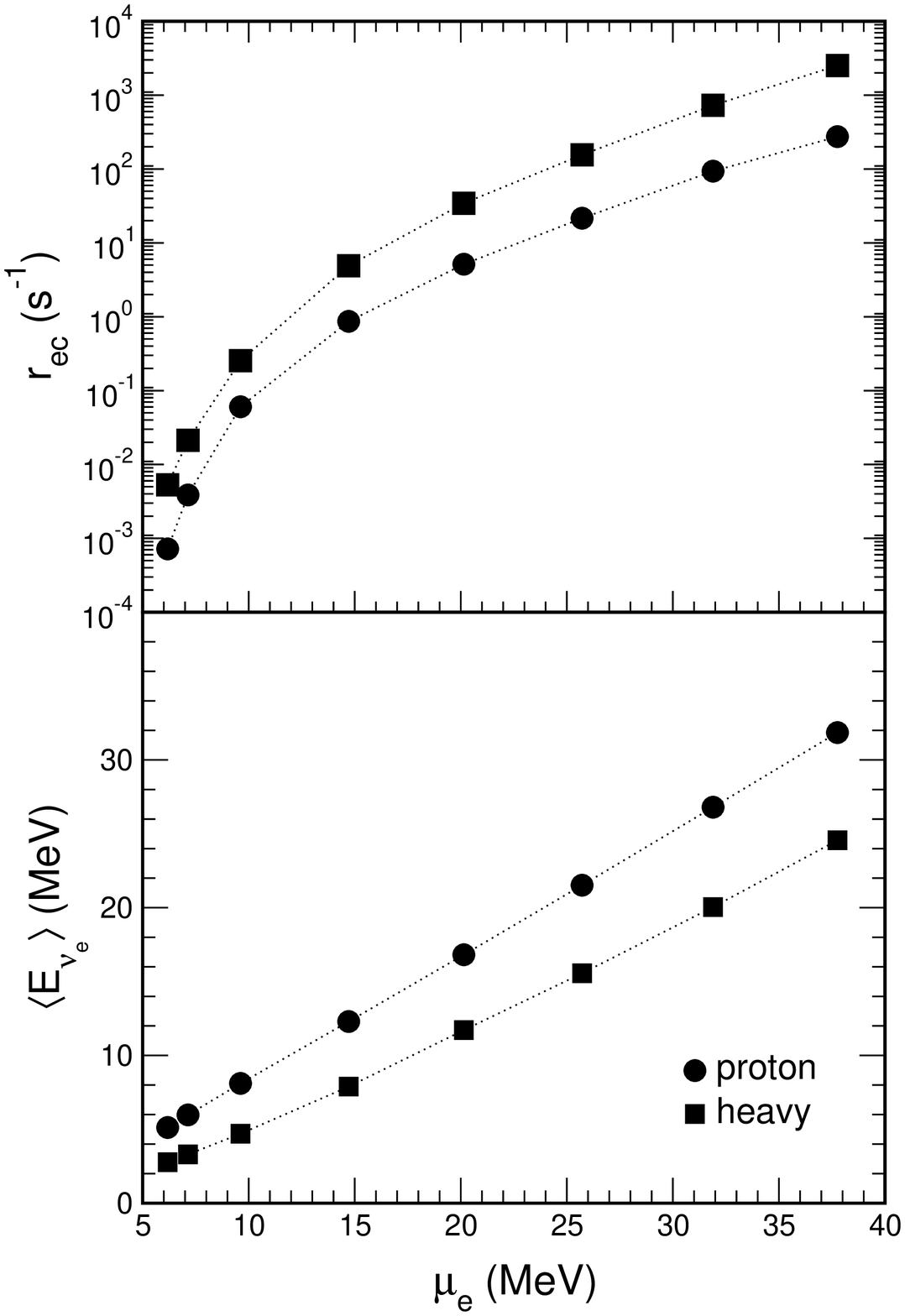}
\caption{\small
The upper panel compares the products of number abundances
and electron capture rates for free protons ($r_p$, circles) and
nuclei ($r_h)$, squares) as functions of electron chemical
potential along a stellar collapse trajectory. The lower panel
shows the related average energy of the neutrinos emitted by
captures on nuclei and protons. The results for nuclei are
averaged over the full nuclear composition. (from \protect\cite{Langanke03})}
\label{fig:average}
\end{center}
\end{figure}

Thus, finite temperature and correlations unquench the GT$_+$
strength noticeably. As electrons have to overcome a larger threshold for
neutron-rich nuclei than for protons, the electron capture rate 
on individual nuclei
is smaller than on free protons. However, what matters is the 
product of abundance times capture rate. As nuclei are much more abundant
in the collapsing core, due to its small entropy of order 1 
$k_{\mathrm{B}}$ per nucleon,
than protons, electron capture on nuclei clearly dominates
as is shown in Fig. \ref{fig:average}.

It is also important to stress that electron capture on nuclei and on
free protons differ quite noticeably in the neutrino spectra they
generate.  Fig.~\ref{fig:average} demonstrates that capture on nuclei
having a mean energy 40-60\% less than that produced by capture on
protons.  These differences are quite relevant as neutrino-matter
interactions, which scale with the square of the neutrino energy
$E_\nu$, are essential for the collapse simulations.  Although capture
on nuclei under stellar conditions involves excited states in the
parent and daughter nuclei, it is mainly the larger $Q$-value which
significantly shifts the energies of the emitted neutrinos to smaller
values.

\begin{figure}
\begin{center}
    \includegraphics[width=0.75\columnwidth]{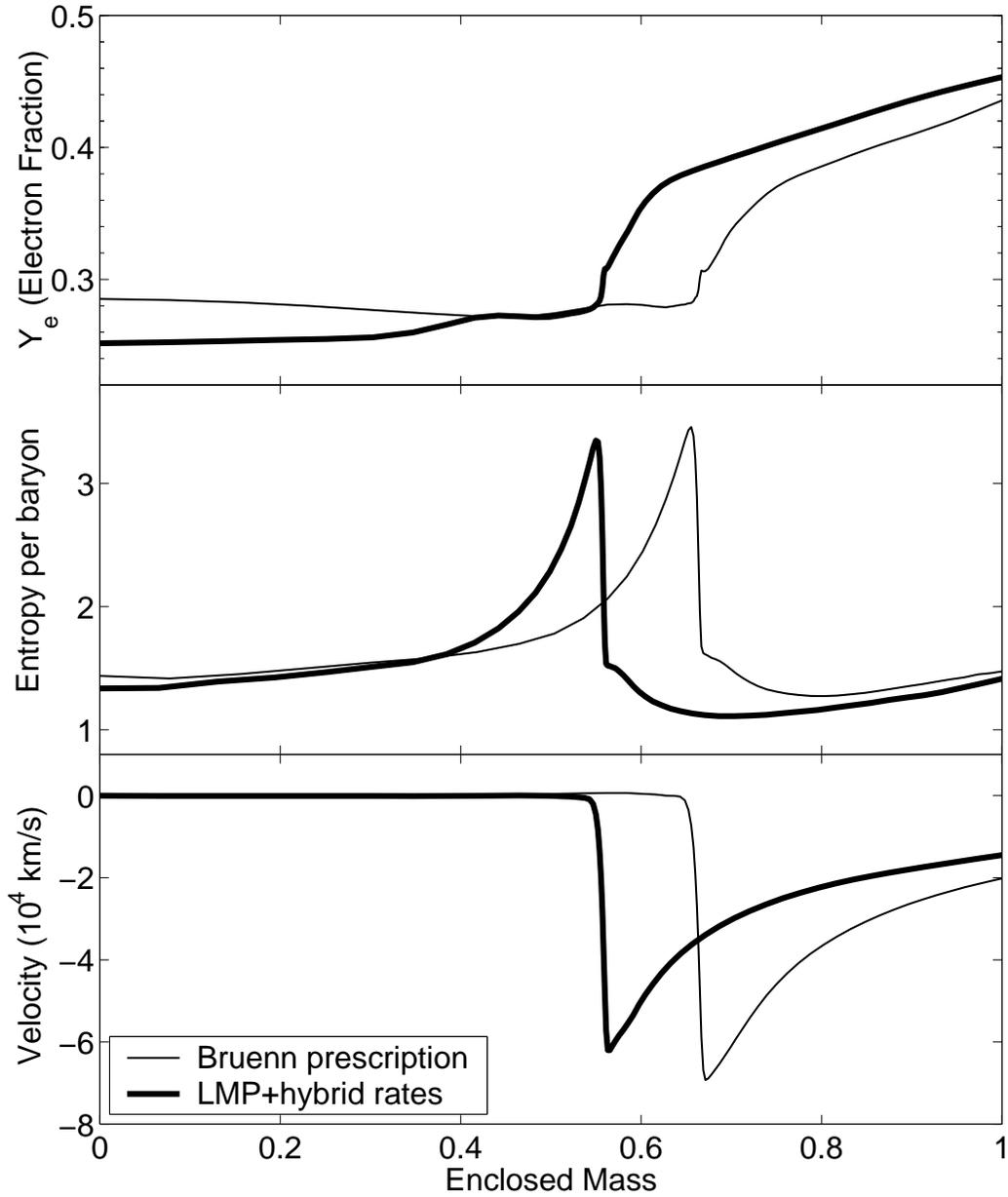}
\caption{\small
The electron fraction, entropy, and velocity
as functions of the
enclosed mass at bounce for a 15 solar mass model \cite{Hix03}.
The thin line is a simulation using the Bruenn parameterization
while the thick line is for a simulation using the LMP and Hybrid
reaction rate sets (from \protect\cite{Hix03}).
  \label{fig:bounce}}
\end{center}
\end{figure}

The effects of this more realistic implementation of electron capture
on heavy nuclei have been evaluated in independent self-consistent
neutrino radiation hydrodynamics simulations by the Oak Ridge and
Garching collaborations \cite{Langanke03,Hix03,Marek06b}.
The changes compared to the previous simulations, which
adopted the IPM rate estimate from Ref. \cite{Bruenn85} and hence
basically ignored electron capture on nuclei, are significant.
Fig.~\ref{fig:bounce} shows a key result: in denser regions,
the additional electron capture on heavy nuclei results in more
electron capture in the new models.  In lower density regions, where
nuclei with $A<65$ dominate, the
shell model rates \cite{Langanke00} result in
less electron capture.
The results of these competing effects can be
seen in the first panel of Figure~\ref{fig:bounce}, which shows the
distribution of $Y_e$ throughout the core at bounce (when the maximum
central density is reached).  The combination of increased electron
capture in the interior with reduced electron capture in the outer
regions causes the shock to form with 16\% less mass interior to it
and a 10\% smaller velocity difference across the shock.
In spite of this mass reduction, the radius from which the shock is
launched is actually displaced slightly outwards to 15.7 km from 14.8 km
in the old models.
Also the altered gradients in
density and lepton fraction  play an important role in the behavior
of the shock. Though also the new models fail to produce explosions in
the spherically symmetric case, the altered gradients allow the shock in
calculations with improved capture rates to reach a somewhat larger 
maximum radius than in the old models.

\subsection{Inelastic neutrino-nucleus scattering}
\label{sec:nuclinel}

While the neutrinos can leave the star unhindered during the
presupernova evolution, neutrino-induced reactions become more and
more important during the subsequent collapse stage due to the
increasing matter density and neutrino energies; the latter are of order
a few MeV in the presupernova models, but increase roughly approximately
to the electron chemical potential~\cite{Bruenn85}.
Coherent and conservative
neutrino scattering off nuclei and scattering on electrons are
the two most important neutrino-induced reactions during the collapse.
The first reaction randomizes the neutrino paths out of the core and, at
densities of about $10^{12}$ g/cm$^3$, the neutrino diffusion timescale
gets larger than the collapse time; the neutrinos are trapped in the
core for the rest of the contraction. Scattering off electrons and
neutrino emission and reabsorption then establishes thermal equilibrium
between the trapped neutrinos and the stellar matter rather quickly.
The inner core with trapped neutrinos collapses with conserved 
lepton number and adiabatically as a homologous unit until it reaches 
densities slightly in excess of nuclear matter, generating a
bounce and launching the supernova shock wave.

Neutrino-induced reactions on nuclei, other than coherent scattering,
can also play a role during the collapse and 
the subsequent explosion phase~\cite{Haxton88}. Note that during the
collapse only $\nu_e$ neutrinos are present. Thus, charged-current
reactions $A(\nu_e,e^-)A'$ are strongly blocked by the large electron chemical
potential~\cite{Bruenn91,Sampaio01}. Inelastic neutrino scattering on
nuclei can compete with $\nu_e+e^-$ scattering at higher neutrino
energies $E_\nu \ge 20$ MeV~\cite{Bruenn91}. Here the cross sections
are mainly dominated by first-forbidden transitions. Finite-temperature
effects play an important role for inelastic $\nu+A$ scattering below
$E_\nu \le 10$ MeV. This comes about as nuclear states get thermally
excited which are connected to the ground state and low-lying excited
states by modestly strong GT transitions and increased phase space. As a
consequence the cross sections are significantly increased
for low neutrino energies at finite temperature and might be comparable
to $\nu_e+e^-$ scattering~\cite{Sampaio02}.
Thus, inelastic neutrino-nucleus scattering, which 
until very recently has been neglected
in collapse simulations, should be implemented in such studies. 
A first result of such an implementation is given below.

\begin{figure}
\begin{center}
   \includegraphics[width=0.75\columnwidth]{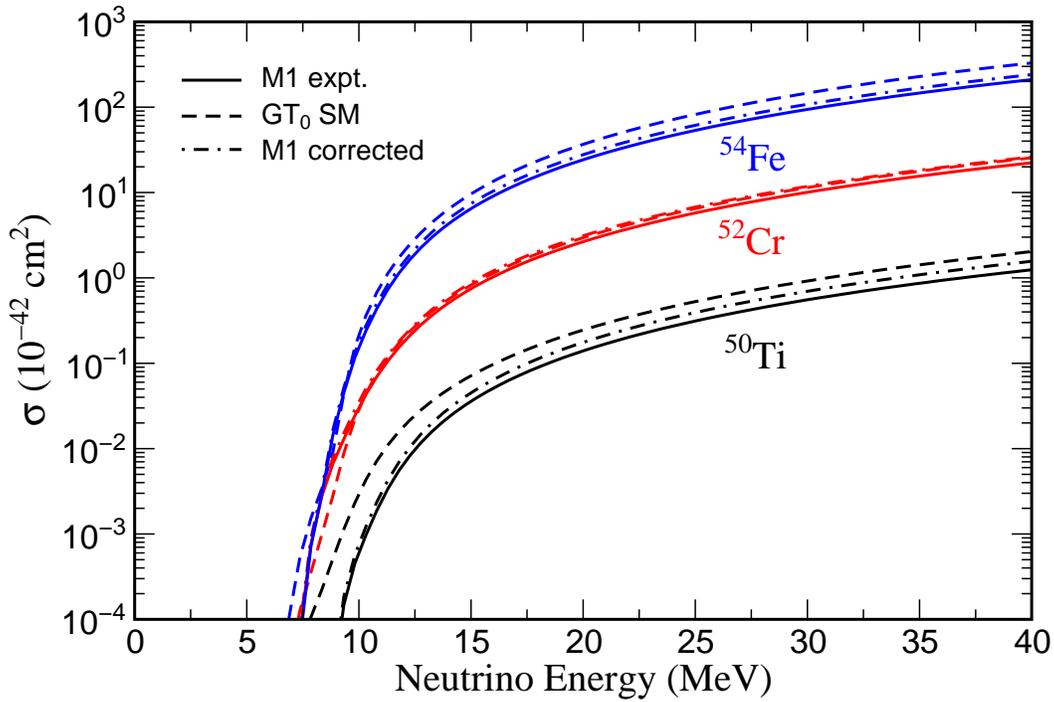}\\
   \caption{
   {\small
    Neutrino-nucleus cross sections calculated from the M1 data
    (solid lines) and the shell-model GT$_0$ distributions
    (dotted) for $^{50}$Ti (multiplied by 0.1), $^{52}$Cr, and
    $^{54}$Fe (times 10). The long-dashed lines show the cross
    sections from the M1 data, corrected for possible strength
    outside the experimental energy window. (from \protect\cite{Langanke04})}}
    \label{fig:M1}
    \end{center}
     \end{figure}

Currently no data for inelastic neutral-current neutrino-nucleus
cross sections are available for supernova-relevant nuclei ($A\sim60$).
However, as has been demonstrated in \cite{Langanke04},
precision M1 data, obtained by inelastic electron scattering, supply
the required information about the
Gamow-Teller GT$_0$ distribution which determines the inelastic
neutrino-nucleus cross sections for supernova neutrino energies.
The argumentation is built on the observation that for M1 transitions
the isovector part dominates and
the respective isovector M1 operator
is given by
\begin{equation}
{\cal O}_{iv} = \sqrt{\frac{3}{4\pi}} \sum_k \left[ {\bf l}(k) {\bf t_z}(k)
+ 4.706 {\bf \sigma}(k) {\bf t_z}(k) \right] \mu_N
\end{equation}
where the sum is over all nucleons and $\mu_N$ is the nuclear magneton,
and that the spin part of the
isovector M1 operator is proportional to the desired
zero-component of the GT operator.
Thus, experimental M1 data yield the needed GT$_0$ information to
determine supernova neutrino-nucleus cross sections, to the extent
that the isoscalar and orbital pieces present in the M1 operator can be
neglected. First, it is wellknown that the major strength
of the orbital and spin M1 responses are energetically well separated.
Furthermore, the orbital part is strongly related to deformation and is
suppressed in spherical nuclei, like $^{50}$Ti, $^{52}$Cr, $^{54}$Fe.
These nuclei have the additional advantage that M1 response data exist
from high-resolution inelastic electron scattering experiments \cite{vNC}.
Satisfyingly, large-scale
shell model calculations reproduce the M1 data quite well, even in
details \cite{Langanke04}. The calculation also confirms that the orbital
and isoscalar M1 strengths are much smaller than the isovector spin strength.
Thus, the M1 data represent, in a good approximation, the needed GT$_0$
information (upto a known constant factor). Fig. \ref{fig:M1}
compares the inelastic
neutrino-nucleus cross sections for the 3 studied nuclei, calculated
from the experimental M1 data and from the shell model GT$_0$ strength.
The agreement is quite satisfactory. It is further improved, if one
corrects for possible M1 strength outside of the explored experimental
energy window.
Also differential
neutrino-nucleus cross sections as functions of initial and final
neutrino energies, calculated from the M1 data and the shell model,
agree quite well \cite{Langanke04}. On the basis of this comparison, one can
conclude that the shell model is validated for the  calculations of
inelastic neutral-current
supernova  neutrino-nucleus cross sections. This model can then also be
used to calculate these cross sections at the finite temperature in the
supernova environment \cite{Langanke04}. Juodagalvis {\rm et al.}
have calculated double-differential inelastic neutrino-nucleus 
cross sections for many nuclei in the $A\sim56$ mass range, 
based on shell model treatment of
the GT transitions and on RPA studies of the forbidden contributions
\cite{Juodagalvis05}. A detailed table of the cross sections as function of
initial and final neutrino energies and for various temperatures
is available from the authors of \cite{Juodagalvis05}.

\begin{figure}
\begin{center}
\center{\includegraphics[height=9cm,angle=-90]{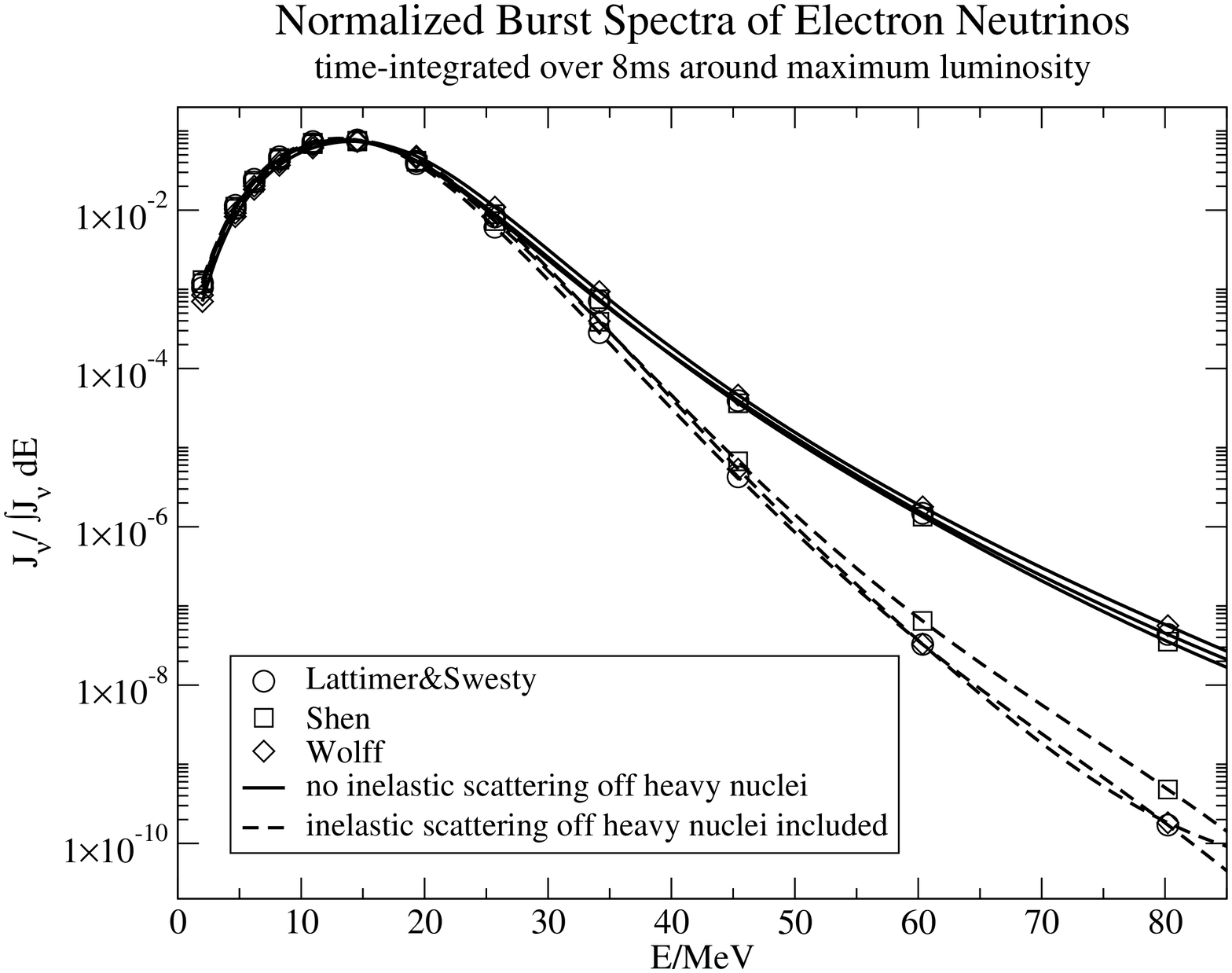}
        \includegraphics[height=9cm,angle=-90]{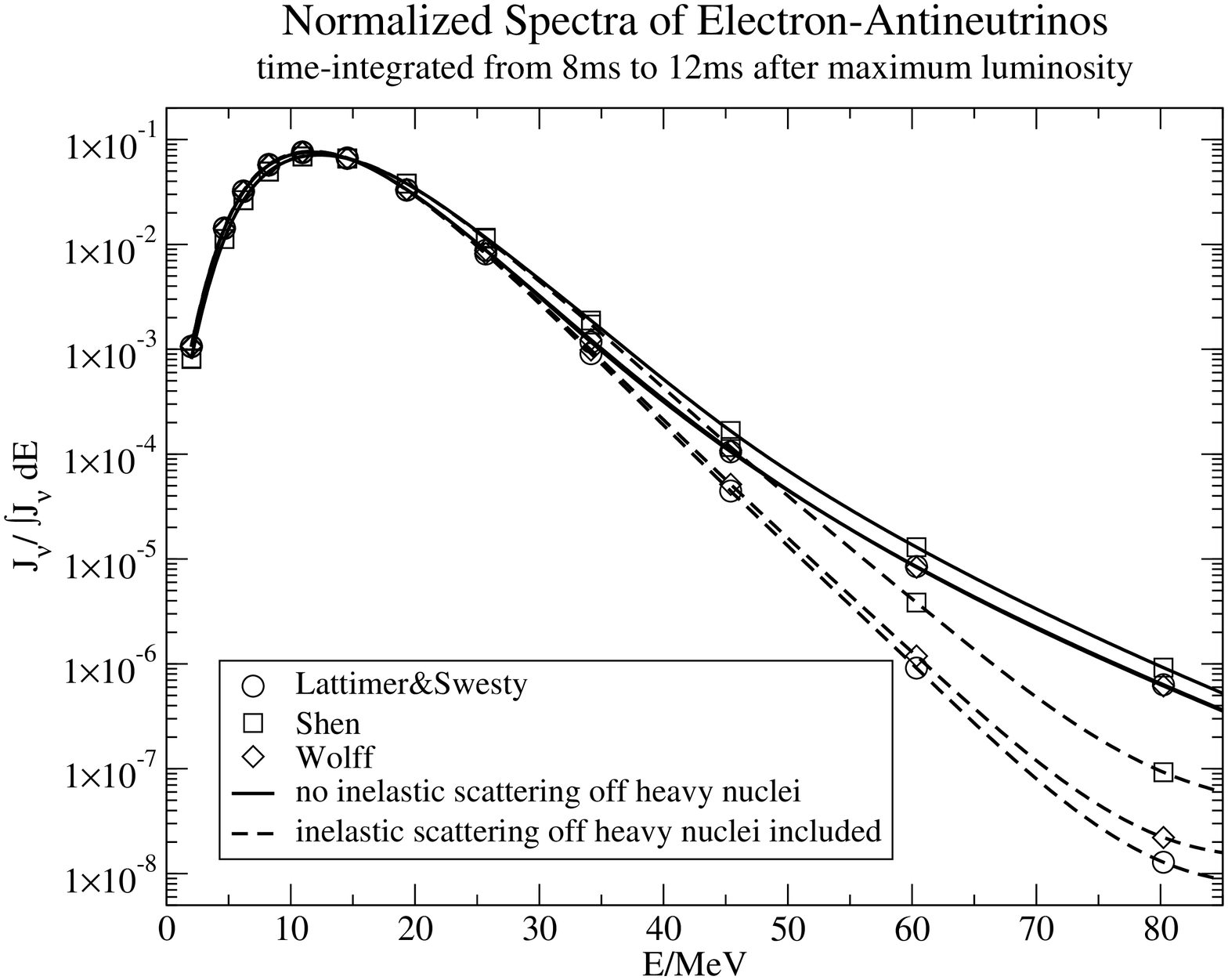}}\\
\center{\includegraphics[height=9cm,angle=-90]{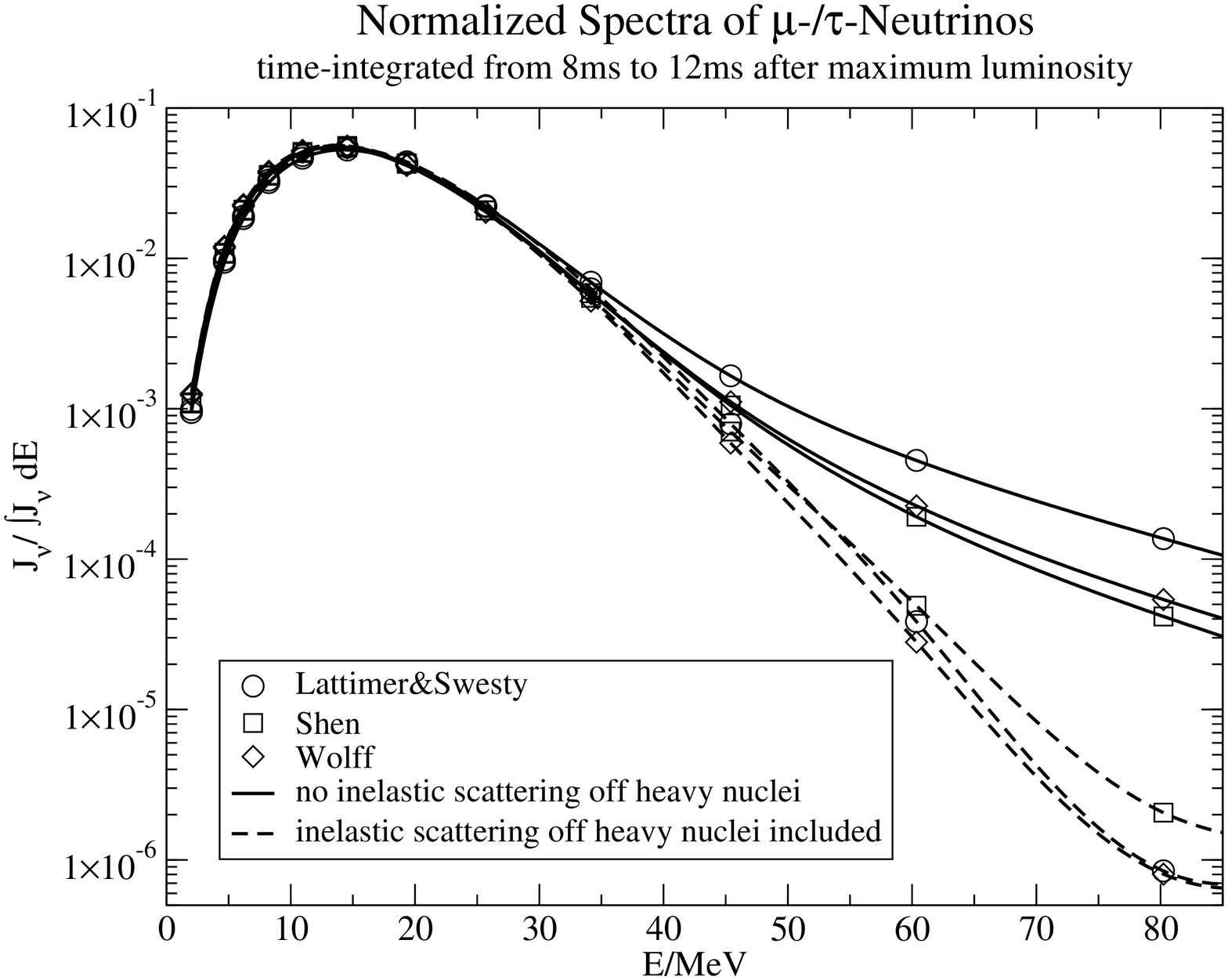}}
      \caption{
      Comparison of the normalized spectra of $\nu_e$, $\bar\nu_e$, and
      $\nu_{\mu,\tau}$ emitted in a time interval around the $\nu_e$ breakout burst
      shortly after bounce, without (solid) and with (dashed) consideration
      of inelastic neutrino-nucleus scattering in supernova simulations with
      the L\&S (open circles), Shen (open squares), and Wolff (open diamonds)
      equations of state (see Sect.~\ref{sec:eos}).
      The flattening of the dashed lines near the high-energy end of 
      the plotted spectra is a consequence of the limited energy range of
      the available rate table for inelastic neutrino-nuclei scatterings and 
      of disregarding these reactions at higher neutrino energies.
      (from \protect\cite{Mueller}).
        \label{fig:nuspectrum}}
	\end{center}
	\end{figure}

Although inelastic neutrino-nucleus scattering contributes to the thermalization
of neutrinos with the core matter, the inclusion of this process
has no significant effect on the collapse trajectories. However,
it increases noticeably the opacity for high-energy neutrinos
after the bounce. As these neutrinos excite the nuclei,
they are down-scattered in energy, in this way significantly reducing the
high-energy tail of the spectrum of emitted supernova neutrinos
(see Fig. \ref{fig:nuspectrum}, \cite{Mueller}).
This makes the detection of supernova
neutrinos by earthbound detectors more difficult, as the neutrino
detection cross section scales with $E_\nu^2$.

\section{Nucleosynthesis in proton-rich supernova ejecta}
\label{sec:nucl}

During the supernova explosion the shells of the star outside
the so-called mass cut are ejected. In this way the elements that
have been produced during the various stellar burning stages
are mixed into the Interstellar Medium. In fact, core-collapse
supernovae are the main contributors of the heavy elements ($A \geq 12$)
in the Universe. For in-depth reviews of nucleosynthesis we refer
to the work of Woosley and collaborators \cite{WW95,Heger02}, and to
\cite{Hashimoto}. These authors also discuss the effect of the
explosion on the nucleosynthesis, as the heating of the material
induced by the passage of the shock wave gives rise to a short
period of very fast (explosive) nuclear reactions which affects
in particular the abundances of elements in the deeper stellar layers.
For many years it has been customary to simulate the explosion and the
effects of the shock wave by igniting a thermal bomb in the star's 
interior or by initiating the explosion by a strong push with a piston.
But recently it has become possible 
\cite{Froehlich05,Pruet05,Froehlich06,Pruet06}
to study the explosive nucleosynthesis using large nuclear networks
coupled to stellar trajectories obtained in state-of-the-art
supernova simulations with sophisticated, energy-dependent
neutrino transport \cite{Janka03,Liebendoerfer} 
(see also Sect.~\ref{sec:1Dmodels})\footnote{Successful explosions 
were obtained in these simulations
by slight modifications of governing parameters. This, however,
did not modify the essential physics that determines the 
nucleosynthesis-relevant neutron-to-proton ratio in the ejecta.}. 
These studies have provided new insights
into the explosive nucleosynthesis and its time evolution, and have led to
the discovery of a novel nucleosynthesis process (the $\nu$p-process)
\cite{Froehlich06a,Pruet06,Wanajo06}, which we will briefly summarize in
this section.

Due to the strong neutrino heating near the newly-formed neutron
star, the ejected matter from the deepest layers of a supernova
has temperatures $T_9 >5$ and is
dissociated into free protons and neutrons. These
building blocks are assembled to nuclei when the matter moves outwards
to cooler regions. This nucleosynthesis process and its final
elemental abundances depend on outflow parameters like
the expansion timescale, the entropy, and the $Y_e$ value of the matter.
As the ejection occurs in extreme neutrino fluences, $Y_e$ is determined
by the competition of electron neutrino and antineutrino absorption on 
neutrons and protons (Eqs.~\ref{eq:nuabs} and \ref{eq:nubarabs}),
respectively. In a rough approximation, the electron fraction that
results from the competition
of these neutron destroying and producing reactions can be expressed by
\begin{equation}
Y_e \sim \eck{1 + \frac
{L_{\bar \nu_e} \langle \epsilon_{\bar \nu_e} \rangle}
{L_{\nu_e} \langle \epsilon_{\nu_e} \rangle}}^{-1}\ ,
\end{equation}
where $\langle \epsilon \rangle$ and $L$ are the average energy 
and the luminosity of the neutrinos or antineutrinos
\cite{Qian93,Qian96}. Although this expression is rather crude
and does not take into account a number of important effects
and corrections, it is still good to capture the basic influence
of neutrino and antineutrino absorptions in setting the 
neutron-to-proton ratio in the outflow\footnote{Fr\"ohlich {\em et al.}\
\cite{Froehlich05}, however, argue that in the early postbounce phase 
the inverse reactions of Eqs.~(\ref{eq:nuabs}) and (\ref{eq:nubarabs})
cannot be ignored and are in fact more important for setting the 
asymptotic $Y_e$-values of the outflowing material.}. 
As anticipated by 
Qian and Woosley \cite{Qian96}, modern supernova simulations show that
the early ejecta are proton-rich ($Y_e > 0.5$), while matter ejected
later may become
neutron-rich, potentially leading to r-process nucleosynthesis
(\cite{Woosley94,Takahashi94,Wanajo01} and references therein).
This might happen seconds later in the baryonic outflow that is 
driven off the surface of the nascent neutron star by the strong heating
of neutrinos radiated from the neutrinosphere.
As the expanding neutron-rich matter in this so-called `neutrino-driven
wind' cools,
nucleons recombine to $\alpha$ particles, some of which can
at even lower temperatures assemble to $^{12}$C by the
reaction sequence
$\alpha(\alpha \mathrm{n},\gamma)^9{\mathrm{Be}}
(\alpha,\mathrm{n})^{12}{\mathrm{C}}$.
The carbon nuclei will then capture additional $\alpha$ particles
and neutrons until iron group nuclei are formed. If free neutrons
are left and the number of neutrons per nucleus is high, the
nuclei can act as ``seed'' for the formation of very heavy elements
by subsequent r-processing in the high-entropy environment of
the neutrino-wind (see Fig.~\ref{fig:snphases}, bottom right panel).

\begin{figure}[tpb!]
\center{\includegraphics[width=0.75\textwidth]{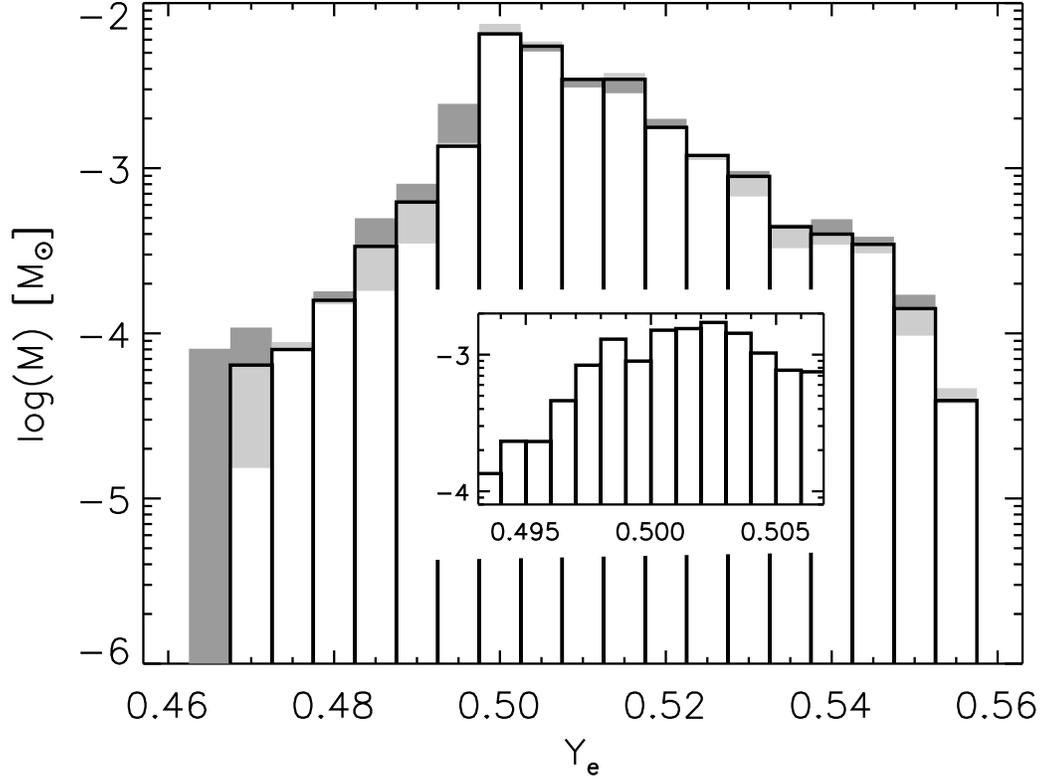}}
\caption{\small
Ejecta mass versus $Y_e$ of neutrino-heated and
processed matter during the convective phase until $\sim$470$\,$ms
post bounce in a 2D simulation of the explosion of a 
15$\,M_\odot$ star \cite{Buras06a}. The plot includes 
`hot bubble ejecta' as well
as some contribution from very early neutrino-wind material.
The insert shows the
region around $Y_e\sim 0.5$ in higher resolution. The
grey shading indicates estimated errors due to the limited spatial
resolution of the two-dimensional simulation (for details, see
\cite{Buras06a}; figure taken from \cite{Pruet05}).} 
\label{fig:MvsYe}
\end{figure}
\begin{figure}[tpb!]
\center{\includegraphics[width=0.49\textwidth]{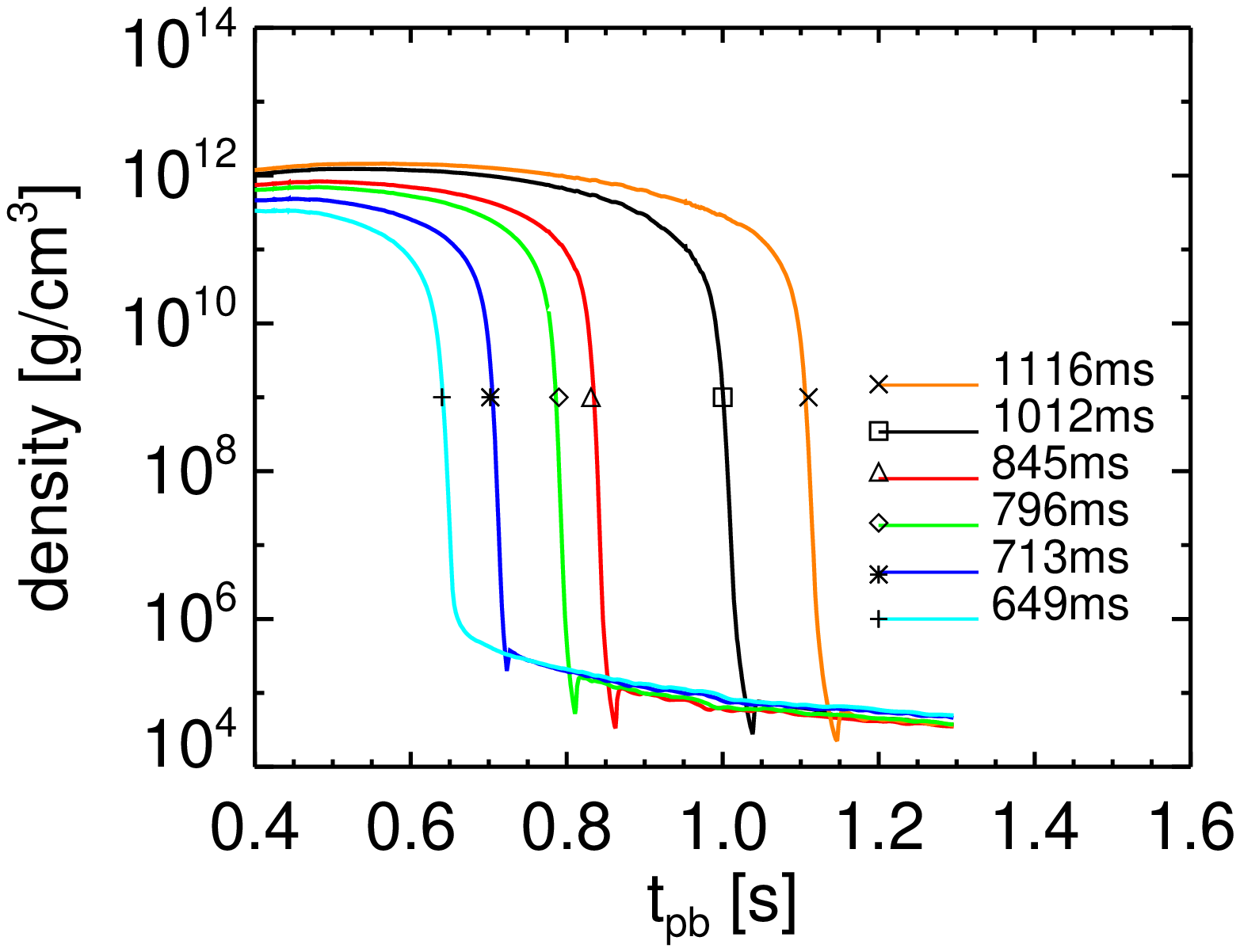}
        \includegraphics[width=0.49\textwidth]{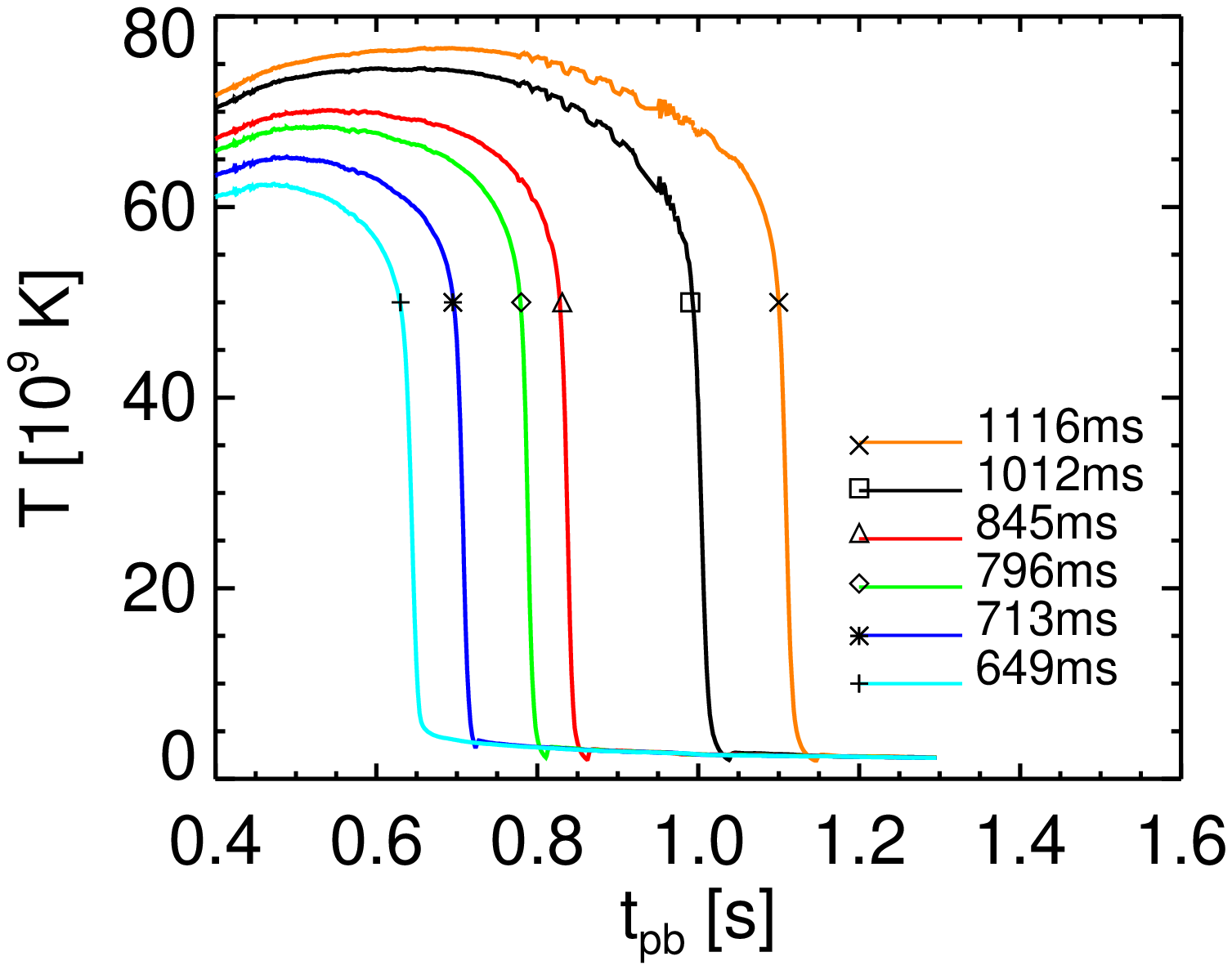}}\\
\center{\includegraphics[width=0.49\textwidth]{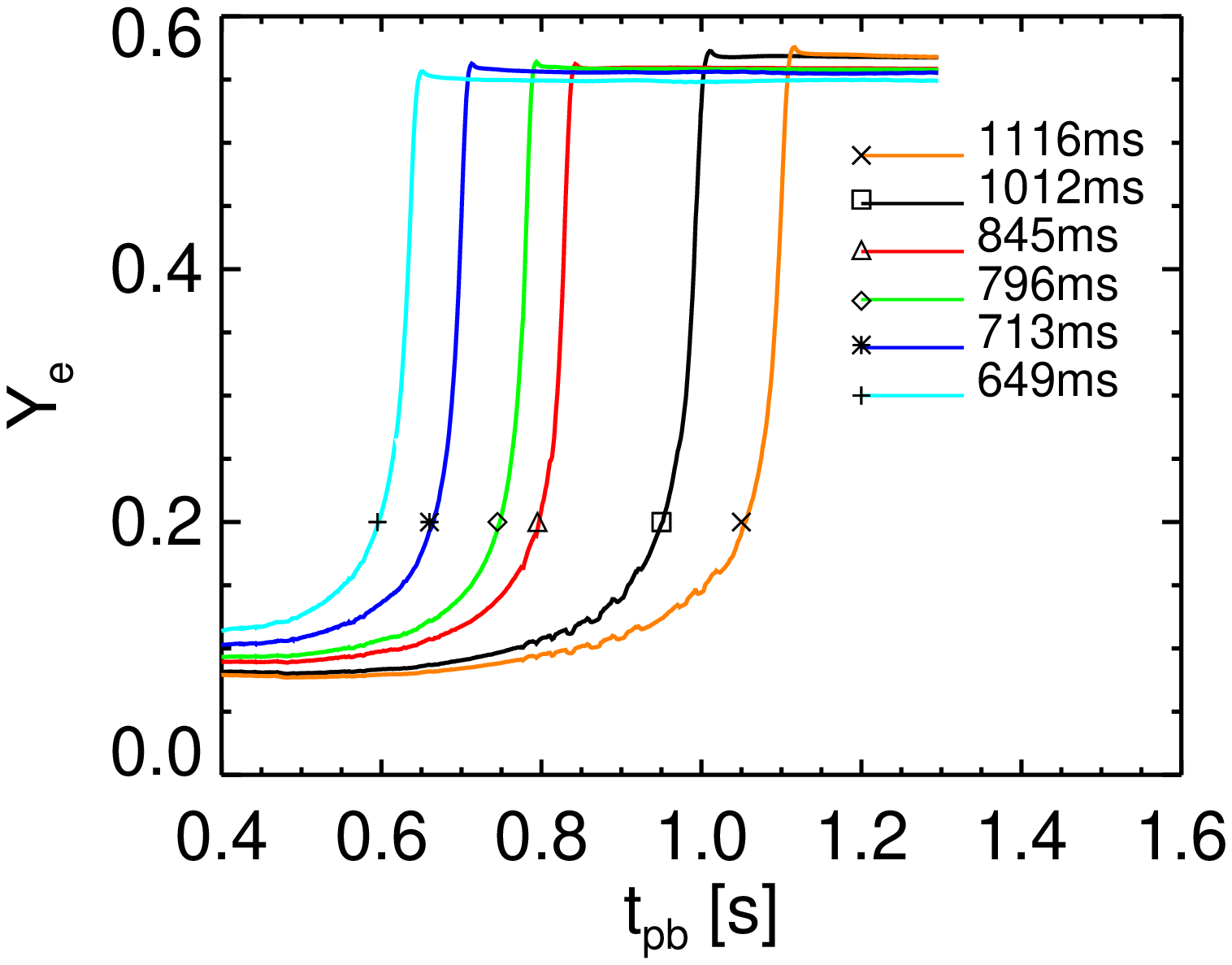}
        \includegraphics[width=0.49\textwidth]{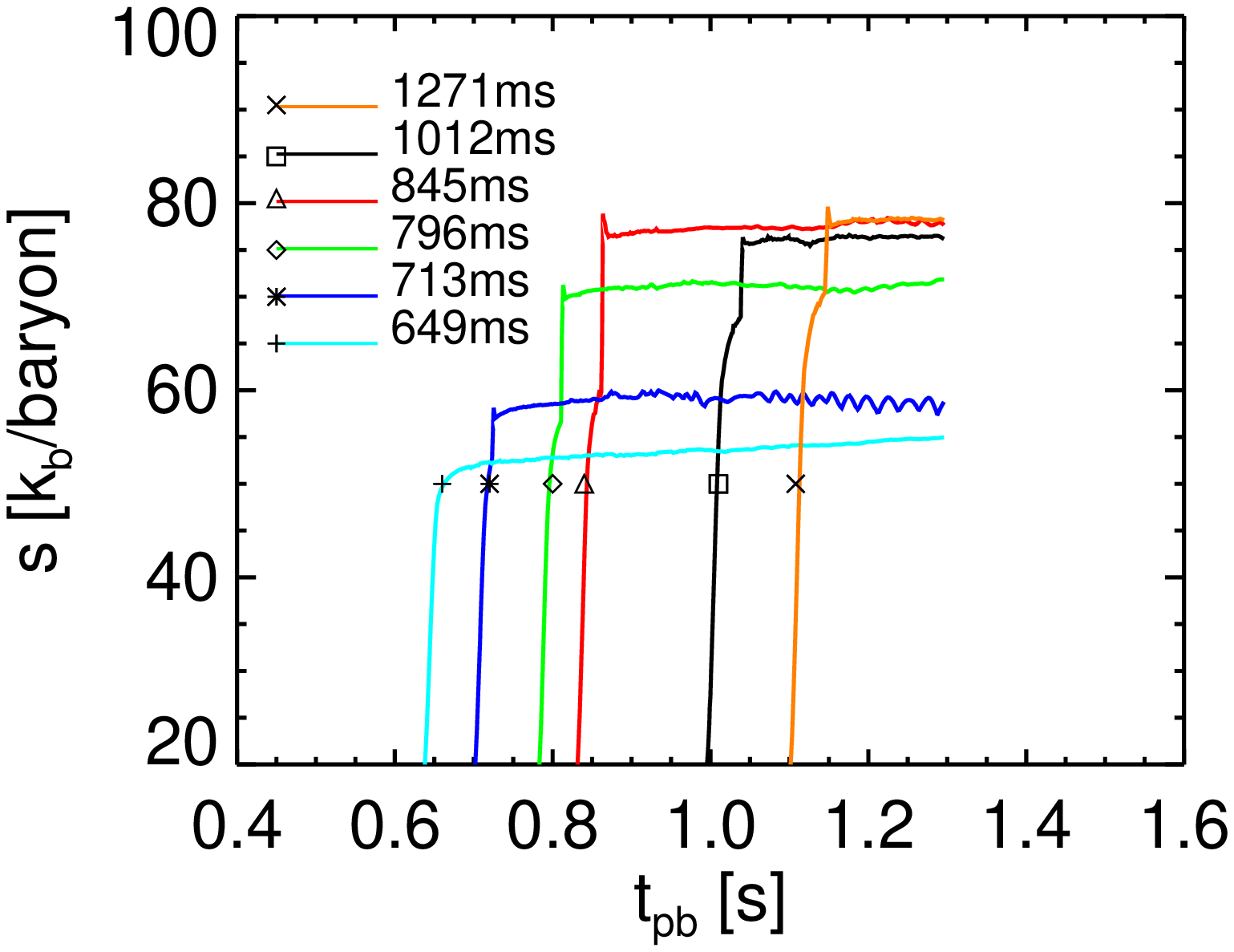}}
\caption{\small
Density, temperature, $Y_e$, and entropy as
functions of postbounce time along the trajectories of mass elements
ejected in the early neutrino-driven wind of the 15$\,M_\odot$
explosion of Fig.~\ref{fig:MvsYe}. The mass cut in this model
develops around an enclosed baryonic mass of 1.41$\,$M$_{\odot}$. 
The elements
first follow the rise of temperature and density in the outer layers
of the contracting neutron star and then enter a phase of very rapid
expansion when they are ejected in the neutrino-driven wind. The
curves are labeled by the time the mass elements
cross a radius of 100$\,$km. The collision with the slower preceding
ejecta occurs through a wind termination
shock and is visible as a non-monotonicity of the density and temperature,
associated with an entropy increase of 10--15$\,k_{\mathrm{B}}$ per 
nucleon (figure taken from \cite{Pruet05}).
}
\label{fig:windshells}
\end{figure}

The early, proton-rich ejecta consist of two components.
On the one hand there is material that comes from the convecting
postshock region and is expelled when the explosion is launched
and the shock accelerates (`hot bubble ejecta'). Much of this
material starts a rather slow expansion from large distances from 
the neutrino-radiating neutron star, is quite dense, has modest 
entropies ($s \sim 15--30\,k_{\mathrm B}$ per nucleon), and is
slightly neutron-rich ($Y_e \ga 0.47$) or moderately proton-rich with 
$Y_e \la 0.52$ \cite{Pruet06} (see Fig.~\ref{fig:MvsYe}).
This material experiences little effect
from neutrino-interactions during nucleosynthesis. This is in
strong contrast to the matter ejected in the second component
that contributes, which is the early neutrino-driven wind
(Fig.~\ref{fig:windshells}). The
wind comes from the surface of the hot neutron star, is strongly
heated by neutrinos, and has to make its way out of the deep
gravitational well of the compact remnant. Therefore the wind
has rather high entropies, short expansion timescales and can
become quite proton-rich ($Y_e \sim 0.57$, \cite{Pruet05,Froehlich06}).

\begin{figure}[htb]
\begin{center}
  \includegraphics[width=0.75\textwidth]{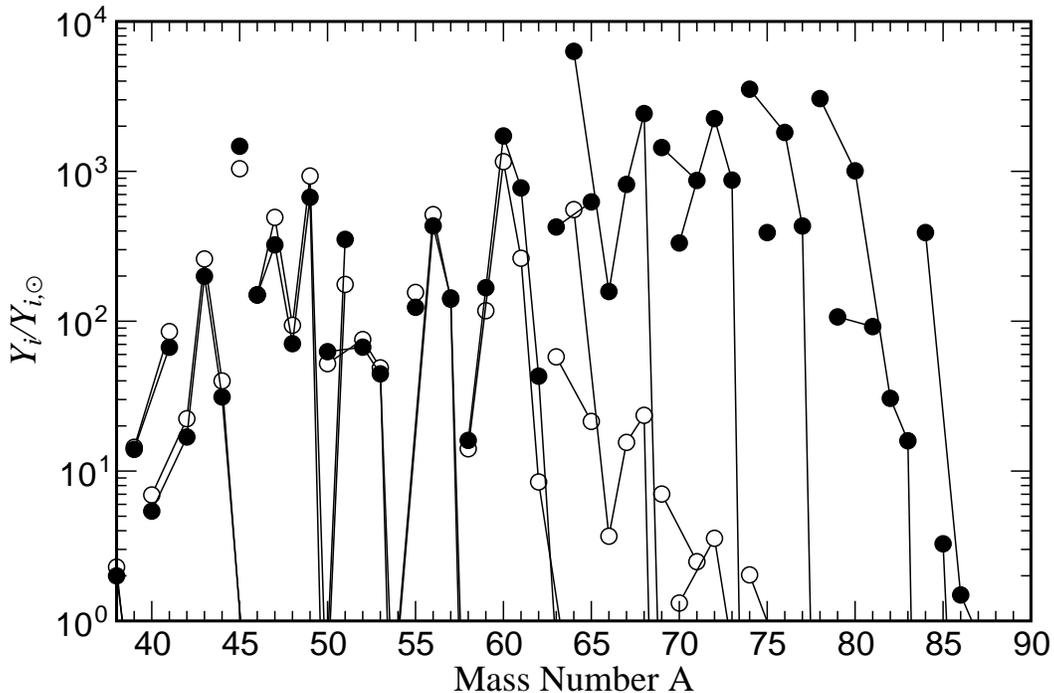}
  \caption{\small
  Elemental abundance yields (normalized to solar)
  for elements produced in the proton-rich environment shortly after
  the supernova shock formation. The matter flow stops at nuclei
  like $^{56}$Ni and $^{64}$Ge (open circles), but can proceed to heavier
  elements if neutrino reactions are included during the network
  (full circles); figure from \cite{Froehlich06}.
  \label{fig:explosive}}
\end{center}
\end{figure}

Moving into cooler regions, protons and neutrons in this wind matter
assemble first into $^{12}$C and then, by a sequence of $(p,\gamma)$,
$(\alpha,\gamma)$ and $(\alpha,p)$ reactions into
even-even $N=Z$
nuclei like $^{56}$Ni, $^{60}$Zn and $^{64}$Ge, with some free protons
left, and with enhanced abundances of
$^{45}$Sc, $^{49}$Ti and $^{64}$Zn solving a longstanding
nucleosynthesis puzzle \cite{Pruet05,Froehlich06}.
In the absence of a sizable neutrino fluence, this nucleosynthesis
sequence resembles explosive hydrogen burning on the surface of an
accreting  neutron star in a binary (the rp-process, \cite{Schatz98})
and matter flow would basically end at $^{64}$Ge as this nucleus has a
$\beta$ halflive ($\approx 64$ s) which is much longer than the expansion
timescale and proton captures are prohibited by the small reaction $Q$ value.
However, the wind material is ejected in the presence of an extreme
flux of neutrinos and antineutrinos. While  $\nu_e$-induced
reactions have no effect as all neutrons are bound in nuclei
with rather large neutron separation energies, antineutrino absorption
on the free protons yield a continuous supply of free neutrons
with a density of free neutrons of $10^{14}$--$10^{15}$~cm$^{-3}$
for several seconds, when the temperatures are in the
range 1--3~GK \cite{Froehlich06a}.
These neutrons, not hindered by Coulomb repulsion, are readily
captured by heavy nuclei in a sequence of ($n,p)$ and ($p,\gamma)$
reactions in this way effectively by-passing the nuclei with long
beta-halflives like $^{64}$Ge and allowing the matter flow
to proceed to heavier nuclei (see Fig. \ref{fig:explosive}).

\begin{figure}[htb]
\begin{center}
\includegraphics[width=0.75\textwidth]{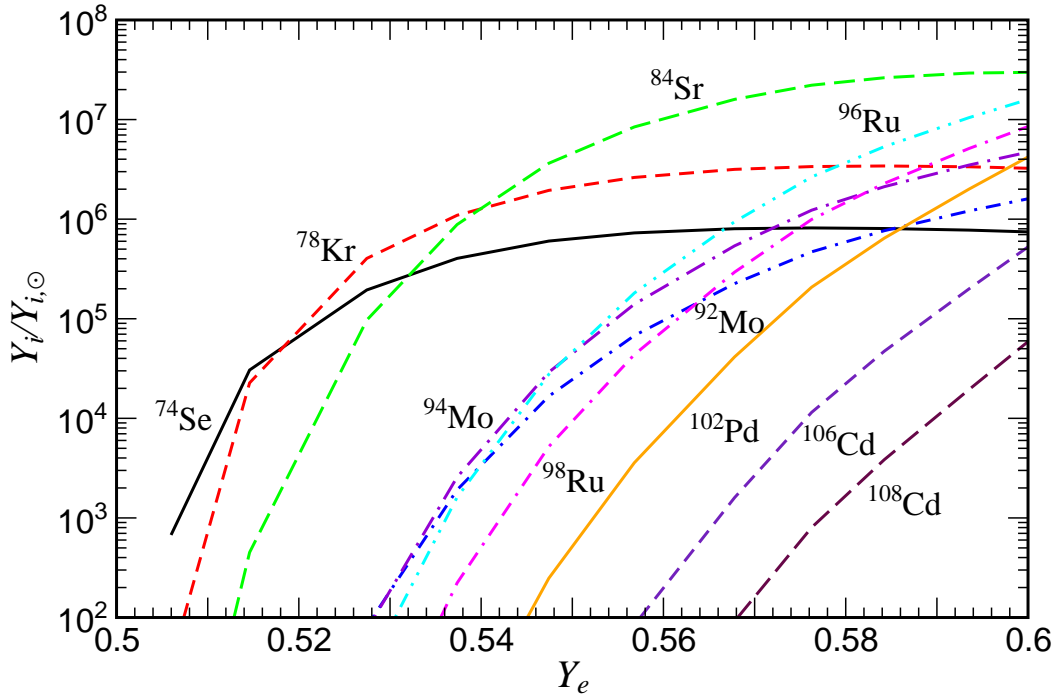}
  \caption{\small
    Light $p$-nuclei abundances in comparison to solar
    abundances as functions of $Y_e$. The $Y_e$-values given are the
    ones obtained at a temperature of 3~GK that corresponds to the
    moment when nuclei are just formed and the $\nu p$-process starts
    to act (from \cite{Froehlich06}).
    \label{fig:ye}}
\end{center}
\end{figure}

Fr\"ohlich {\it et al.} argue that all core-collapse supernovae will eject
hot, explosively processed matter subject to neutrino irradiation and that
this novel nucleosynthesis process (called $\nu$p-process) will operate
in the innermost ejected layers producing neutron-deficient nuclei above
$A>64$ \cite{Froehlich06a}.
However, how far the mass flow within the $\nu$p-process
can proceed, strongly depends on the environment conditions,
most noteably on the $Y_e$ value of the matter
\cite{Pruet06,Froehlich06,Wanajo06}. Obviously the larger
$Y_e$, the larger the abundance of free protons which can be
transformed into neutrons by antineutrino absorption.
(The reservoir of free neutrons is also larger if the
luminosities and average energies of antineutrinos are larger or the
wind material expands more slowly.)
Figure~\ref{fig:ye} shows the dependence of the $\nu$p-process
abundances on the $Y_e$ value of the
ejected matter (similar results are presented in \cite{Pruet06,Wanajo06}).
Nuclei heavier than $A=64$ are only
produced for $Y_e>0.5$, showing a very strong dependence on $Y_e$ in
the range 0.5--0.6.  A clear increase in the production of the light
$p$-nuclei, $^{92,94}$Mo and $^{96,98}$Ru, is observed as $Y_e$ gets
larger. Thus the $\nu$p process offers the explanation for the
production of these light $p$-nuclei, which was yet unknown.
However,  simulations fail to reproduce the observed
abundance of $^{92}$Mo, the most abundant p-nucleus in nature,
which might be related to current uncertainties  in the nuclear
physics involved \cite{Pruet06}. It is, however, observed
that $^{92}$Mo is significantly enhanced in slightly neutron-rich
winds with
$Y_e$ values between 0.46 and 0.49 as they might be found in a later
phase of the explosion (a few seconds after bounce) \cite{Wanajo06}.
Here the $\alpha$-rich freeze-out overabundantly produces $^{90}$Zr
from which some matter flow is carried to $^{92}$Mo by successive proton
captures \cite{Pruet06,Wanajo06}.
The $\nu$p process might also explain the presence of strontium
in the extremely
metal-poor, and hence very old, star HE 1327-2326 \cite{Frebel05}.

\section{Conclusions}
\label{sec:conclusion}

Where does supernova research stand some 15 years after Hans Bethe's
seminal review \cite{Bethe90}? This review was already inspired by 
the once-in-a-hundred-years event of Supernova
1987A, the closest supernova for more than 300 years and certainly the
best observed one ever. Not only did SN1987A impressively confirm our
general picture of stellar core collapse and neutron star birth by
the historical detection of two dozens of the emitted neutrinos. It also
provided us with unambiguous evidence for the importance of hydrodynamic 
instabilities during the explosion, which triggered large-scale radial
mixing in the star and destroyed the spherical onion-shell structure
of chemical composition layers in the progenitor star. But it 
was not clear at the time Hans Bethe wrote his paper, how strongly
this observation was linked to the processes in the deep interior
of the star and causal for the explosion.

Most of the general ideas and concepts outlined in 
Hans Bethe's review are of fundamental relevance and therefore have
retained their validity. Better numerical models and progress in 
theoretical understanding, however, have sharpened our view of many
of the ingredients that play a role in stellar explosions. In fact, 
these most powerful and most violent explosive events in
the universe turn out to be more fascinating and richer the closer
we look. 

New horizons were opened by the discovery of extremely energetic
explosions of massive stars, so-called hypernovae, in connection with
long-duration gamma-ray bursts (for a recent review on this subject, 
see \cite{WooBloo06}). In the collapsar model \cite{Woosley93,MacFad99},
the production of a gamma-ray burst is linked to the collapse of a 
rapidly rotating massive star, in whose core a black hole is formed
and disk accretion releases the energy that drives a pair of 
ultrarelativistic jets along the rotation axis. These jets are responsible
for the production of the gamma-ray burst and its afterglow. The major
part of the released energy powers the off-axis ejection of 
nonrelativistic material in the hypernova explosion. While this scenario
applies only for a small fraction of stellar core collapses, it defines
the factors that might discriminate events leading 
to hypernovae from those producing ordinary supernovae: massive stellar
iron cores, black hole formation, very rapid rotation, and the build-up
of strong magnetic fields in an accretion disk on the one hand and
smaller cores, neutron star formation, slow rotation, and possibly
a non-magnetohydrodynamic explosion mechanism on the other. Nature,
of course, may differentiate less strictly and might produce a variety
of intermediate events in which a combination of processes is at work.

The neutrino-driven explosion mechanism, originally proposed by 
Colgate \& White \cite{ColWhi66}, discovered in its modern version
by Wilson \cite{Wilson}, casted into a framework of equations
that capture its physics by Bethe \& Wilson \cite{BetWil85}
and Bethe in a sequence of papers \cite{Bethe93}, and 
interpreted as a global radial instability of the accretion envelope
between the nascent neutron star and the stalled
shock in the presence of sufficiently strong neutrino heating 
by Burrows \& Goshy \cite{BurrGosh93}, is still considered as
the standard paradigm for explaining how supernova explosions 
begin. But still a convincing demonstration
of the viability of this mechanism and of its robustness in detailed
numerical models has not been achieved. 
An exception may be progenitors stars in the
8--10$\,M_\odot$ range, where new simulations with significantly
improved and now very sophisticated neutrino transport have recently
been able to confirm \cite{Kitaura06}, at least qualitatively, the 
neutrino-driven explosions seen by Mayle \& Wilson \cite{MayWil88}.

The undoubtedly greatest discovery in supernova theory after 1990
was the importance of hydrodynamic instabilities in the supernova 
core already during the very early moments of the explosion
\cite{Her92,Her94,Bur95,JanMue96}. While convection and 
mixing instabilities in supernovae and forming
neutron stars were discussed earlier (e.g.,
\cite{Bethe87,BurrLat88,Burrows87,WilMay88,WilMay93}), the violence
and the implications of
convective instabilities in the neutrino-heated postshock layer
had not been anticipated and was only recognized when the 
first two-dimensional and meanwhile
even three-dimensional \cite{FryWar02,FryWar04} simulations
became available.
The presence of nonradial flow in the layer between gain radius and
shock was found to significantly strengthen the neutrino-energy
deposition there and to support the outward motion of the shock.
It is now generally accepted to play a pivotal role for 
understanding the nature of the explosion mechanism,
although the convectively supported neutrino-driven explosions 
obtained in the early multi-dimensional models with simple, 
grey neutrino diffusion 
\cite{Her94,Bur95,Fry99} could not be reproduced by recent
two-dimensional models with more elaborate, energy-dependent 
neutrino transport \cite{Buras06a,Burrows06a}.
In contrast, the increase of the neutrinospheric luminosities
and mean neutrino energies by convection below the neutrinosphere,
which had been suggested to be the key ingredient for robust 
explosions \cite{Bethe87,Burrows87,BurrFryx92}, 
has not been found to be of great relevance in 
state-of-the-art multi-dimensional models 
\cite{Buras06b,Dessart06,Burrows06a,Burrows06b}. 

More recently, theorists have recognized that not only convection
takes place behind the stalled supernova shock, but that the 
accretion shock can be generically unstable to nonradial deformation
\cite{Blondin03,Scheck04,Blondin06,Foglizzo06,Ohnishi06}.
This so-called SASI 
was found to lead to the preferred growth of low-mode deformation
and violent pulsational shock motion. The corresponding
sloshing mode and anisotropic shock expansion does not only have
the potential of giving support to the onset of neutrino-driven
explosions more strongly than convection alone \cite{Buras06b}.
It was also claimed to instigate strong g-mode oscillations of the
neutron star core that provide acoustic flux and thus power 
acoustically-driven explosions \cite{Burrows06a,Burrows06b}.
In both cases the 
associated dipole and quadrupole asymmetries imprinted on the
developing explosion have the potential
to explain the origin of the observed high space velocities of
pulsars \cite{Scheck04,Scheck06} and of the strong radial 
mixing and high nickel velocities observed to be present in
Supernova 1987A \cite{Kifonidis06}. A three-dimensional 
variant of the SASI may help giving neutron stars their natal
spins \cite{BlondinMezz06}.

A final breakthrough in our understanding of how supernova explosions
work, generally accepted and based on self-consistent models with 
all relevant physics included, however, has not been
achieved yet. Still even fundamental constraining parameters
and ingredients are controversial. Do we understand the neutrino physics
sufficiently well? Are our models correct in predicting the luminosities
and mean energies of the radiated neutrinos? How important is
rotation in the collapsing core? Do magnetohydrodynamic effects
play a crucial role, tapping a large reservoir of free energy of
rotation? Maybe the identification of such key aspects in the
explosion mechanism will require observations that yield more
direct evidence of what is going on in the supernova core
than can be provided by explosion asymmetries, pulsar kicks or 
nucleosynthesis yields. The measurement of neutrino signals
and gravitational waves will be able to yield such information,
but that will require a galactic supernova to happen.

Meanwhile, waiting for such an event, theorists will continue to 
work on improved models, driven by the goal to advance their 
simulations to where the real world is, namely to the
third dimension, and including the effects of general relativity
in full beauty \cite{Shibata05,Ott06}.
Due to the decisive role played by neutrinos in the supernova
core in governing the cooling and neutronization of the nascent
neutron star, determining the thermodynamic conditions around
the mass cut, and in setting the neutron-to-proton ratio in the
supernova ejecta, an increasingly sophisticated incorporation
of the neutrino physics and neutrino transport in future models
is indispensable. In spherical symmetry, neutrino transport can
now be treated by solving the multi-group Boltzmann transport 
problem \cite{Liebendoerfer01,Thomp03,Rampp00}. This has led to
the discovery that proton-rich conditions exist in the early 
neutrino-heated ejecta and thus p-processing can take place
\cite{Pruet05,Froehlich05,Buras06a,Froehlich06,Pruet06} as
reviewed in this paper. In two-dimensional
models ``ray-by-ray-plus'' transport can now handle the full
energy-dependence of the problem with approximations to the
non-radial transport \cite{Buras06a,Buras06b} or, 
alternatively, the transport in all spatial
directions is described by making use of the diffusion 
approximation \cite{Swesty06} but sacrificing the detailed
treatment of the energy redistribution \cite{Burrows06b}. 
Ultimately, these shortcomings will have to be removed and
modeling therefore faces the challenge of solving a five-dimensional,
time-dependent transport problem in 2D and a 6-dimensional such
problem in 3D \cite{Cardall05}.

Further improvements are also mandatory in the important 
microphysics that determines neutrino-matter interactions and
the thermodynamic properties of the dense plasma in the stellar 
core. Weak magnetism corrections, nucleon recoil and thermal
motions in neutrino-nucleon neutral-current and charged-current 
reactions (for a review of these effects, see \cite{Horowitz02}),
as well as $\nu\bar\nu$ pair production in nucleon-nucleon
bremsstrahlung and flavor-coupling neutrino-pair reactions
turned out to have a noticeable impact on the emitted neutrino
spectra, in particular leading to lower mean energies of the 
muon and tau neutrinos radiated during the 
supernova evolution \cite{Keil03,Buras03b}.

In their landmark paper Bethe, Brown, Applegate, and Lattimer
\cite{BBAL} pointed
out the dominance of electron captures on nuclei during the supernova
collapse. In recent years the theoretical and numerical tools
have become available to calculate the rates for the relevant
weak-interaction processes at supernova conditions and, importantly,
the theoretical model calculations could be constrained and guided
by experimental data. As electron capture rates at supernova
conditions are dominated by Gamow-Teller transitions, it has 
been particularly helpful that such transitions for several
relevant nuclei could be measured by (n,p) and (d,$^2$He) 
charge-exchange experiments (e.g. \cite{Alford,Frekers}), 
the latter reaction with
impressive energy resolution. As shell model calculations reproduce
these data --- and the nuclear energy spectra at low excitation 
energies ---
quite well, this many-body model has been applied to calculate
the weak-interaction rates for nuclei that are abundant in the early
phase of the collapse \cite{Langanke00}. 
For heavier nuclei the rates have been 
obtained by another variant of the
shell model (Shell Model Monte Carlo, \cite{Langanke03}), 
which is capable of describing
nuclear properties at non-zero temperatures considering the most 
important nucleon-nucleon correlations. These studies show that
electron captures 
on nuclei are faster than on free protons during the infall stage,
leading to significant quantitative modifications of stellar core
collapse and core bounce, shock formation, and 
early shock propagation \cite{Langanke03,Hix03,Marek06}.

We have also briefly discussed the relevance of the equation of state
for a variety of aspects in the supernova, ranging from the 
composition and deleptonization during collapse, the strength of
the prompt shock, the compactness and temperatures and thus 
neutrino emission properties of the nascent neutron star, to the
stagnation radius of the stalled shock. Both nuclear EoSs currently
widely in use, the Lattimer \& Swesty \cite{Lattimer91} and 
Shen et al.\ \cite{Shen98} EoSs, were derived on the basis of 
nuclear mean-field models. Comparisons between versions with 
soft and stiff phases, respectively, above 
$\sim 10^{14}\,$g$\,$cm$^{-3}$,
and with the Hartree-Fock EoS of Hillebrandt \&
Wolff \cite{Hillebrandt} in 1D as well as 2D supernova simulations
reveal sizable quantitative
differences although such variations do not appear to be decisive
for success or failure of the explosion. However, more work 
needs to be done along these lines, extending the 
simulations to later evolution stages and to a larger space of 
high-density EoS-possibilities.

Thus, improved microphysics has led to a more reliable description of
supernova dynamics. Despite this progress, further improvements are
desirable. Except for the famous $^{12}$C($\alpha,\gamma$)$^{16}$O rate,
which, as Willy Fowler phrased it in his Nobel lecture \cite{Fowler84},
is of 'paramount importance' for the carbon-to-oxygen ratio in the Universe
and for the evolution of stars, including their final fate as supernovae,
there is probably no single nuclear datum which single-handedly
influences the supernova dynamics. Rather it is an overwhelming wealth of
nuclear input required for supernova simulations. This is not necessarily
a disadvantage as it is impossible to directly measure nuclear rates under
supernova conditions, mainly due to the non-zero temperature environment.
Thus, individual nuclear rate uncertainties might average out in the
supernova matter composition, provided such uncertainties are not
systematic (like those discovered for the electron rates as discussed in some
details in this review). Moreover, there is considerable hope that such
systematic inaccuracies, if present, will be detected in the near future by
a concerted effort of improved nuclear models and novel experimental
tools and facilities. In fact, we are currently experiencing
decisive advances in nuclear modelling with the development of
many-body models, which, based on microscopic nuclear wave functions
and realistic (or reasonable) nucleon-nucleon interactions,
describe the low-energy nuclear spectra and properties. Furthermore,
attempts are on the way to combine the progress in nuclear structure
with the dynamics of reaction models. Certainly these models benefit
from ever growing computational power, but they will decisively improve
once the next generation of radioactive ion-beam facilities like
RIBF at RIKEN and the Facility for Antiproton and Ion Research FAIR at GSI
are operational. These facilities will allow the experimental determination
of the properties of many of the unstable nuclei that are essential in many
explosive astrophysical scenarios, including core-collapse supernovae.
In particular, these facilities will advance our understanding 
of the isospin degree of freedom in nuclei and, more specifically for one
of our major themes in this review, experiments with radioactive ion-beams
will improve and constrain the nuclear models required to calculate
the weak-interaction rates on nuclei during the core collapse by
determining the single-particle structure and ground-state Gamow-Teller
distributions for neutron-rich nuclei. 

While radioactive ion-beam facilities will only indirectly contribute to
the nuclear physics during the collapse, they will provide direct and
essential nuclear data for supernova nucleosynthesis studies.
Such crucial improvements range from precise mass measurements of
heavy nuclei with $N \sim Z$, which will determine whether the $\nu p$ process
is indeed capable of producing the light 
p-nuclides \cite{Froehlich06,Pruet06}, to mass and halflive measurements
for many r-process nuclei, hopefully removing the largest uncertainties
in simulations of r-process nucleosynthesis. Certainly we see the break of
dawn of a new and exciting era in nuclear astrophysics in general and
in supernova modelling in particular.

\bigskip\noindent
{\bf Acknowledgements.} 
Our research presented in this review has strongly benefitted from
collaborations with R. Buras, E. Caurier, D.J. Dean, C. Fr\"ohlich, A. Heger,
W. Hillebrandt,
W.R. Hix, R. Hoffman, A. Juodagalvis, M. Liebend\"orfer, O.E.B. Messer,
A. Mezzacappa, E. M\"uller, P. von Neumann-Cosel, F. Nowacki, J. Pruet, 
M. Rampp, 
A. Richter, J.M. Sampaio, F.-K. Thielemann, and S.E. Woosley.
The research in Garching was supported by 
Son\-der\-for\-schungs\-be\-reich~375
on ``Astro-\-Particle Physics'' and 
Son\-der\-for\-schungs\-be\-reich-Trans\-re\-gio~7
on ``Gravitational Wave Astronomy'' of the 
Deut\-sche For\-schungs\-ge\-mein\-schaft. 
Computer time grants at the John von Neumann Institute for 
Computing (NIC) in J\"ulich, the H\"ochst\-leistungs\-re\-chen\-zentrum
of the Stuttgart University (HLRS), the Leib\-niz-Re\-chen\-zentrum
M\"unchen, and the RZG in Garching are acknowledged.

\end{document}